\documentclass[11pt,a4paper]{article}
\pdfoutput=1                    

\usepackage{jheppub}
\graphicspath{{./Pictures/}}

\usepackage[T1]{fontenc} 
\usepackage{import}
\usepackage{bigints}
\usepackage{enumitem}
\usepackage{multirow}
\usepackage{multicol}
\usepackage{subfigure} 
\usepackage{relsize} 
\usepackage{float} 
\usepackage{subfigure} 
\usepackage{import}
\usepackage{bigints}
\usepackage{ifthen}
\usepackage{tkz-euclide}
\usepackage[mode=buildnew]{standalone}
\usepackage{etoolbox}
\patchcmd{\thebibliography}{\section*{\refname}}{}{}{}

\newcommand{\beq}{\begin{equation}}
\newcommand{\eeq}{\end{equation}}
\newcommand{\bba}{\begin{align}}
\newcommand{\eea}{\end{align}}
\newcommand{\Fred}{{\color{red}F}}
\newcommand{\Fblue}{{\color{blue}F}}
\newcommand{\Fmagenta}{{\color{magenta}F}}

\title{
Perturbative Four-Point Functions In Planar $\mathcal{N}=4$ SYM From Hexagonalization 
}
\author[1,2,3]{Frank Coronado}\emailAdd{fcoronado@pitp.ca}

\affiliation[1]{Perimeter Institute for Theoretical Physics, Waterloo, ON N2L 2Y5, Canada
}
\affiliation[2]{Department of Physics $\&$ Astronomy, University of Waterloo, Waterloo, ON N2L 3G1, Canada
}
\affiliation[3]{ICTP South American Institute for Fundamental Research, S\~ao Paulo, SP Brazil 01440-070
}

\abstract{
We use hexagonalization to compute four-point correlation functions of long BPS operators with special R-charge polarizations. We perform the computation at weak coupling and show that at any loop order our correlators can be expressed in terms of single-valued polylogarithms with uniform maximal transcendentality. As a check of our results we extract nine-loop OPE data and compare it against sum rules of (squared) structures constants of non-protected exchanged operators described by hundreds of Bethe solutions. 
}

\begin{document}
\maketitle
\addtocontents{toc}{\protect\setcounter{tocdepth}{2}}

\section{Introduction and Summary}

In a conformal field theory (CFT), four-point correlation functions of local operators are  observables encoding a rich amount of dynamical information and interesting kinematical limits. They give access to scaling dimensions and structure constants of an infinite number of operators through the operator product expansion.  Besides their space-time dependence is not completely fixed by conformal symmetry and allows us to study various kinematical limits. These include the light-cone limit, Regge limit and the bulk-point limit (in holographic CFT) \cite{LightCone,Regge,BulkPoint}, each of them giving access to different physics of the correlator. In particular in holographic CFTs these limits can be mapped to limits of the S-matrix of AdS modes. 

In order to study these kinematical limits it is important to find analytic expressions for these correlators. Therefore many efforts are made to improve perturbative techniques or come up with new non-perturbative methods for these computations. A special theory for which there has been much progress in this endeavor is  $\mathcal{N}=4$ SYM. Specially in the planar limit (large rank of the gauge group $N\to \infty$) and  in particular for the simplest non-trivial correlators: four-point functions of one-half BPS single-trace operators $\langle\mathcal{O}_{K_{1}}\mathcal{O}_{K_{2}}\mathcal{O}_{K_{3}}\mathcal{O}_{K_{4}}\rangle$ with arbitrary  integer scaling dimensions $K_{i}$. 

In the planar perturbative regime, small `t Hooft coupling $g^{2}\to 0$, these correlators are studied using conformal symmetry, dual conformal symmetry and analytic bootstrap conditions. Their results  are given as linear combinations of conformal integrals with known coefficients and integrands but in most cases with unknown integrated expressions, specially at four loops and higher. Up to three loops the integrand was found in \cite{AllThreeLoop} and the recent paper \cite{AllFiveLoop}\footnote{To completely fix the five-loop integrand this reference used the results of our present paper.} provides the planar integrand up to five loops\footnote{At the level of the integrand the case $K_{i}=2$ (stress-tensor multiplet) is known up to 10 loops \cite{10loopsK2}}. 

Alternatively hexagonalization \cite{HexagonalizationI}{\footnote{See also \cite{Cushions} for similar ideas on tesselations introduced at tree level.}} provides a non-perturbative method, any value of the `t Hooft coupling, which relies on the integrability of the two-dimensional effective world-sheet of $\mathcal{N}=4$ SYM.  In this paper we will use this approach to compute some four-point functions explicitly. However, in order to achieve this we will have to restrict to the asymptotic limit $K_{i}\gg 1$ at weak `t Hooft coupling $g^{2}\to 0$. As it will become clear this is the regime where  this method is most efficient.

\begin{figure}[ht]
\centering
\raisebox{-277787sp}{\resizebox{0.34\totalheight}{!}{\includestandalone[width=.8\textwidth]{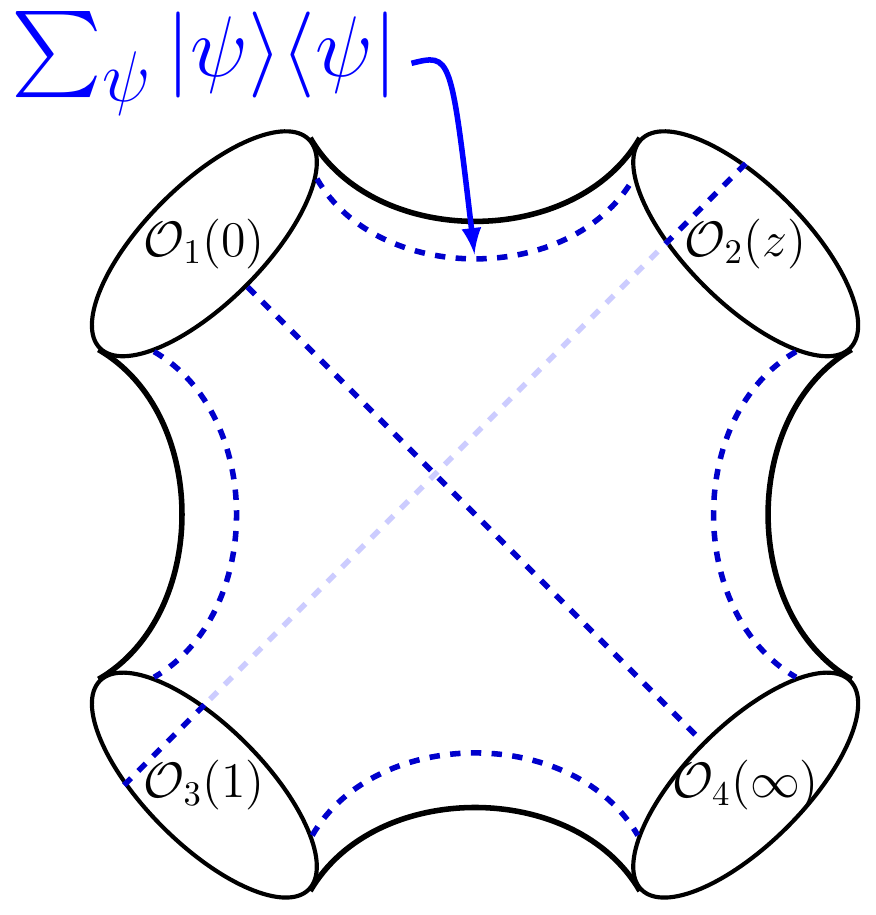}}}
	 \resizebox{2.6\totalheight}{!}{\includestandalone[width=.8\textwidth]{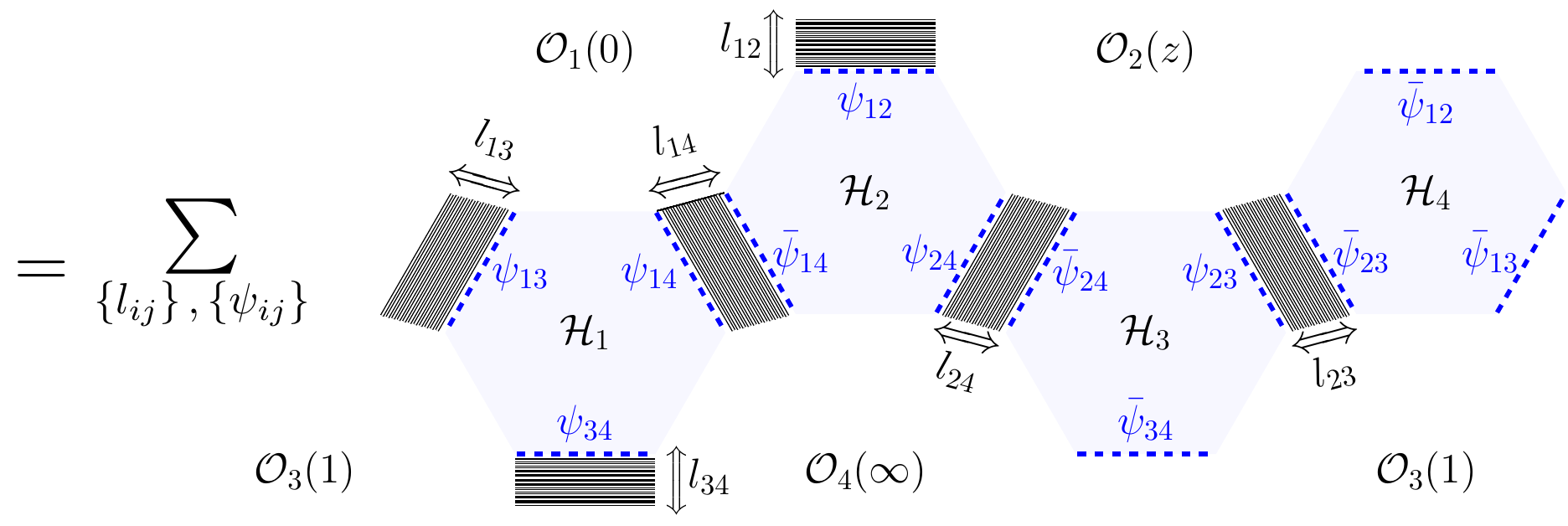}}
	 \caption{On the left, the effective world-sheet that resums all planar Feynman graphs. On the right, hexagonalization: we start with the tree level planar graphs characterized by the Wick contractions or bridge lengths $\{l_{ij}\}$ constrained by $K_{1}=l_{12}+l_{13}+l_{14}$ (and likewise for the other three operators).  Each of these graphs  is tessellated into four hexagons by inserting a complete basis of 2D intermediate states $\mathsf{1}=\sum|\psi\rangle\langle \psi|$ on six mirror cuts performed along each bridge $l_{ij}$. The coupling $(g^{2})$ dependence  is incorporated by the hexagon form factors $\langle \mathcal{H} | \psi_{ij}\rangle$ and chemical potentials such as the Boltzmann factor $e^{-E_{\psi_{ij}}\,l_{ij}}$. To get back the planar four-point function we need to sum over all mirror states $\{\psi_{ij}\}$ for each graph $\{l_{ij}\}$.} 
\label{fig:Hexagonalization}
\end{figure}

In this integrability approach we use a map of the 4D planar four-point function into a finite volume 2D correlator of four hexagon operators\footnote{A n-point correlator in  the 4D space is mapped to a finite volume correlator  of $2(n-2)$ hexagon operators in the 2D effective world-sheet}, with the volume determined by the external scaling dimensions $K_{i}$:
\beq
\langle \mathcal{O}_{K_{1}}\,\mathcal{O}_{K_{2}}\,\mathcal{O}_{K_{3}}\,\mathcal{O}_{K_{4}}\rangle_{4D} \Longrightarrow \langle \mathcal{H}_{1}\, \mathcal{H}_{2} \, \mathcal{H}_{3} \, \mathcal{H}_{4}\rangle_{2D} 
\eeq
This 2D correlator is then computed by a spectral decomposition whose ingredients, the 2D mirror spectrum and the hexagon form factors, are known at finite coupling thanks to super-symmetry and integrability.  This provides a series expansion where each element gives a finite coupling contribution but there are an infinite number of them and a resummation has to be performed to get back the four-point function with finite $K_{i}$.  See figure \ref{fig:Hexagonalization} for a representation of hexagonalization and the parameters entering this form factor expansion.

The resummation of this form factor series is very challenging at finite coupling. However this changes as we go to weak coupling where hexagonalization becomes suitable for the perturbative study of correlators of long operators. This is evident when analyzing the Boltzmann factors that weight the contributions of the intermediate 2D states $\psi_{ij}$. These depend on the energy $E_{\psi_{ij}}$ and the bridge length $l_{ij}$ or number of Wick contractions  between operators $\mathcal{O}_{i}$ and $\mathcal{O}_{j}$: $e^{-E_{\psi_{ij}}\,l_{ij}}$. At weak coupling these weights scale as $\left(g^{2}\right)^{l_{ij}}$ for the non-trivial lowest states and even more suppressed for heavier states. As a consequence at a given loop order and for large enough $l_{ij}$ all intermediate states are projected out and only the vacuum propagates on this bridge. In the strict limit $l_{ij}\to \infty$ this remains valid at any loop order. Thus the only contributions come from intermediate states propagating through small bridges. 

\begin{figure}[ht]
\centering
\raisebox{-277787sp}{\resizebox{1.3\totalheight}{!}{\includestandalone[width=.8\textwidth]{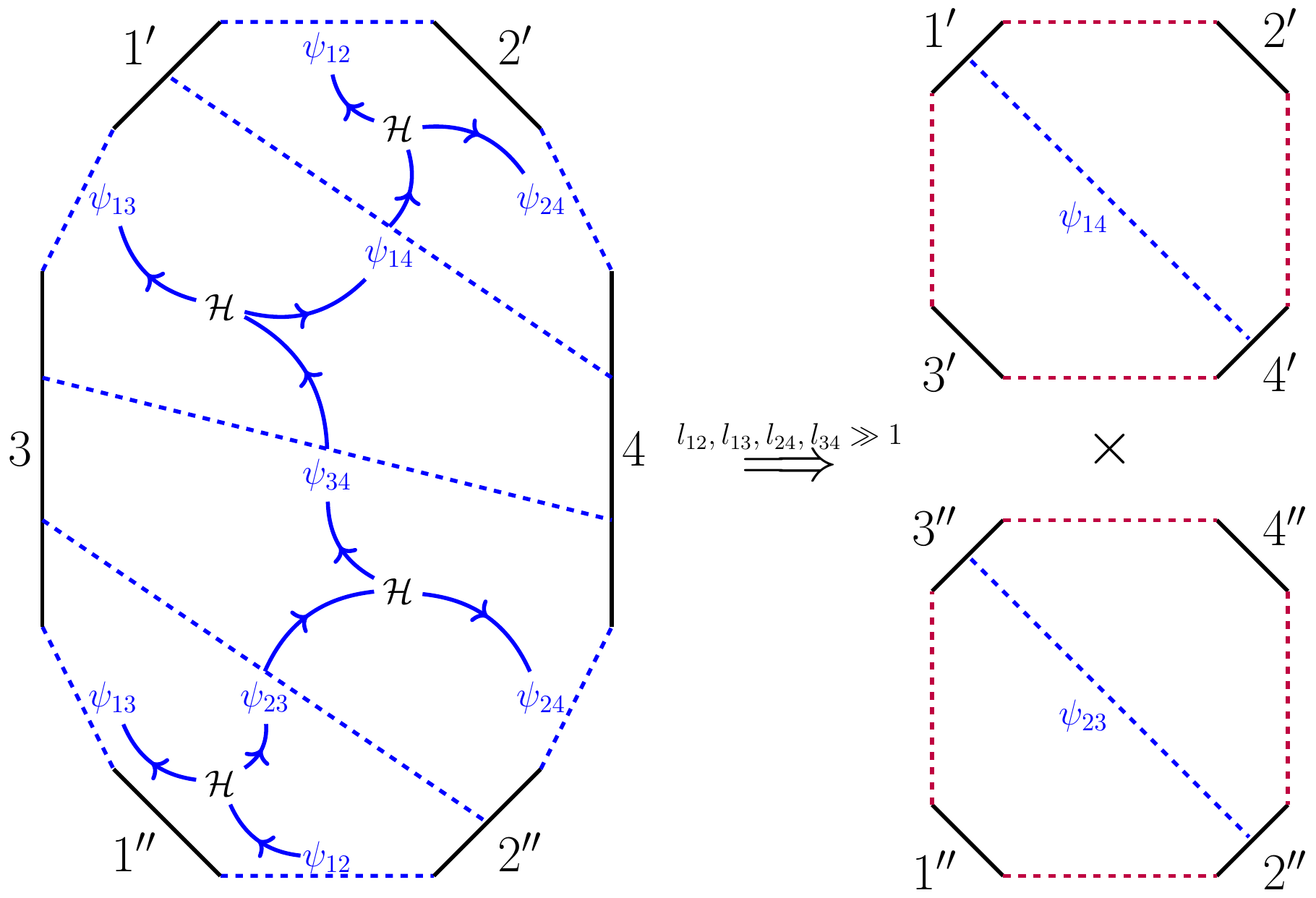}}}\caption{Sequence of transitions cause by acting with hexagon operators $\mathcal{H}$. Each state $\psi_{ij}$ picks a  Boltzmann factor $e^{-E_{\psi_{ij}}\,l_{ij}}$ when propagating through the bridge connecting operators $i$ and $j$. A sequence of transitions such as $\psi_{12}\overset{\mathcal{H}}{\to} \psi_{23} \overset{\mathcal{H}}{\to} \psi_{34} \overset{\mathcal{H}}{\to} \psi_{14}\overset{\mathcal{H}}{\to} \psi_{12}$  wraps around operators $\mathcal{O}_{1}$ and $\mathcal{O}_{3}$ as well as $\mathcal{O}_{2}$ and $\mathcal{O}_{4}$. It is in general given by  a complicated contraction of the four tensors $\langle \psi_{ij}|\mathcal{H}\rangle$. This simplifies when we have large bridges $l_{12},l_{13},l_{24},l_{34}\gg 1$  which only let the vacuum propagate through them. This leads to a factorization of the present polygon into two octagons $\mathbb{O}_{l_{14}}$ and $\mathbb{O}_{l_{23}}$, composed by the two top and the two bottom hexagons respectively. } 
\label{fig:TwoOctagons}
\end{figure}

In this paper we accompany the limit $K_{i}\gg 1$ with specific choices of $R$-charge polarization for the external operators. This allows us to achieve the greatest simplification in the perturbative regime, see figure \ref{fig:TwoOctagons}. This consists on fixing a large $R$-charge  flowing between pairs of operators in order to enforce a large bridge stretching between them. Such that we are left only with two non-adjacent small bridges, the only two that can host non-trivial intermediate states.

We consider two instances of four-point functions of polarised operators with  equal dimension $K$. At tree level these correlators are given schematically by the Wick contractions:
\\\\
\beq\label{eq:IntroFour}
\mathbb{S}^{(0)}\sim \raisebox{-2877787sp}{   \resizebox{.25\totalheight}{!}{\includestandalone[width=.8\textwidth]{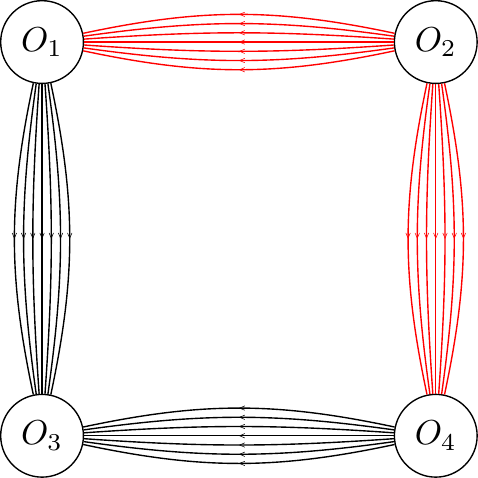}}  } \qquad \text{and}\qquad \mathbb{A}^{(0)}\sim \sum_{l=0}^{K/2}\raisebox{-4277787sp}{  \resizebox{.42\totalheight}{!}{\includestandalone[width=.8\textwidth]{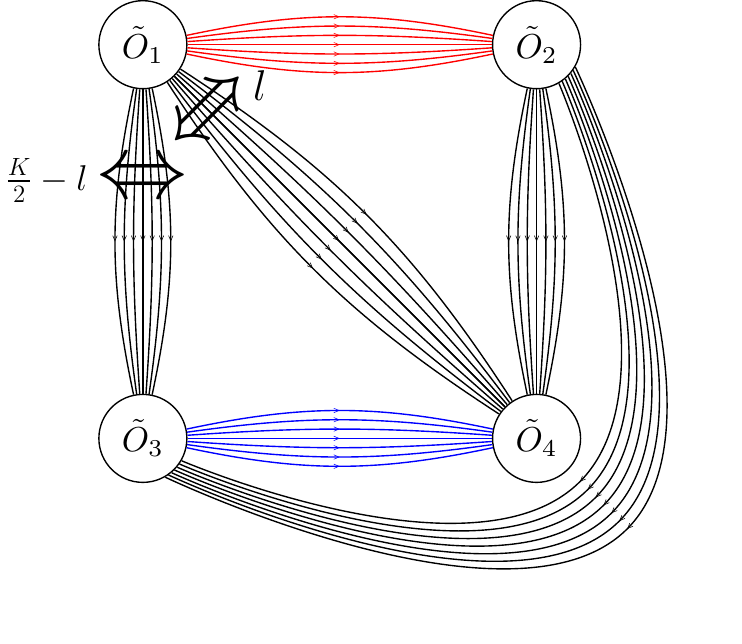}} }
\eeq
where the colors represent the complex scalar $R$-charges {\color{red}$X$}, {\color{blue}Y} and $Z$.

As hexagonalization prescribes, when turning on the coupling we dress the tree level graph(s) with mirror particles. In the limit $K\gg 1$, our polarized correlators only receive contributions from graphs where the simplification described in figure \ref{fig:TwoOctagons} applies. Schematically this gives the results:
\beq
\mathbb{S} \sim \mathbb{O}^{2}_{l=0} \qquad \text{and} \qquad \mathbb{A} \sim \sum_{l=0}^{K/2}\,\mathbb{O}_{l}^{2}
\eeq
where $\mathbb{O}_{l}$ denotes the octagon form factors depicted on the right panel of figure \ref{fig:TwoOctagons}. These are composed of two hexagons glued by summing over all intermediate states along a common edge.

The main goal of this paper is to first provide a finite coupling representation for the octagon form factor $\mathbb{O}_{l}$ and second, at weak coupling, a useful representation for the octagon and the polarized correlators in \eqref{eq:IntroFour},  in terms of analytic functions at arbitrary loop order. To achieve this we take the following steps:

\subsubsection*{Outline}
\qquad In section \ref{sec:TheOctagon} we construct the finite coupling octagon form factor by gluing two hexagons. We show how the matrix parts of the hexagon form factors simplified when contracted. This allow us to express the final result in terms of $n$-particle contributions, each of them containing $n$ sums and $n$ integrals to perform on an abelianized integrand. 

In section \ref{sec:WeakOctagon} we take the weak coupling limit ($g^{2}\to0$) of the octagon and show the structure it takes when expressed in terms of well known Ladder integrals at each loop order. In appendix \ref{app:EfficientResidue} we provide technical details on how to efficiently perform the $n$-particle integrals by residues. We also provide explicit expressions for the non-vanishing integrals up to nine loops in appendix \ref{app:NineLoopIntegrals}. 

In section \ref{sec:WeakPolarisedFour} we use hexagonalization to construct the polarized four-point functions. We start by reviewing the hexagonalization prescription focusing on the skeleton graphs. Then we work out in detailed how to express the \textit{simplest} and \textit{asymptotic} four point functions in terms of the octagons of sections \ref{sec:TheOctagon} and \ref{sec:WeakOctagon}. In this way we obtain expressions for these correlators up to arbitrary loop order provided the external scaling dimensions are large enough. In appendix \ref{app:CrossRatios} we provide a  comprehensive list of cross ratios we use throughout  this section and the next.

In section \ref{sec:OPEcheckNineLoops} we test our results up to nine loops. For this we extract nine-loop OPE data from  our octagon prediction for the \textit{asymptotic} correlator. Then we reproduce these OPE coefficients by solving Beisert-Staudacher equations and using the BKV hexagon formula for three-point functions. The agreement is perfect. 

This section is supplemented by appendix \ref{app:Blocks} containing the 4D long super-conformal blocks and their OPE expansion in radial coordinates, appendix \ref{app:WeakOPE} that shows the organization of the OPE data into sum rules at weak coupling, appendix \ref{app:LargeNineOPE} containing some explicit nine-loop sum rules and appendix \ref{app:BetheSolutions} about  $\mathfrak{sl}(2)$ Bethe solutions.

We conclude in section \ref{sec:Conclusions} with a discussion of our results and  a list of ongoing and future directions.

\section{The octagon}\label{sec:TheOctagon}

In this section we construct the octagon form factor which will serve as a building block of our polarized four-point functions.  We consider an octagon with four physical and four mirror edges with the corresponding BMN vacuum at each edge. 
\begin{figure}[ht]
\centering
\resizebox{.8\totalheight}{!}{\includestandalone[width=.8\textwidth]{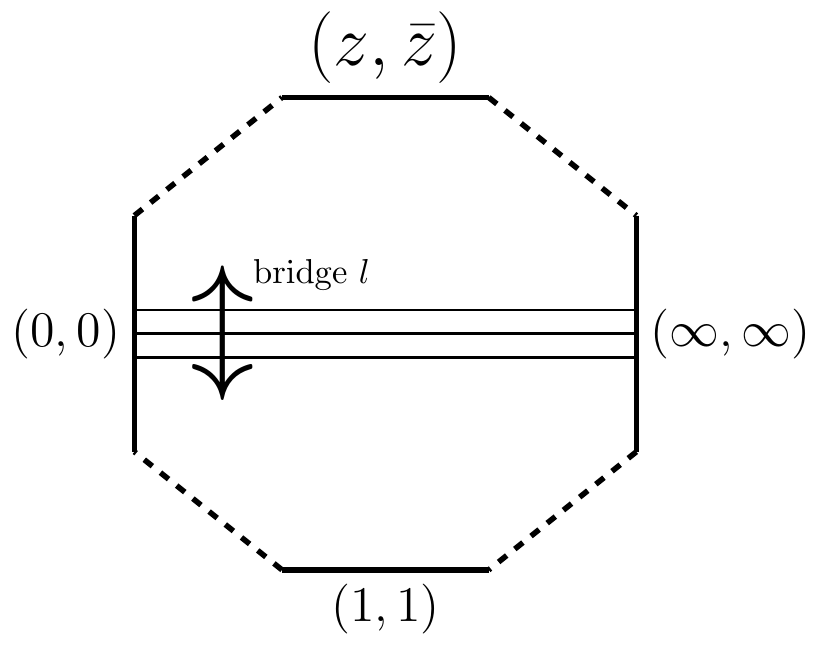}}
\caption{octagon}
\label{fig:Octagon}
\end{figure}
This octagon depends on the bridge length $l$ given by the number of Wick contractions between physical edges at $(0,0)$ and $(\infty,\infty)$. It also depends on the spacetime cross ratios $(z,\bar{z})$ and  the $R$-charge cross ratios $(\alpha,\bar{\alpha})$, which in figure \ref{fig:Octagon} come as the coordinates of the top physical edge. 

The octagon can be decomposed into two hexagons by means of the insertion of a complete set of states along a mirror cut. We consider this cut to stretch between the operators connected  by the bridge length $l$. Gluing back the two hexagons by resumming all intermediate states $\psi$ we recover the octagon as:
\bba\label{eq:OctagonSum1}
\mathbb{O}_{l}\, &= \sum_{\psi}\quad \raisebox{-6077787sp}{\resizebox{.4\totalheight}{!}{\includestandalone[width=.8\textwidth]{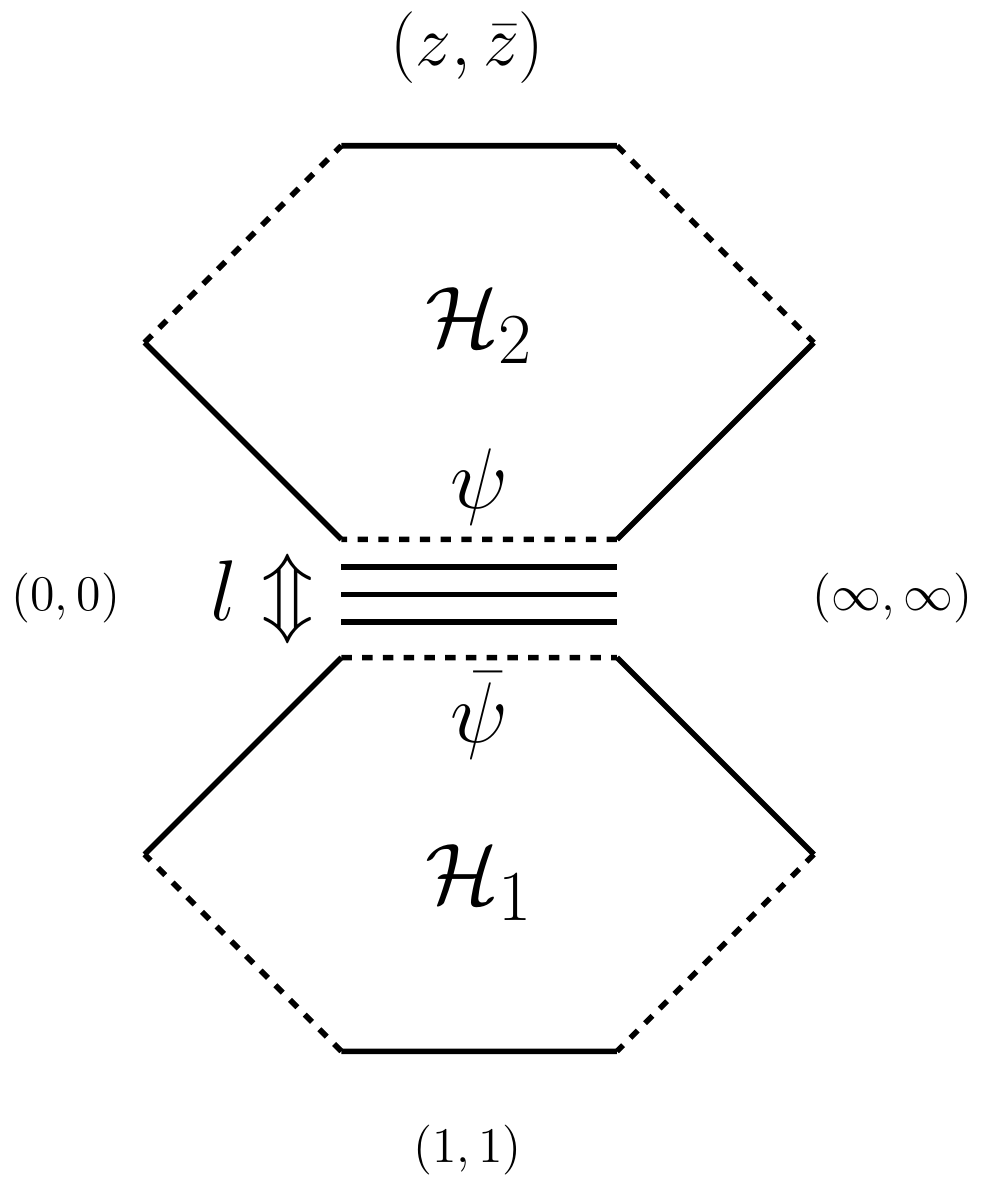}}} \quad = \,\sum_{\psi}\, \langle \mathcal{H}_{2}\,|\psi\rangle\,\mu_{\psi}\, e^{-E_{\psi} l}\,\langle \, \psi |\,\mathcal{H}_{1} \rangle
\end{align}
The measure $\mu_{\psi}$ gives the cost to create the state $\psi$ at the mirror cut. The Boltzmann factor $e^{-E_{\psi}\,l}$ which controls this expansion weighting each state according to its energy $E_{\psi}$ and the bridge length $l$ it propagates through.  We also use the short-hand notation $\langle\mathcal{H}_{i}|\psi \rangle$ (or the conjugate $\langle \psi | \mathcal{H}_{i}\rangle$) for the hexagon form factors which have a single non-trivial state $\psi$. 

Both hexagon operators $\mathcal{H}_{1}$ and $\mathcal{H}_{2}$ can be brought to the standard hexagon $\mathcal{H}$ defined in appendix \ref{app:hexagon}, which is independent of the cross ratios and only depends on the coupling. This is achieved by means of a similarity transformation which when acted upon the mirror states brings new chemical potentials:
\bba\label{eq:OctagonSumPsi}
\mathbb{O}_{l}(z,\bar{z},\alpha,\bar{\alpha})\, &= \,\sum_{\psi}\, \langle \mathcal{H}\,|\psi\rangle\,\mu_{\psi}\, e^{-E_{\psi}\,l}\,e^{i\,p_{\psi}\,\log(z\bar{z})}\,e^{i\,L_{\psi}\phi}\,e^{i \,R_{\psi}\, \theta}\,e^{i \,J_{\psi}\, \varphi}\langle \, \psi |\,\mathcal{H}\rangle
\end{align}
The cross ratios now enter through the angle variables:
\beq
\phi =-\frac{i}{2}\,\log\left(\frac{z}{\bar{z}}\right) \qquad \theta=-\frac{i}{2}\,\log\left(\frac{\alpha}{\bar{\alpha}}\right) \qquad \varphi = \frac{1}{2}\, \log\left(\frac{\alpha\bar{\alpha}}{z\bar{z}}\right)
\eeq
and the correspondent conjugate charges are: the angular momentum $L_{\psi}$, the R-charges $R_{\psi}$ and $J_{\psi}$. Including also the momentum $p_{\psi}$ conjugate to translation. 

Using the details about the multi-particle mirror basis $\psi$ and the hexagon form factors $\langle\mathcal{H}|\psi\rangle$ provided in appendix \ref{app:MoreOctagon}, we can express the octagon as a sum over the number of particles $n$. Including an integral over the rapidity $u_{i}$ and a sum over the bound state number  $a_{i}$ for each particle. More precisely this is:
\beq\label{eq:FinalOctagon}
\mathbb{O}_{l}(z,\bar{z},\alpha,\bar{\alpha}) \,=\, 1+\sum_{n=1}^{\infty}\mathcal{X}_{n}(z,\bar{z},\alpha,\bar{\alpha}) \times \mathcal{I}_{n,l}(z,\bar{z})
\eeq
where the unity stands for the vacuum contribution and the factor $\mathcal{X}_{n}$ that we name the character is given by:
\beq\label{eq:Xn}
\mathcal{X}_{n}(z,\bar{z},\alpha,\bar{\alpha}) = \frac{\left(\mathcal{X^{+}}\right)^{n}+\left(\mathcal{X^{-}}\right)^{n}}{2}  
\eeq
with:
\beq\label{eq:characters}
\mathcal{X}^{+}=-\frac{(z-\alpha)(\bar{z}-\alpha)}{\alpha} \quad \text{and} \quad \mathcal{X}^{-}=-\frac{(z-\bar{\alpha})(\bar{z}-\bar{\alpha})}{\bar{\alpha}}
\eeq

 The $n$-particle  sum and integral $\mathcal{I}_{n,l}$ is given by:
\bba\label{eq:Inl}
\mathcal{I}_{n,l}(z,\bar{z}) &= \frac{1}{n!}\sum_{a_{1}=1}^{\infty}\cdots \sum_{a_{n}=1}^{\infty}\,\int du_{1}\cdots \int du_{n}\, \prod_{j=1}^{n}\bar{\mu}_{a_{j}}(u_{j},l,z,\bar{z}) \times \prod_{j<k}^{n}\,P_{a_{j}a_{k}}(u_{j},u_{k})
\end{align}
The integrand contains the coupling dependence and is composed as follows: 
\begin{itemize}
\item The one-particle effective measure $\bar{\mu}$ where we package the chemical potentials for each particle:
\beq\label{eq:Emeasure}
\bar{\mu}_{a}(u,l,z,\bar{z}) = 
\frac{1}{\sqrt{z\bar{z}}}\,\frac{\sin a\phi}{\sin\phi}\times\,\mu_{a}(u) \times e^{-E_{a}(u)\,l}\times (z\bar{z})^{-i\,p_{a}(u)}
\eeq
where the one-particle measure $\mu_{a}(u)$,energy  $E_{a}(u)$ and  momentum $p_{a}(u)$ are defined in \eqref{eq:weakMeasure}. 
\item The(abelian) symmetric product of  two-particle hexagon form factors $P_{ab}(u,v)$  defined in \eqref{eq:symProd}.
\end{itemize}

Two comments are in order regarding the simplicity of the integrand \eqref{eq:Inl} and the structure of the character \eqref{eq:Xn}:
\begin{itemize}
\item \textbf{The matrix part simplifies:} The hexagon form factors are in general complicated tensors with as many $\mathfrak{su}(2|2)^{2}$   flavour indexes as the number of particles. For a $n$-particle state this matrix part is constructed multiplying $n$ copies of the $\mathfrak{su}(2|2)$ Beisert's $S$-matrix (see appendix \ref{app:hexagon}). Fortunately when contracting $\langle \psi |\mathcal{H}\rangle$ and $\langle \mathcal{H} | \psi\rangle$ these tensors simplified as shown in figure \ref{fig:3characters}, thanks to the unitarity of the $S$-matrix:
\beq\label{eq:Sunit}
\raisebox{-3577787sp}{\resizebox{.7\totalheight}{!}{\includestandalone[width=.8\textwidth]{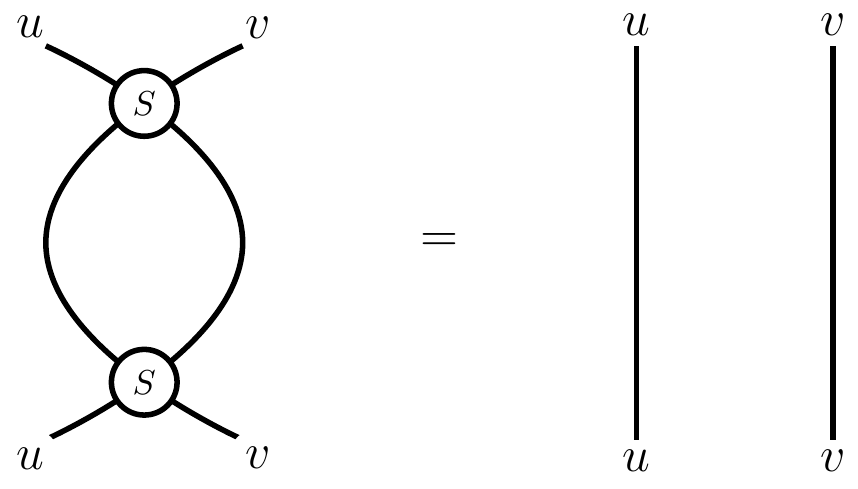}}}
\eeq
This simplification does not happen when the hexagons have other non-trivial states on different edges. These non-trivial cases will not show up in our polarized four-point functions of section \ref{sec:WeakPolarisedFour}. 
\item \textbf{Prescription for character $\mathcal{X}_{n}$:}  A complete knowledge of the mirror basis is an essential ingredient to carry on with hexagonalization. However its construction as representations of $\mathfrak{psu}(2|2)^{2}$ comes with ambiguities associated to the action of super-charges which, in principle, can introduce an arbitrary number of so-called $Z$-markers (see section 5.1 in \cite{HexagonalizationI}). These markers represent the insertion or deletion of the vacuum field $Z$ and change the $R$-charge $J_{\psi}$ in \eqref{eq:OctagonSumPsi}.

 We still lack a physical interpretation of how to correctly take them into account in the mirror basis. Nevertheless an empirical prescription was proposed in \cite{HexagonalizationI} for the one-particle states and generalized to multi-particle states in \cite{HexagonalizationII}. We are instructed to include these markers in two different ways and then take the average. This results in the average that takes place in the definition of the character $\mathcal{X}_{n}$, see equation \eqref{eq:Xn}.

This prescription has so far only been tested at one loop and it is important to remark that at this order  other prescriptions could work. In order to get rid of this ambiguity, in this paper, we will perform a test as far as nine loops using our polarized four-point functions. At this order we receive for the first time a contribution from  $\mathcal{X}_{n=3}$ and this will give us strong evidence for this prescription. 
\begin{figure}[ht]
\centering
\resizebox{4\totalheight}{!}{\includestandalone[width=.8\textwidth]{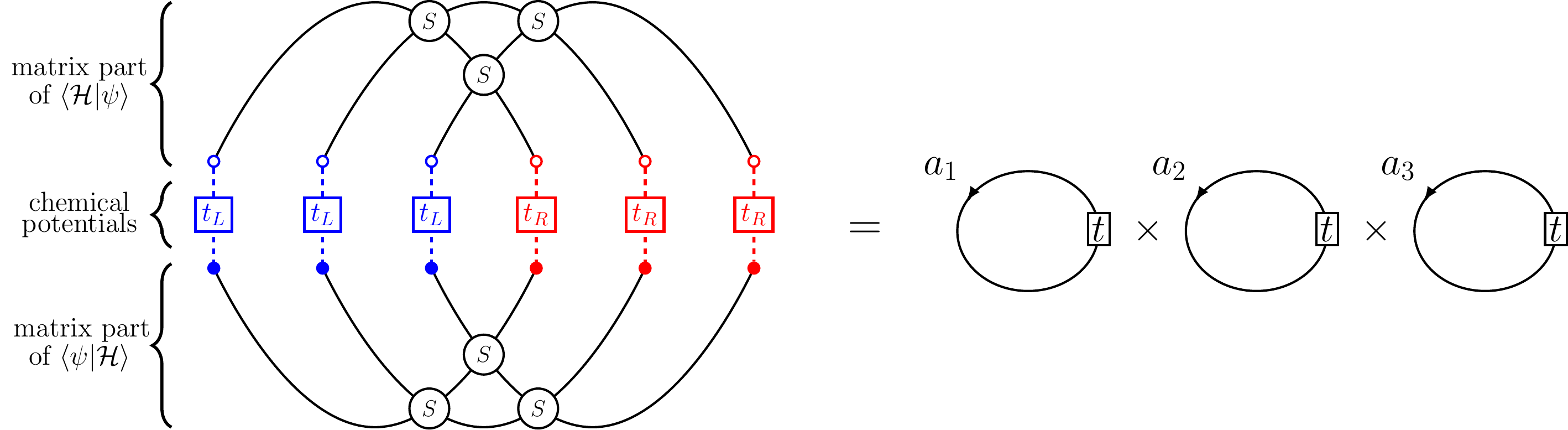}}
\caption{On the left the tensorial contraction of two hexagon form factors of three mirror particles with bound state numbers $a_{i}$. The dashed lines denote the sum over $\mathfrak{psu}(2|2)_{L}\times\mathfrak{psu}(2|2)_{R}$ flavour indexes on each representation $a_{i}$. The twists $t_{L,R}$ account for the chemical potentials which act on  each copy $\mathfrak{psu}(2|2)_{L}$ and $\mathfrak{psu}(2|2)_{R}$. Using the invariance of $\mathfrak{psu}(2|2)$ Beisert's $S$-matrix under the twists and unitarity \eqref{eq:Sunit} we simplify this contraction. The result is a product of three twisted  transfer matrices (with twist $t=t_{L}t_{R}$) on representations $a_{i}$. }
\label{fig:3characters}
\end{figure}
\end{itemize}

\section{Octagons at weak coupling}\label{sec:WeakOctagon}

In this section we take the weak coupling limit of the mirror integrals \eqref{eq:Inl}. 
  The coupling enters in the components of the integrand exclusively through the Zhukovsky variable:
\beq\label{eq:Zhukovsky}
x^{[\pm a]}(u)= x(u\pm\frac{i\,a}{2})\,\qquad \text{and}\qquad x(u) = \frac{u+\sqrt{u^{2}-4g^{2}}}{2g}
\eeq
This  square-root branch cut whose size is controlled by the magnitude of the coupling develops in to a series of poles in the limit     $g\to 0$:
\beq\label{eq:Xpansion}
x^{[\pm a]} = \frac{u\pm\frac{i}{2}\,a}{g} - \frac{g}{u\pm\frac{i}{2}\,a} - \frac{g^{3}}{(u\pm\frac{i}{2}\,a)^{3}} - \frac{2\,g^{5}}{(u\pm\frac{i}{2}\,a)^{5}} - \frac{5 \, g^{7}}{(u\pm\frac{i}{2}\,a)^{7}}\, + \mathcal{O}(g)^{9}
\eeq
These poles lie on the imaginary axis $\pm \frac{i}{2}\,a$ and their degree match the exponent of $g$.

This simple structure of poles is inherited by the mirror integrand. In more detail, the integrand only contains poles on single variables  with generic form $(u_{j}\pm\frac{i}{2}\,a_{j})^{\#}$. This can be appreciated on the leading order expansion of the one-particle component:
\bba
\mu_{a}(u) &=
\, a\,\,\frac{g^{2}}{\left(u^{2}+\frac{a^{2}}{4}\right)^{2}} \,+\,\mathcal{O}(g)^{4}\nonumber\\
e^{-E_{a}(u)} &=\,\frac{g^{2}}{\left(u^{2}+\frac{a^{2}}{4}\right)} \, + \, \mathcal{O}(g)^{4}\qquad\text{and}\quad p_{a}(u) =\, u + \mathcal{O}(g)^{2}
\end{align}
and the two-particle interaction:
\bba\label{eq:weakP}
P_{ab}(u,v) \,
&= g^{4}\, \frac{\left((u-v)^{2}+\frac{(a-b)^{2}}{4}\,\right)\left((u-v)^{2}+\frac{(a+b)^{2}}{4}\,\right)}{(u^{2}+\frac{a^{2}}{4})^{2}(v^{2}+\frac{b^{2}}{4})^{2}} + \mathcal{O}(g)^{6}
\end{align}
This latter component contains  differences between rapidities $(u-v)$, but these only appear in the numerator and can be easily expanded out. By doing so we are able to disentangle the integration variables and with that the integrals that we need to evaluate. 

The upshot of this analysis  is that the multivariable mirror integrals can be expanded at weak coupling into sums of products of one-variable integrals. In appendix \ref{app:EfficientResidue} we identify a basis of these one-variable integrals and use it to algorithmically find the mirror integrals, in principle, at arbitrary loop order. The result is explicilty obtained in terms of polylogarithms.  

Another observation that adds to the simplicity of this expansion is the delay of the $n$-particle state to start contributing only at $n(n+l)$-loop order:
\beq
 \prod_{j=1}^{n}\,\bar{\mu}_{a_{j}}(u_{j})\times \prod_{j<k}\,P_{a_{j}a_{k}}(u_{j},u_{k}) = \mathcal{O}(g^{2})^{n(n+l)}
\eeq
 For instance we would have to reach nine loops to have a first contribution from the three-particle mirror state (when $l=0$).  This is the reason we go to this high loop order in section \ref{sec:OPEcheckNineLoops} to test our results.

Using a $\mathsf{Mathematica}$ implementation of the algorithm described in appendix \ref{app:EfficientResidue} we made an explicit computation of the mirror integrals up to $n=4$ and up to $17$ loops for bridges $l=0,1,2$. In all these cases we were able to express our results as:
\beq\label{eq:SumLadders}
\mathcal{I}_{n,l} \, = \sum_{j=n(n+l)}^{\infty}\,\left(g^{2}\right)^{j}\sum_{k_{1}+\cdots k_{n}=j}\, d_{l;k_{1},\cdots, k_{n}}\,F_{k_{1}}\cdots F_{k_{n}}
\eeq
where we only consider positive integer partitions $\{k_{1},\cdots,k_{n}\}\in\mathbb{Z}^{+}$ and the basis of singled-valued conformal ladder integrals is given by \cite{LadderOld}:
\beq
F_{L}(z,\bar{z})  =\frac{1}{z-\bar{z}}
 \sum_{m=0}^{L} \frac{(-1)^{m}\,(2 L-m)!}{L!(L-m)!\,m!}\,\left(\log(z\bar{z})\right)^{m}\,\left(\text{Li}_{2L-m}(z) - \text{Li}_{2L-m}(\bar{z})\right)
\eeq
The coefficients $d_{l;k_{1},\cdots, k_{n}}$ depend on the bridge length $l$ and could be zero for some integer partitions. We know them explicitly up to high loop orders for the cases aforementioned,  but we were unable to find them in closed form.  

This way of expressing our results makes manifest single-valuedness in the Euclidean regime and also the uniform and maximal transcedentality at each loop order.

In \eqref{eq:OctagonBridge0} we present the octagon form factor with bridge length $l=0$ up to nine loops. We highlight the $n=1,2,3$  mirror integrals $\mathcal{I}_{n,l=0}$ in {\color{red}red},{\color{blue}blue} and {\color{magenta}magenta} respectively, they go dressed with the correspondent character $\mathcal{X}_{n}$ as:

{\smaller{\bba\label{eq:OctagonBridge0}
\mathbb{O}_{l=0}&= 1  + g^{2}\,\left(\mathcal{X}_{1}\Fred_{1}\right) + g^{4}\,\left(-2\,\mathcal{X}_{1}\Fred_{2}\right) + g^{6}\,\left(6\,\mathcal{X}_{1}\Fred_{3}\right)+ g^{8}\,\left(-20\,\mathcal{X}_{1}\Fred_{4}\,+\,\mathcal{X}_{2}\left(-\frac{\Fblue_{2}^{2}}{3}+{\color{blue}F}_{1}\Fblue_{3}\right)\right)\nonumber\\
   &\quad + g^{10}\,\left(70\,\mathcal{X}_{1}\Fred_{5}+\mathcal{X}_{2}\left(\Fblue_{2}\Fblue_{3}-6\Fblue_{1}\Fblue_{4}\right)\right)\nonumber\\
    &\quad + g^{12}\,\left(-252\,\mathcal{X}_{1}\Fred_{6}\,+\,\mathcal{X}_{2}\left(-\frac{9\Fblue_{3}^{2}}{5}+\frac{4\Fblue_{2}\Fblue_{4}}{5}+28\Fblue_{1}\Fblue_{5}\right)\right)\nonumber\\
     &\quad + g^{14}\,\left(92\,\mathcal{X}_{1}\Fred_{7}\,+\,\mathcal{X}_{2}\left(\frac{36\Fblue_{3}\Fblue_{4}}{5}-16\Fblue_{2}\Fblue_{5}-120\Fblue_{1}\Fblue_{6}\right)\right)\nonumber\\
     &\quad + g^{16}\,\left(-3432\,\mathcal{X}_{1}\Fred_{8}\,+\,\mathcal{X}_{2}\left(-\frac{486\Fblue_{4}^{2}}{35}-\frac{9\Fblue_{3}\Fblue_{5}}{7}+\frac{690 \Fblue_{2}\Fblue_{6}}{7}+495\Fblue_{1}\Fblue_{7}\right)\right)\nonumber\\
     &\quad + g^{18}\,\left(12870\, \mathcal{X}_{1}\Fred_{9}\,+\,\mathcal{X}_{2}\left(\frac{465\Fblue_{4}\Fblue_{5}}{7}-\frac{1203\Fblue_{3}\Fblue_{6}}{14}-\frac{979\Fblue_{2}\Fblue_{7}}{2}-2002\Fblue_{1}\Fblue_{8}\right) \right.\nonumber\\
 &\qquad\qquad\qquad\qquad\qquad\qquad \left. +\mathcal{X}_{3}\left(-\frac{\Fmagenta_{3}^{3}}{20}+\frac{\Fmagenta_{2}\Fmagenta_{3}\Fmagenta_{4}}{5}-\frac{3\Fmagenta_{1}\Fmagenta_{4}^{2}}{5}-\frac{\Fmagenta_{2}^{2}\Fmagenta_{5}}{3}+\Fmagenta_{1}\Fmagenta_{3}\Fmagenta_{5}\right)\right)\nonumber\\
 &\quad+\mathcal{O}(g)^{20}
\end{align}}}

This octagon is the building block of the $simplest$ four-point function introduced in section \ref{sec:Simplest}. We have similar expression for other octagons with different bridge lengths which are relevant for the $asymptotic$ four point function introduce in \ref{sec:Asymptotic}. In appendix \ref{app:NineLoopIntegrals} we provide all non-vanishing  mirror integrals up to nine loops. 


\section{Hexagonalization of polarized four-point functions}\label{sec:WeakPolarisedFour}

In this section we use hexagonalization to compute some polarised four-point functions of protected operators with equal scaling dimension. We consider a limit of large R-charges which leads to a factorization of the correlator into octagon form factors.  We first review the starting point of the hexagonalization prescription: the skeleton graphs. Then in sections \ref{sec:Simplest} and \ref{sec:Asymptotic} we construct the projected correlators named as \textit{simplest} and \textit{asymptotic} using octagons.
\subsection{Skeleton graphs and hexagonalization}
As prescribed in \cite{HexagonalizationI} to compute a planar four-point function we first need to identify the skeleton graphs. We start by considering a four-point function of scalar  $\frac{1}{2}$-BPS operators with generic $\mathfrak{so}(6)$ polarizations and scaling dimension $K$:
\beq\label{eq:BPS}
\mathcal{O}(x_{i},y_{i}) = \text{Tr}\,\left(y_{i}.\Phi(x_{i})\right)^{K}
\eeq
where $x_{i}$ gives the space-time position and $y_{i} $ is a six dimensional null vector $y_{i}.y_{i}=0$ that specifies the $\mathfrak{so}(6)$ R-charge when contracted with the vector of six real scalars $\Phi=\left(\phi_{1}\;\phi_{2}\;\phi_{3}\;\phi_{4}\;\phi_{5}\;\phi_{6}\right)$. 

For the planar connected four-point function of operators \eqref{eq:BPS} the skeleton graphs are given by the Wick contractions in the tree level correlator\footnote{ In the general unpolarized case disconnected graphs may need to be included as skeleton graphs, see appendix F in \cite{NonPlanar}. However for our polarized correlators they will not play a role.}
\bba\label{eq:skeleton}
\langle \mathcal{O}(x_{1},y_{1})\,\mathcal{O}(x_{2},y_{2})\,\mathcal{O}(x_{3},y_{3})\,\mathcal{O}(x_{4},y_{4})\rangle^{(0)}  \,=\qquad\qquad\qquad\qquad\qquad\qquad\qquad\qquad\nonumber\\
\, \sum_{l_{12}=1}^{K-1} \,\raisebox{-2077787sp}{  \resizebox{.17\totalheight}{!}{\includestandalone[width=.8\textwidth]{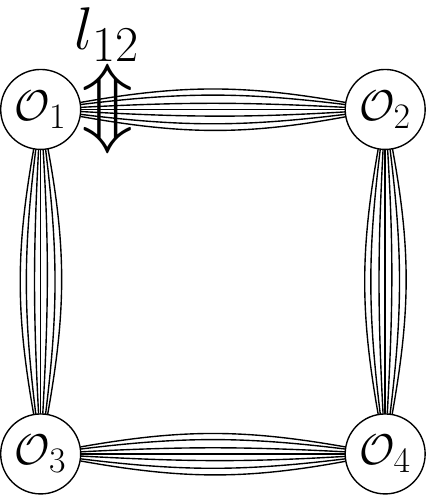}} }\,+\, \sum_{l_{14}=1}^{K-1}\left( \,\raisebox{-3377787sp}{  \resizebox{.3\totalheight}{!}{\includestandalone[width=.8\textwidth]{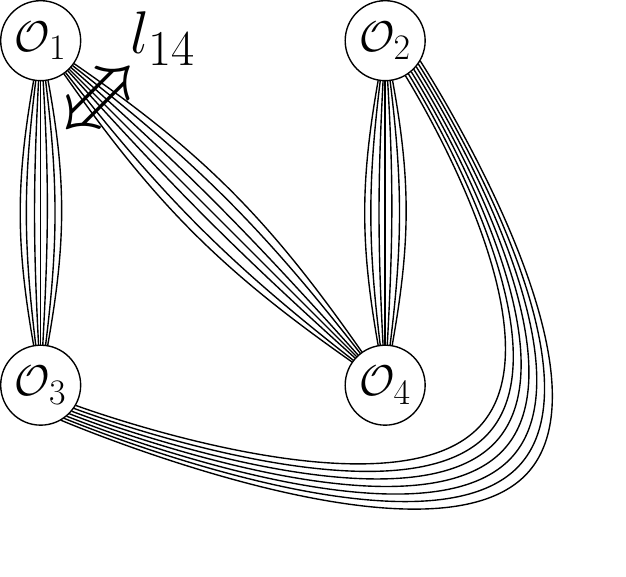}} }+\,\raisebox{-3377787sp}{  \resizebox{.3\totalheight}{!}{\includestandalone[width=.8\textwidth]{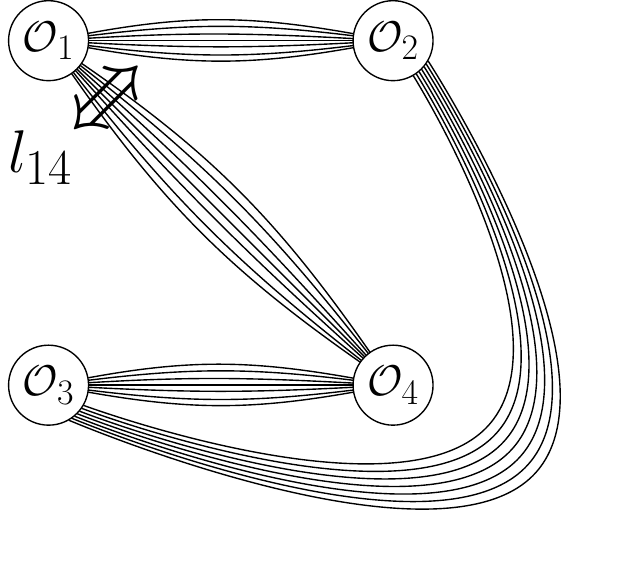}}}\right)\nonumber\\
+\,\sum_{l_{14}=1}^{K-2}\sum_{l_{12}=1}^{K-l_{14}-1} \left( \,\raisebox{-3377787sp}{  \resizebox{.26\totalheight}{!}{\includestandalone[width=.8\textwidth]{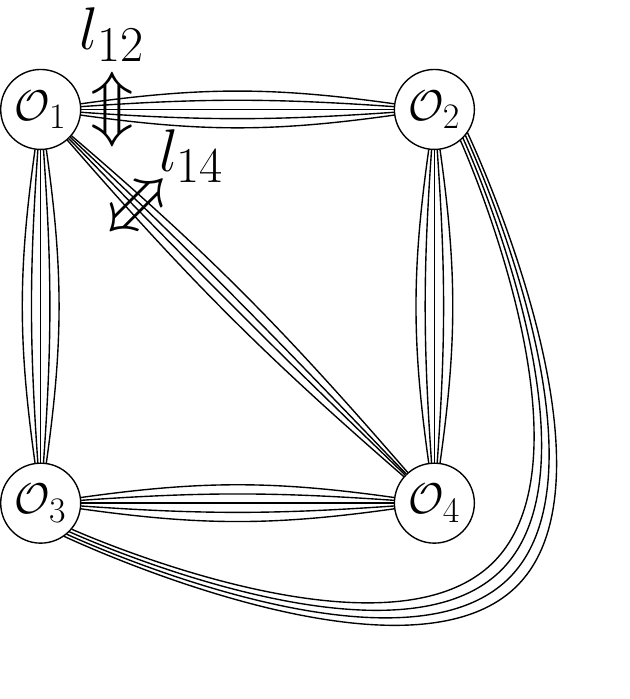}} } \,+\, \,\raisebox{-3377787sp}{  \resizebox{.26\totalheight}{!}{\includestandalone[width=.8\textwidth]{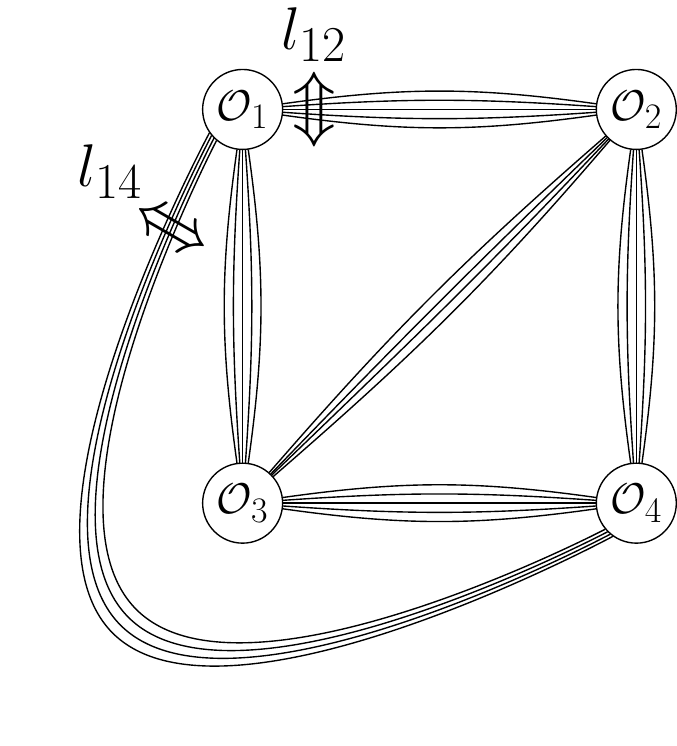}} }\right)\qquad
\end{align}
where each line connecting two operators $\mathcal{O}_{i}$ and $\mathcal{O}_{j}$ represents a tree level propagator $\frac{y_{i}.y_{j}}{x_{ij}^{2}}$ and their  number is given by the bridge length $l_{ij}$. The bridges left implicit are fixed by the condition $K=K_{i}=\sum_{j\neq i}^{4}l_{ij}$. In particular this sets the identifications: $l_{12}=l_{34}$, $l_{13}=l_{24}$, $l_{14}=l_{23}$ for all skeleton graphs.

It is useful to make manifest conformal invariance by defining a reduced correlator after stripping out a simple factor:
\beq
\mathcal{G}_{K}(u,v,\sigma,\tau) = \left(\frac{x_{12}^{2}x_{34}^{2}}{(y_{1}.y_{2})(y_{3}.y_{4})}\right)^{\frac{K}{2}}\,\langle \mathcal{O}(x_{1},y_{1})\,\mathcal{O}(x_{2},y_{2})\,\mathcal{O}(x_{3},y_{3})\,\mathcal{O}(x_{4},y_{4})\rangle
\eeq
This reduced four-point function only depends on the spacetime cross ratios:
\beq
u =z \bar{z} = \frac{x_{12}^{2} x_{34}^{2}}{x_{13}^{2}x_{24}^{2}}\quad\text{and}\quad v=(1-z)(1-\bar{z})=\frac{x_{14}^{2}x_{23}^{2}}{x_{13}^{2}x_{24}^{2}}
\eeq
and $R$-charge cross ratios:
\beq
\sigma = \alpha\bar{\alpha}=\frac{\left(y_{1}.y_{2}\right)\,\left(y_{3}.y_{4}\right)}{\left(y_{1}.y_{3}\right)\,\left(y_{2}.y_{4}\right)}\quad \text{and} \quad  \sigma = (1-\alpha)(1-\bar{\alpha})=\frac{\left(y_{1}.y_{4}\right)\,\left(y_{2}.y_{3}\right)}{\left(y_{1}.y_{3}\right)\,\left(y_{2}.y_{4}\right)}
\eeq
At tree level  the reduced four-point function can be expressed as:
\beq\label{eq:reducedTreeLevel}
\mathcal{G}_{K}^{(0)}(u,v,\sigma,\tau)  = \sum_{r=1}^{K-1}\,\left(G_{0,r,K-r} + G_{r,0,p-r}+G_{r,K-r,0}\right) \, +\, \sum_{r=2}^{K-1}\,\sum_{s=1}^{r-1} G_{s,r-s,K-r} 
\eeq
where our notation $G_{l_{14},l_{13},l_{12}}$ serves to identify a skeleton graph through  the non-identical bridges and also provides the tree level Wick contractions in terms of cross ratios as:
\beq\label{eq:bwick}
G_{l_{14},l_{13},l_{12}}\equiv
\begin{cases}
2\,\left(\frac{u}{\sigma}\right)^{l_{14}+l_{13}}\,\left(\frac{\tau}{v}\right)^{l_{14}}\,, \qquad \text{if}\quad l_{12}\neq 0\,\&\,l_{13}\neq 0\,\&\,l_{14}\neq 0\,,\\
\left(\frac{u}{\sigma}\right)^{l_{14}+l_{13}}\,\left(\frac{\tau}{v}\right)^{l_{14}} \,, \qquad\quad \text{otherwise}
\end{cases}
\eeq
The first three terms in \eqref{eq:reducedTreeLevel} correspond to the three type of skeleton graphs at the top of \eqref{eq:skeleton}. The fourth term represents the two types at the bottom of \eqref{eq:skeleton}, which have the same spacetime dependence and can be simply combined explaining the factor of two in \eqref{eq:bwick}.

To incorporate the coupling dependence \cite{HexagonalizationI} prescribes to dress the skeleton graphs \eqref{eq:skeleton} with four hexagons which meet along six mirror cuts where we pose multi-particle mirror states. To recover the full-four-point function we must sum over all intermediate states on each cut on each skeleton graph:
\bba
\langle \mathcal{O}(x_{1},y_{1})\,\mathcal{O}(x_{2},y_{2})\,\mathcal{O}(x_{3},y_{3})\,\mathcal{O}(x_{4},y_{4})\rangle  \,=\qquad\qquad\qquad\qquad\qquad\qquad\qquad\qquad\nonumber\\
\, \sum_{\{\psi_{ij}\}} \,\raisebox{-4077787sp}{  \resizebox{.3\totalheight}{!}{\includestandalone[width=.8\textwidth]{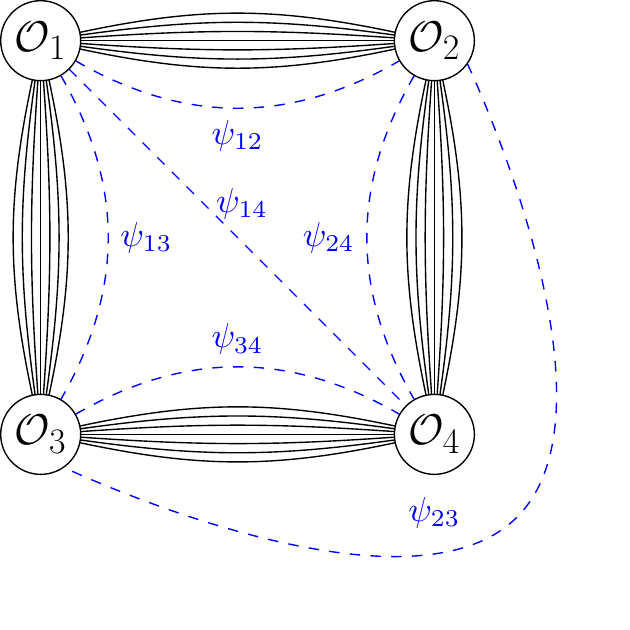}} }\,+\,\cdots 
+\,\sum_{\{\psi_{ij}\}} \,\raisebox{-4077787sp}{  \resizebox{.3\totalheight}{!}{\includestandalone[width=.8\textwidth]{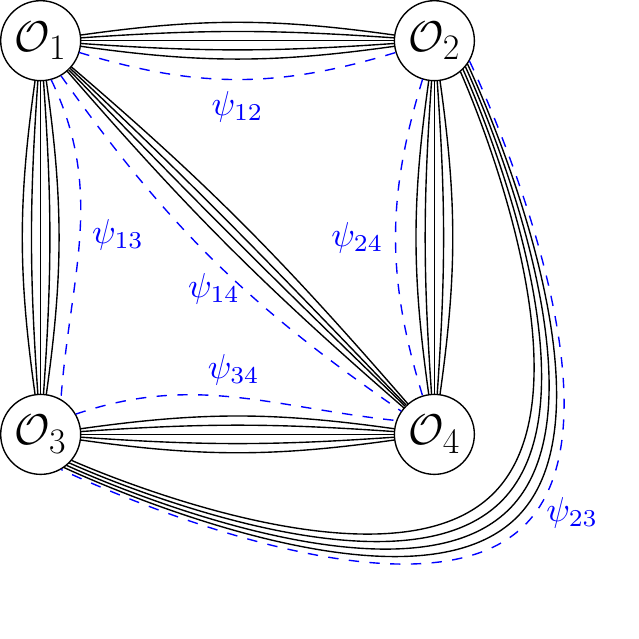}} } 
\end{align}
where the ellipsis accounts for the other skeleton graphs in \eqref{eq:skeleton}.

When dealing with generic polarizations we need to consider every possible configuration of mirror states $\{\psi_{ij}\}$ on a tesselation.
 Technically speaking the taming of each of these contributions comes with different degrees of difficulty.
 
  The simplest cases are given by configurations where only one mirror cut hosts non-trivial states and the other five only host the vacuum or when two non-adjacent cuts, such as $1$-$4$ and $2$-$3$, host non-trivial states and the other four only the vacuum.  The former case is accounted by an octagon and the latter by the product of two octagons. As we pointed out in section \ref{sec:TheOctagon}, the octagon has a simple structure thanks to the outstanding simplicity of the tensor contraction of the two hexagons form factors in figure \ref{fig:3characters}.

\begin{figure}[ht]
\centering
\resizebox{3\totalheight}{!}{\includestandalone[width=.8\textwidth]{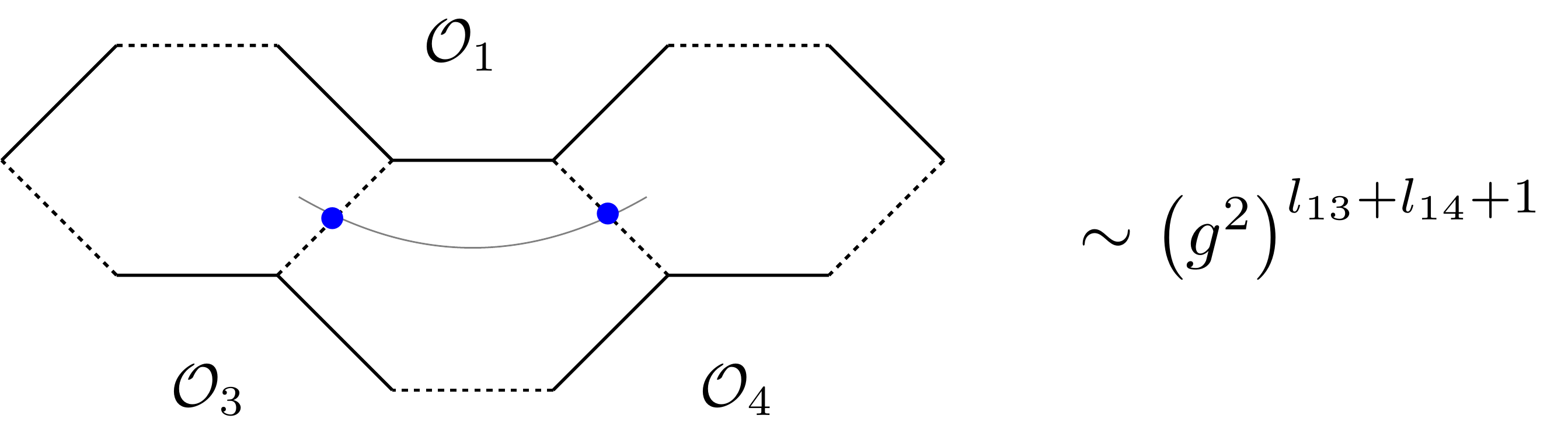}}
\caption{Two-particle string}
\label{fig:TwoString}
\end{figure}

 On the other hand the configurations that include non-trivial states on different  edges of the same hexagon are significantly harder to deal with.  These show up in the non-trivial gluing of three or four hexagon form factors. The simplest of this type of contributions corresponds to the two-particle string in figure \ref{fig:TwoString}. The tensor contraction in this case does not simplify and its complexity grows with the bound state number of the particles \cite{HexagonalizationII}\footnote{In this reference the two-particle string  was tamed at leading order in the context of five-point functions.}.  Having non-trivial states turn on in all six mirror cuts is  the most challenging case and would demand a huge effort even just to find its leading order contribution.
 
We would like to consider a regime where the contributions of these ``strings" of mirror particles can be neglected. This can be partially achieved in the limit of large external dimension $K\gg 1$. As this limit enforces large bridges in many skeleton graphs delaying their contributions to high loop orders. However it is not simple enough yet, as there are still some graphs which allow strings to kick in at low loop orders, such as $G_{l_{14}=0,\,l_{13}=K-1,\,l_{12}=1}$. In order to completely get rid of these strings we further consider special choices of $R$-charge external polarizations which project out these troublesome graphs.  This will be such that by $R$-charge conservation only graphs containing four large bridges and two non-adjacent small bridges will contribute to the correlator. 

In the next sections we present two instances of such polarized four-point functions which can be computed by composition of the octagons described in the previous section.   We name these correlators as the $simplest$ and	 $asymptotic$ four-point functions. 

Let us conclude this section by stressing we will be working in the weak coupling regime $g\to 0$ so we are considering an expansion of the four-point function as:
\beq\label{eq:GeneralFour}
\mathcal{G}_{K}(u,v,\sigma,\tau) \,=\, \mathcal{G}^{(0)}_{K}(u,v,\sigma,\tau)  \, +\,\sum_{m=1}^{\infty} \, \left(g^{2}\right)^{m}\, \mathcal{G}_{K}^{(m)}(u,v,\sigma,\tau) 
\eeq

\subsection{The \textit{simplest} four-point function}\label{sec:Simplest}
\begin{figure}[ht]  
\centering       
\resizebox{1\totalheight}{!}{\includestandalone[width=.8\textwidth]{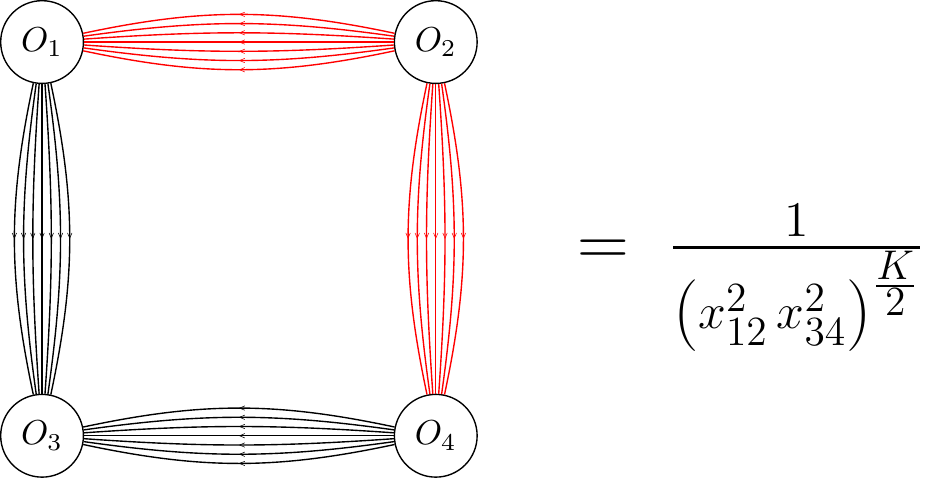}}
\caption{The \textit{simplest} four-point function with projected external operators $O_{1}(0) = \text{Tr}\left(Z^{\frac{K}{2}}\,{\color{red}\bar{X}}^{\frac{K}{2}}\right)$, $O_{2}(z) = \text{Tr}\left({\color{red}X}^{K}\,\right)$, $O_{3}(1) = \text{Tr}\left(\bar{Z}^{K}\right)$ and $O_{4}(\infty) = \text{Tr}\left(Z^{\frac{K}{2}}\,{\color{red}\bar{X}}^{\frac{K}{2}}\right)$. The Wick contractions form a perimeter with four bridges of width $\frac{K}{2}$. }
\label{fig:SimplestTreeLevel}
\end{figure}

Our simplest choice of polarised external operators consists of:
\bba\label{eq:SimplestOperators}
O_{1} &= Tr(Z^{\frac{K}{2}}\,{\color{red}\bar{X}}^{\frac{K}{2}}) + \text{cyclic permutations} &
O_{2} &= Tr({\color{red}X}^{K})\nonumber \\
O_{3} &= Tr(\bar{Z}^{K}) &\qquad 
O_{4} &= Tr(Z^{\frac{K}{2}}\,{\color{red}X}^{\frac{K}{2}})+ \text{cyclic permutations} 
\end{align}
where cyclic permutation means that we only keep one instance of all permutations that are cyclically identified. At tree level this correlator is given by a single graph as shown in figure \ref{fig:SimplestTreeLevel}.

This \textit{simplest} four-point can be extracted from the general correlator by choosing the polarization vectors as:
\bba\label{eq:SimplestYs}
y_{1} & = \frac{1}{\sqrt{2}}(1,i,0,0,\beta_{1}, i\,\beta_{1}) &
y_{2} & =\frac{1}{\sqrt{2}} (0,0,0,0,1,-i)\nonumber \\
y_{3} & =\frac{1}{\sqrt{2}} (1,-i,0,0,0,0 ) &\qquad 
y_{4} & = \frac{1}{\sqrt{2}}(1, i,0,0,\beta_{4}, i\,\beta_{4} )
\end{align}
performing $\frac{K}{2}$ derivatives with respect to each auxiliary parameter $\beta_{1}$ and $\beta_{4}$, to then set them to zero:
\bba\label{eq:SimplestProjection}
\langle O_{1}(x_{1})O_{2}(x_{2})O_{3}(x_{3})O_{4}(x_{4})\rangle &=\nonumber\\
&\!\!\!\!\!\!\!\!\!\!\!\!\left(\frac{\partial}{\partial\beta_{1}}\frac{\partial}{\partial\beta_{4}}\right)^{\frac{K}{2}}\,\langle \mathcal{O}(x_{1},y_{1})\mathcal{O}(x_{2},y_{2})\mathcal{O}(x_{3},y_{3})\mathcal{O}(x_{4},y_{4})\rangle\,\bigg{|}_{\beta_{1},\beta_{4}=0}
\end{align}
We can as well define this projection as an operation on the $R$-charge cross ratios $\sigma,\tau$ of the super-correlator. For this we first define the reduce \textit{simplest} four-point function as:
\beq
\mathbb{S}_{K}(z,\bar{z})= \left(x_{12}^{2} x^{2}_{34}\right)^{\frac{K}{2}}\,\langle O_{1}(x_{1})O_{2}(x_{2})O_{3}(x_{3})O_{4}(x_{4})\rangle
\eeq
From this definition it follows we can effectively perform the projection on the reduced super-correlator by extracting the coefficient of $\sigma^{-\frac{K}{2}}$ and setting $\tau=0$ : 
\beq\label{eq:Sprojection}
\mathbb{S}_{K}(z,\bar{z}) = \mathcal{G}_{K}(u,v,\sigma,\tau)\big{|}_{\text{Coefficient of }\sigma^{-\frac{K}{2}},\tau\to 0}
\eeq
At tree level this gives:
\beq
\mathbb{S}^{(0)}_{K}  \,= u^{\frac{K}{2}} = \left(z\bar{z}\right)^{\frac{K}{2}}
\eeq
We are interested in finding the loop corrections for this four-point function in the limit $K\gg 1$.  Following hexagonalization, in this regime, we expect to have only the vacuum state on the large bridges $l_{12},l_{13},l_{24},l_{34}\gg 1$. With this simplification we now only need to consider the effect of the polarization on the skeleton graph(s) when dressed  by two octagons,  with mirror particles along cuts $1$-$4$ and $2$-$3$: 
\beq
\raisebox{-4077787sp}{  \resizebox{.3\totalheight}{!}{\includestandalone[width=.8\textwidth]{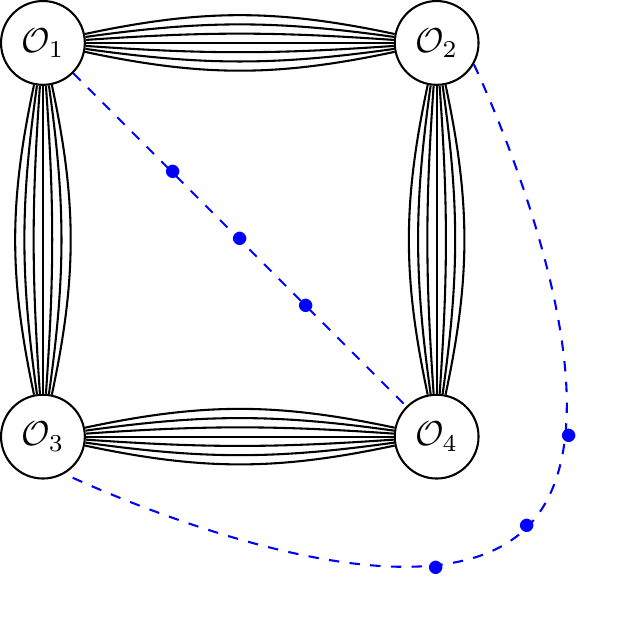}} } \text{\Huge$\bigg{|}$}_{\text{Coefficient of }\sigma^{-\frac{K}{2}}\,,\tau\to0} \;\; \overset{?}{\neq}\;\; 0
\eeq
For this we need to  focus on the character $\mathcal{X}_{n}$, which is responsible for changing the original $R$-charge dependence of a skeleton graph and allow it to survive our \textit{simplest} projection. In order to find out which are these surviving graphs we rewrite the character in terms of cross ratios $\sigma$ and $\tau$:
\bba
\mathcal{X}_{n} &= \frac{\left(\frac{-(z-\alpha)(\bar{z}-\alpha)}{\alpha}\right)^{n}+\left(\frac{-(z-\bar{\alpha})(\bar{z}-\bar{\alpha})}{\bar{\alpha}}\right)^{n}}{2}= \,\sum_{k=-n}^{n}\,\sigma^{k}\,f_{n,k}(z,\bar{z},\tau)
\end{align}
In these variables the $n$-particle character can be written as a finite series on $\sigma$ with integer exponents ranging from $-n$ to $n$ and coefficients given by the polynomials $f_{n,k}(z,\bar{z},\tau)$. From this we learn  the character $\mathcal{X}_{n}$ can change the exponent of $\sigma$ in a skeleton graph by a  shift of at most $\pm n$. Later these shifts are further enhanced when considering the product of the two octagons. On the other hand, due to the polynomiality of $f_{n,k}$ in $\tau$, the effect of the character over graphs with $\tau$ dependence such as those with $l_{14},l_{23}\neq 0$  is to maintain it or increase its positive exponent. Thus we do not need to include this type of graphs as they will vanish when taking $\tau\to 0$ in our projection. 

\begin{figure}[ht]  
\centering    
\resizebox{2\totalheight}{!}{\includestandalone[width=.8\textwidth]{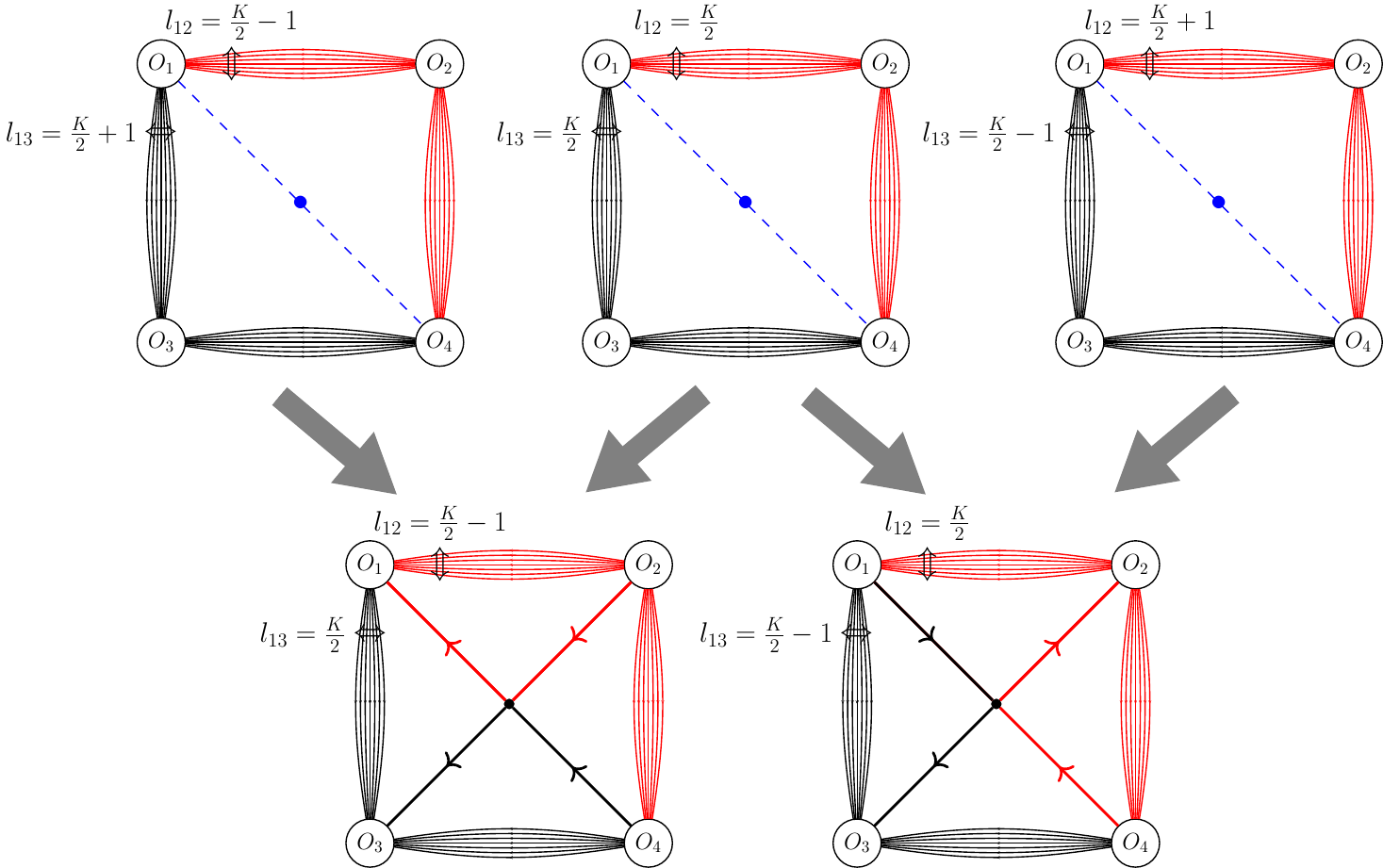}} 
\caption{At the top the skeleton graphs that we need to hexagonalize and add up to obtain the one-loop \textit{simplest} four-point function. At the bottom an interpretation of the  neighboring graphs in terms of Feynman diagrams.}
\label{fig:OneLoopProjection}
\end{figure}

In conclusion, the graphs that survive the \textit{simplest} projection have $l_{14}=l_{23}=0$ and can have some shifts on the other bridge lengths  compare to the length $\frac{K}{2}$ in the original tree level graph of figure \ref{fig:SimplestTreeLevel}. These shifts depend on the loop order as this truncates the number of mirror particles $n$. For instance at one-loop  only one-particle states show up, exclusively on cut $1$-$4$ or  cut $2$-$3$, and only the three skeleton graphs in figure \ref{fig:OneLoopProjection} contribute. These are the original tree level graph and two neighbors which survive thanks to the character $\mathcal{X}_{1}$.  At two-loops we can have a one-particle state in each  mirror cut $1$-$4$ and $2$-$3$, which gives $\mathcal{X}_{1}^{2}$ and with that a maximal of shift of $\pm2$ on the bridges at the perimeter.     
As we go to higher loops the admissible number of particles increases and with that also the number of neighboring skeleton graphs. 
\begin{table}[ht]
\centering
\begin{tabular}{|c|c|c|c|c|c|c|c|c|c|c|c|c|c|c|}
\hline
loop order $\Lambda$ & 1 & 2 & 3 & 4 & 5 & 6 &  7 & 8 & 9 &  10 & 11 & 12 & 13 \\ 
\hline
$m_{\Lambda}$ & 1 & 2 & 2 & 2 & 3 & 3 & 3 & 4 & 4 &  4 & 4 & 4 & 5  \\
\hline
\end{tabular}
\caption{Maximal shifts $m_{\Lambda}$ at each loop order $\Lambda$}
\label{tab:ShiftSimplest}
\end{table}

In table \ref{tab:ShiftSimplest} we present the  maximal shift $\pm m_{\Lambda}$ at loop order $\Lambda$. At this order we need to include skeleton graphs with $\sigma$ powers going from $\sigma^{-K-m_{\Lambda}}$ to $\sigma^{-K+m_{\Lambda}}$. Thus the hexagonalization of the \textit{simplest} correlator at the first loop orders goes schematically as:

\bba\label{eq:5loopsSimplest}
\mathbb{S}^{(0)}_{K} &= u^{\frac{K}{2}}
 \nonumber\\
\mathbb{S}^{(1)}_{K} &= \left(\frac{u}{\sigma}\right)^{\frac{K}{2}} \left(\frac{1}{\sigma}+1+\sigma\right)\,\left(1+\mathcal{X}_{1}\,\mathcal{I}_{1,0}\right)^{2} \bigg{|}_{\text{coefficient of }\left(g^{2}\,\sigma^{-K/2}\right)\,,\tau\to0}\nonumber\\
\mathbb{S}^{(2)}_{K} &=\left(\frac{u}{\sigma}\right)^{\frac{K}{2}} \left(\frac{1}{\sigma^{2}}+\frac{1}{\sigma}+1+\sigma + \sigma^{2}\right)\,\left(1+\mathcal{X}_{1}\,\mathcal{I}_{1,0}\right)^{2} \bigg{|}_{\text{coefficient of }\left(g^{4}\,\sigma^{-K/2}\right)\,,\tau\to0}\nonumber\\
&\quad\vdots\nonumber\\
\mathbb{S}^{(5)}_{K} &= \left(\frac{u}{\sigma}\right)^{\frac{K}{2}}\left(\frac{1}{\sigma^{3}}+\frac{1}{\sigma^{2}}+\frac{1}{\sigma}+1+\sigma + \sigma^{2}+ \sigma^{3}\right)\,\left(1+\mathcal{X}_{1}\,\mathcal{I}_{1,0}+\mathcal{X}_{2}\,\mathcal{I}_{2,0}\right)^{2} \bigg{|}_{\text{coefficient of }\left(g^{10}\,\sigma^{-K/2}\right)\,,\tau\to0}\nonumber\\
\end{align}

The contributions of these new skeleton graphs can be resummed at each loop order and then repackaged to obtain a compact expression for the \textit{simplest} correlator, valid at any loop order for sufficiently large $K$:
\beq\label{eq:AllLoopSimplest}
\mathbb{S}_{K} = u^{\frac{K}{2}}\times\mathbb{O}_{l=0}^{2}\,\big{|}_{\mathcal{X}_{n}\to\,\mathcal{X}^{(s)}_{n}}
\eeq
This is simply the original tree level correlator times the squared of the octagon in \eqref{eq:OctagonBridge0} with the character replaced by the polarized effective character:
\beq
\mathcal{X}^{(s)}_{n}=(-v)^{n} \,=\,\left(-(1-z)(1-\bar{z})\right)^{n}
\eeq
Given the simplicity of \eqref{eq:AllLoopSimplest}, with the dependence on $K$ totally factorized in the prefactor, we are tempted to claim this expression should be valid at finite coupling provided $K\to \infty$. 

\textbf{The regime of validity for finite $K$}:
We can have an estimate for the smallest value of $K$ for which  \eqref{eq:AllLoopSimplest} holds up to a given loop order $\Lambda$. In our criteria we demand that all neighboring skeleton graphs have bridge lengths  $l_{12}$ or $l_{13}$ larger than $\Lambda$. This guarantees we can neglect including mirror particles along those bridges.

Considering the smallest bridge is given by $\frac{K}{2}-m_{\Lambda}$, see table  \ref{tab:ShiftSimplest}, our criteria sets the lower bound:
\beq\label{eq:Sbound}
K \geq 2\,\Lambda +2\,m_{\Lambda}
\eeq 
Below this bound, for instance $K=2\,\Lambda+2\,m_{\Lambda} -2$ (we only consider even $K$), at loop order $\Lambda$ we need to start  including mirror particles over the non-zero bridges $l_{12}$ and $l_{13}$ and also the two-particle strings of figure \ref{fig:TwoString} over bridges $l_{14}$-$l_{13}$ or $l_{14}$-$l_{12}$.

This criteria might not be optimal and our estimated lower bound maybe too large. To be more rigorous we would need to take into account the R-charge dependence of the two-particle strings and see at which loop order they survive the \textit{simplest} projection. We leave this for future work and for now we just stick to the possibly exaggerated bound \eqref{eq:Sbound}. 

\subsection{The \textit{asymptotic} four point function}\label{sec:Asymptotic}
In this section we consider the four-point function already studied in \cite{AsymptoticPaper} from the OPE point of view,  but now using hexagonalization. This consists of external operators:
\bba\label{eq:AsymOperators}
\tilde{O}_{1} &= Tr(Z^{\frac{K}{2}}\,{\color{red}\bar{X}}^{\frac{K}{2}}) + \text{cyclic permutations} &
\tilde{O}_{2} &= Tr({\color{red}X}^{\frac{K}{2}}\,Z^{\frac{K}{2}})+ \text{cyclic permutations}\nonumber \\
\tilde{O}_{3} &= Tr(\bar{Z}^{\frac{K}{2}}\,{\color{blue}\bar{Y}}^{\frac{K}{2}})+ \text{cyclic permutations}  &\qquad 
\tilde{O}_{4} &= Tr({\color{blue}Y}^{\frac{K}{2}}\,\bar{Z}^{\frac{K}{2}})+ \text{cyclic permutations}
\end{align}
The correspondent $R$-charge polarization are:
\bba\label{eq:AsymYs}
y_{1} & = \frac{1}{\sqrt{2}}(1,i,0,0,\beta_{1},-i\,\beta_{1}) &
y_{2} & =\frac{1}{\sqrt{2}}(1,i,0,0,\beta_{2},i\,\beta_{1})\nonumber \\
y_{3} & =\frac{1}{\sqrt{2}}(1, -i,\beta_{3}, -i\,\beta_{3},0,0) &\qquad 
y_{4} & = \frac{1}{\sqrt{2}}(1, -i,\beta_{4}, i\,\beta_{4},0,0)
\end{align}
and the \textit{asymptotic} four-point is obtained by differentiating the super-correlator respect to the parameters $\beta_{i}$: 
\bba
\langle 
\tilde{O}_{1}(x_{1})\tilde{O}_{2}(x_{2})\tilde{O}_{3}(x_{3})\tilde{O}_{4}(x_{4})\rangle &=\nonumber\\
&\!\!\!\!\!\!\!\!\!\!\!\!\!\!\!\!\!\!\!\!\!\!\!\!\!\!\!\!\!\!\!\!\!\!\!\!\!\!\!\! \left(\frac{\partial}{\partial\beta_{1}}\frac{\partial}{\partial\beta_{2}}\frac{\partial}{\partial\beta_{3}}\frac{\partial}{\partial\beta_{4}}\right)^{\frac{K}{2}}\langle \mathcal{O}(x_{1},y_{1})\mathcal{O}(x_{2},y_{2})\mathcal{O}(x_{3},y_{3})\mathcal{O}(x_{4},y_{4})\rangle\,\bigg{|}_{\beta_{i}=0}
\end{align}
It is convenient to define the reduced correlator which only depends on the conformal cross ratios:
\beq
\mathbb{A}_{K}(z,\bar{z})=\left(x^{2}_{12}x^{2}_{34}\right)^{\frac{K}{2}}\,\langle \tilde{O}_{1}(x_{1})\tilde{O}_{2}(x_{2})\tilde{O}_{3}(x_{3})\tilde{O}_{4}(x_{4}) \rangle
\eeq
Then the R-charge projection can be effectively performed over \eqref{eq:GeneralFour} as:
\beq\label{eq:AsymProjection}
\mathbb{A}_{K}(z,\bar{z})\,=\, \mathcal{G}_{K}(u,v,\sigma,\tau)\,\big{|}_{\text{Coefficient of }\sigma^{-K/2},\tau\,\to 1}\qquad 
\eeq
At tree level the graphs surviving this projection satisfy $l_{12}=\frac{K}{2}$ and $l_{13}+l_{14}=\frac{K}{2}$, see figure \ref{fig:AsymptoticTree}. In the notation of \eqref{eq:bwick}  this is:
\beq\label{eq:AsymTree}
\mathbb{A}^{(0)}_{K}\, = \sum_{l=0}^{K/2}\,G_{l,\frac{K}{2}-l,\frac{K}{2}} = u^{\frac{K}{2}} + 2\,\sum_{l=1}^{\frac{K}{2}-1}\, \frac{u^{\frac{K}{2}}}{v^{l}} +\frac{u^{\frac{K}{2}}}{v^{\frac{K}{2}}} 
\eeq
\begin{figure}[ht]  
\centering    
\resizebox{.4\totalheight}{!}{\includestandalone[width=.8\textwidth]{AsymptoticTree}}
\caption{ A tree level graph of the \textit{asymptotic} four-point function with projected external operators $\tilde{O}_{1}(0) = \text{Tr}\left(Z^{K/2}\,{\color{red}X}^{K/2}\right)$, $\tilde{O}_{2}(z) = \text{Tr}\left({\color{red}\bar{X}}^{K/2}\,Z^{K/2}\right)$, $\tilde{O}_{3}(1) = \text{Tr}\left(\bar{Z}^{K/2}\,{\color{blue}Y}^{K/2}\right)$ and $\tilde{O}_{4}(\infty) = \text{Tr}\left(\bar{Z}^{K/2}\,{\color{blue}\bar{Y}}^{K/2}\right)$.}
\label{fig:AsymptoticTree}
\end{figure} 

\begin{figure}[ht]
\centering
\resizebox{1.2\totalheight}{!}{\includestandalone[width=.8\textwidth]{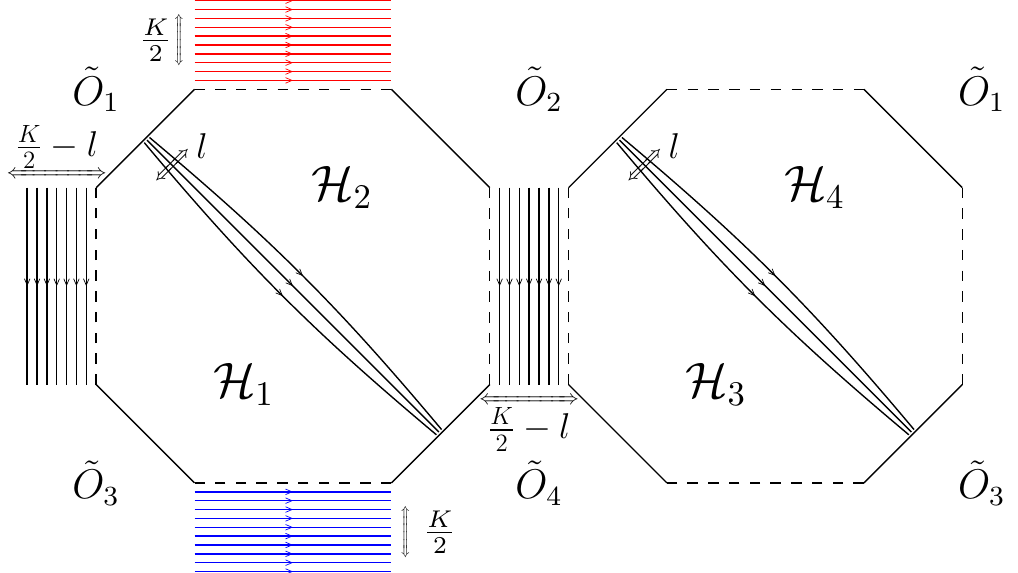}}
\caption{At a given loop order for sufficiently large $K$ and small bridge $l$ we have two decouple octagons: $\mathcal{H}_{1}\cup \mathcal{H}_{2}$ and $\mathcal{H}_{3}\cup \mathcal{H}_{4}$. When $l\sim \frac{K}{2}$, then the small bridge is $\frac{K}{2}-l$ and the decouple octagons are $\mathcal{H}_{2}\cup \mathcal{H}_{3}$ and $\mathcal{H}_{1}\cup \mathcal{H}_{4}$.}
\label{fig:AsymptoticOctagons}
\end{figure}

In the limit $K\gg 1$, to find the loop corrections, we need to identify the  small bridge lenghts which can host mirror particles. In this regime, we can have skeleton graphs with $l_{14}$($l_{23}$) small and $l_{13}$($l_{24}$)$\sim \frac{K}{2}$ large or the other way around but not both small at the same time, see figure \ref{fig:AsymptoticOctagons}. Thus the \textit{asymptotic} correlator receives two types of contributions. The first(second) type  comes from mirror particles on small bridges $l_{14}$  and $l_{23}$ ($l_{13}$ and $l_{24}$). The first type is accounted by the product of two octagons with bridge parameters $l=l_{14}=l_{23}$ and cross ratios $z$,$\alpha$ (and conjugates). While the second type is also given by two octagons but now with bridge parameter  $l=l_{13}=l_{24}$ and cross ratios $\frac{z}{z-1}$,$\frac{\alpha}{\alpha-1}$\footnote{This change of cross ratios follow from the different relative positions of the physical edges, see figure 5 in \cite{HexagonalizationI}}(and conjugates). Then the loop corrections are obtained by simply adding up these two types of independent contributions.

As explained in the previous section, at loop level we can have new skeleton graphs  with $l_{13}+l_{14}\neq \frac{K}{2}$ which survive the \textit{asymptotic} projection \eqref{eq:AsymProjection}  when dressed with mirror particles.  We repeat the same exercise as in equation \eqref{eq:5loopsSimplest}  identifying all skeleton graphs that survive at each loop order. Resumming those contributions we find the \textit{asymptotic} correlator can be expressed as a sum over the original tree level graphs in \eqref{eq:AsymTree}, each dressed by a new effective  octagon squared. We define this octagon as the \textit{asymptotic} octagon:
\beq
\tilde{\mathbb{O}}_{l}(z,\bar{z}) \,=\mathbb{O}_{l}\big{|}_{\mathcal{X}_{n}\to\tilde{\mathcal{X}}_{n}} =1+\sum_{n=1}^{\infty}\,\tilde{\mathcal{X}}_{n}(z,\bar{z}) \;\mathcal{I}_{n,l}(z,\bar{z})  
\eeq
with the correspondent effective character:
\beq 
\tilde{\mathcal{X}}_{n}(z,\bar{z})=\left(1-v\right)^{n} \, = \, \left(z+\bar{z}-z\bar{z}\right)^{n}
\eeq
Using this definition the \textit{asymptotic} four-point function is  expressed in terms of the $z$-channel with mirror particles on $l_{14}=l_{13}=l$, and the $\frac{z}{z-1}$-channel with mirror particles on $l_{13}=l_{24}=\frac{K}{2}-l$:
\bba\label{eq:AsymFourAllLoop}
\mathbb{A}_{K}(z,\bar{z})  &=\,\sum_{l=0}^{\frac{K}{2}}\,G_{l,\frac{K}{2}-l,\frac{K}{2}}\big{|}_{\sigma\to 1\atop\tau\to 1 }\, \left(\tilde{\mathbb{O}}^{2}_{l}(z,\bar{z})\,+\,\tilde{\mathbb{O}}^{2}_{\frac{K}{2}-l}\left(\text{\small$\frac{z}{z-1}$},\text{\small$\frac{\bar{z}}{\bar{z}-1}$}\right)\,-1\right)
\end{align}
or under a simple change of summation index:
\bba\label{eq:AsymFourAllLoop2}
\mathbb{A}_{K}(z,\bar{z})  &=\,\sum_{l=0}^{\frac{K}{2}}\left(G_{l,\,\frac{K}{2}-l,\,\frac{K}{2}}\big{|}_{\sigma\to 1\atop\tau\to 1 }\times \tilde{\mathbb{O}}^{2}_{l}(z,\bar{z})\,+\,G_{\frac{K}{2}-l,\,l,\,\frac{K}{2}}\big{|}_{\sigma\to 1\atop\tau\to 1 }\times\tilde{\mathbb{O}}^{2}_{l}\left(\text{\small$\frac{z}{z-1}$},\text{\small$\frac{\bar{z}}{\bar{z}-1}$}\right)\,\right) - \mathbb{A}_{K}^{(0)}(z,\bar{z})
\end{align}
where the (-1) in \eqref{eq:AsymFourAllLoop} and $(-\mathbb{A}^{(0)}_{K})$ in \eqref{eq:AsymFourAllLoop2} remove the overcounting of tree level graphs.

Prediction \eqref{eq:AsymFourAllLoop2} is only valid for loop orders below $\frac{K}{2}$. So in practice various octagons there can just be set to one, such as: $\tilde{\mathbb{O}}_{l=\frac{K}{2}} = 1 + O(g^{K+2})$.

\subsection{An example for finite $K$ and nine loops}\label{sec:Example9}
Here we choose the minimum value of $K$ for which prediction \eqref{eq:AsymFourAllLoop} is valid up to nine loops. For this we need to make sure the skeleton graphs surviving the \textit{asymptotic} projection only admit particles on bridges $l_{14}$ and $l_{23}$ or exclusively on $l_{13}$ and $l_{24}$. Under this criteria we set the minimum value as $K=22$. In this case, at nine loops, the skeleton graphs to take into account are in the range:
\beq\label{eq:Range9}
9\leq l_{13}+l_{14} \leq 15
\eeq
This lower bound on $l_{13}+l_{14}$ guarantees that we can ignore two-particle strings on $l_{13}$-$l_{14}$, as these would kick in at order $\left(g^{2}\right)^{l_{13}+l_{14}+1}$. There is yet a subtlety  as this range also includes  graphs with $l_{12}=7$ and $l_{12}=8$, which can host one-particle states at nine loops. However we have checked that these contributions do not survive the projection, so we ignore them.

Resumming the contributions from the skeleton graphs in \eqref{eq:Range9} and using the \textit{asymptotic} octagon  we express the four point function in terms of graphs satisfying $l_{13}+l_{14}=11$ only:
\beq\label{eq:OctagonNinePrediction}
\mathbb{A}_{22}(z,\bar{z})=\,\sum_{l=0}^{11}\,G_{l,\,11-l,\,11}\big{|}_{\sigma\to 1\atop\tau\to 1 }\,\left(\tilde{\mathbb{O}}^{2}_{l}(z,\bar{z})\,+\,\tilde{\mathbb{O}}^{2}_{11-l}\left(\text{\small$\frac{z}{z-1}$},\text{\small$\frac{\bar{z}}{\bar{z}-1}$}\right)\,-1\right)
\eeq
This prediction can only be trusted up to nine loops. In this respect the corrections to $\mathbb{A}_{22}^{(0)}$  come more explicitly from:
\bba\label{eq:OctagonNinePrediction2}
\mathbb{A}_{22}(z,\bar{z})-\mathbb{A}_{22}^{(0)}(z,\bar{z})\,&= \, u^{11}\,\left( \tilde{\mathbb{O}}^{2}_{l=0}(z,\bar{z})-1\right)\,+2\sum_{l=1}^{8}\,\frac{u^{11}}{v^{l}}\,\left(\tilde{\mathbb{O}}^{2}_{l}(z,\bar{z})-1\right)\, \nonumber\\
&\qquad\quad +\, \left(z\to \text{\small$\frac{z}{z-1}$}\,,\bar{z}\to\text{\small$\frac{\bar{z}}{\bar{z}-1}$}\right)
\end{align}
where we have only kept octagons which contribute non-trivially up to nine loops and  $\left(z\to \text{\small$\frac{z}{z-1}$}\,,\bar{z}\to\text{\small$\frac{\bar{z}}{\bar{z}-1}$}\right)$ denotes a contribution analog to the first line on the right hand side with the cross ratios changed as indicated.

In the following section we use this prediction to perform a OPE test up to nine loops.  In appendix \ref{app:NineLoopIntegrals} we provide the relevant mirror integrals that contribute to \eqref{eq:OctagonNinePrediction2}.

\section{The OPE expansion: a test of the hexagonalization prediction}\label{sec:OPEcheckNineLoops}
An alternative and more standard way of decomposing a four-point correlator is through the conformal operator product expansion. This is also realized through the insertion of a complete set of states but now along a physical cut as depicted in \eqref{eq:figOPE}. In the effective world-sheet these are realized as closed string states and they correspond to the local operators in the spectrum of our 4D CFT, the eigenstates of the dilatation operator. This decomposition effectively expands the four-point function into products of three point functions, depicted as pair of pants in \eqref{eq:figOPE}, of the exchanged operators and the external protected operators.
\beq\label{eq:figOPE}
\centering 
\raisebox{-4077787sp}{ \resizebox{0.20\totalheight}{!}{\includestandalone[width=1\textwidth]{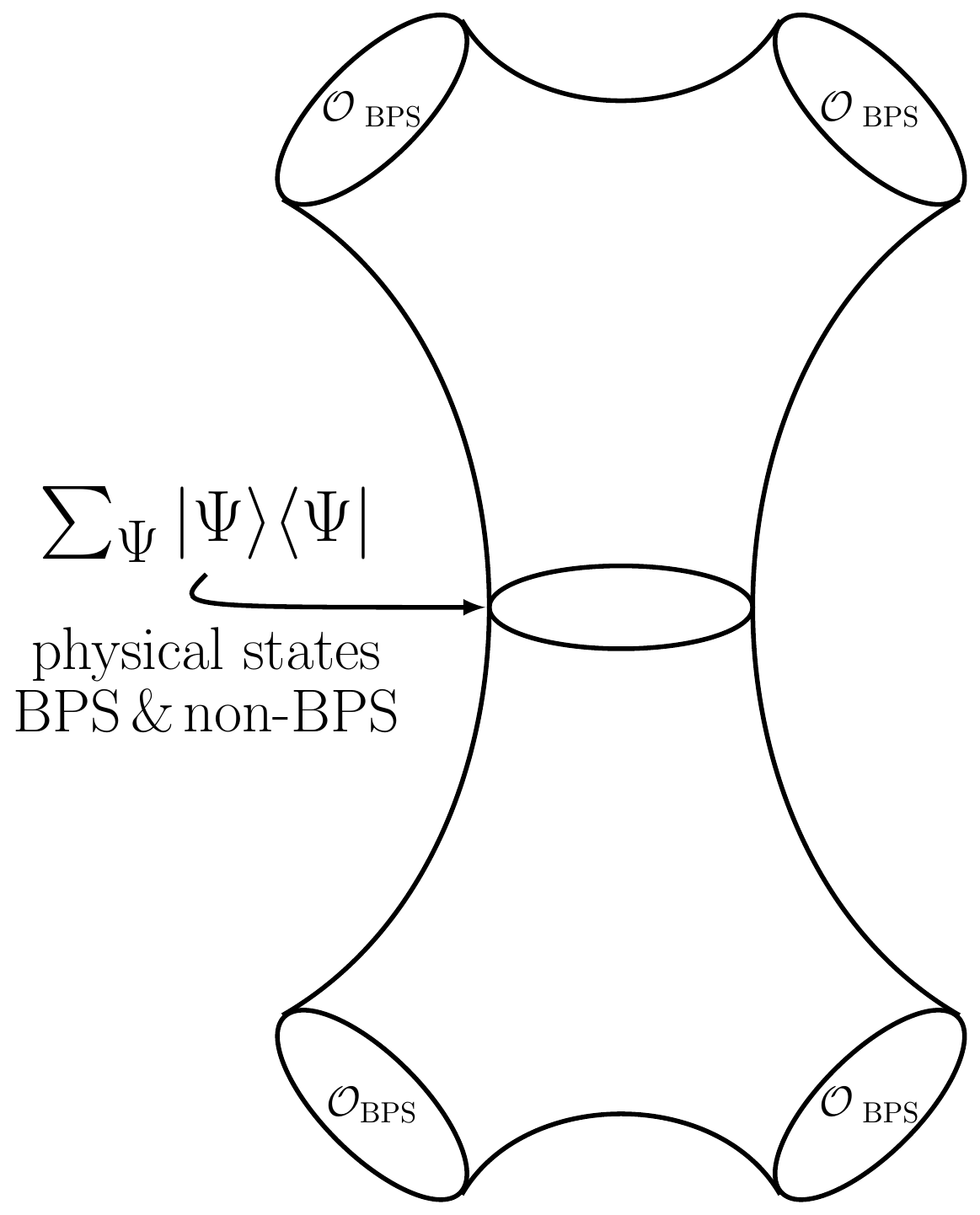}} }  = \;\sum_{ {\text{\tiny Bethe} \atop \text{\tiny solutions} }}
\left(\text{\small $\sum_{ {\text{\tiny Bethe} \atop \text{\tiny roots} }=\alpha\cup\bar{\alpha}}$ }\,\raisebox{-2877787sp}{\resizebox{0.4\totalheight}{!}{\includestandalone[width=.8\textwidth]{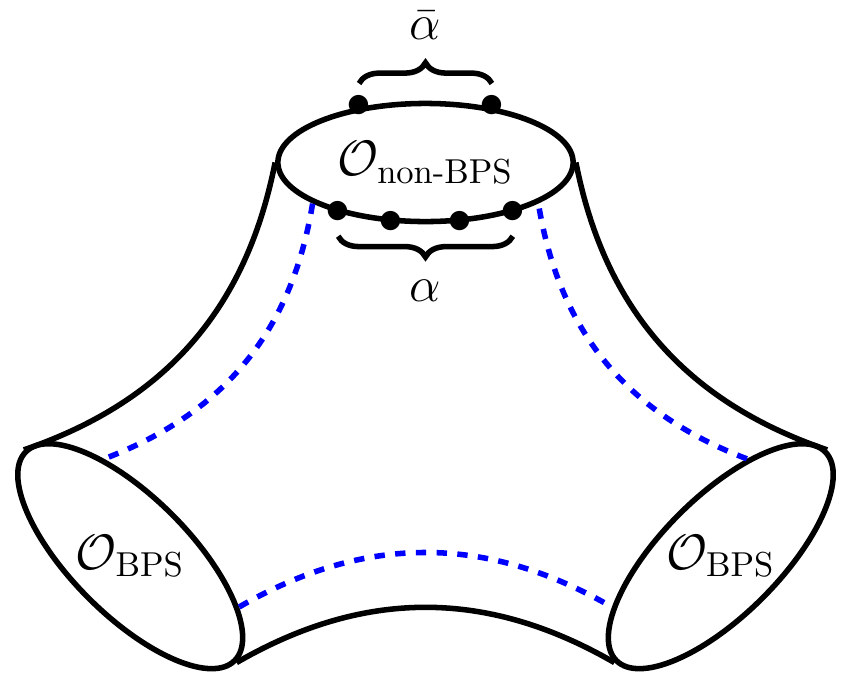}}}\right)^{2}\; \times \;{ \text{\footnotesize(super)}\atop {\text{conformal}  \atop \text{block}}}
\eeq
In \cite{AsymptoticPaper} we prescribed how to build all three-point functions necessary to, in principle, reconstruct the \textit{asymptotic}  four-point function $\mathbb{A}_{K}$ at weak coupling in the limit of $K\gg 1$. There we performed a one-loop check of this prescription by constructing the three-point functions of the lowest lying operators dominating the OPE expansion ($z,\bar{z}\to 0$) and comparing them against the coefficients in the OPE series of the one-loop correlator known from the literature. 

We now endevour to perform the same exercise but now up to nine-loops, comparing the OPE reconstruction  against our hexagonalization nine-loop prediction \eqref{eq:OctagonNinePrediction} for the \textit{asymptotic} four-point function. We perform a minimal check in the sense that we will only match the leading OPE coefficients, but due to mixing, this already requires summing over hundreds of exchanged super-primaries. So we consider that this still constitutes a very non-trivial  consistency check of the integrability methods we used to realized both mirror (hexagonalization) and physical (OPE) decompositions.

\begin{figure}[ht]
\centering
\resizebox{1.65\totalheight}{!}{\includestandalone[width=1\textwidth]{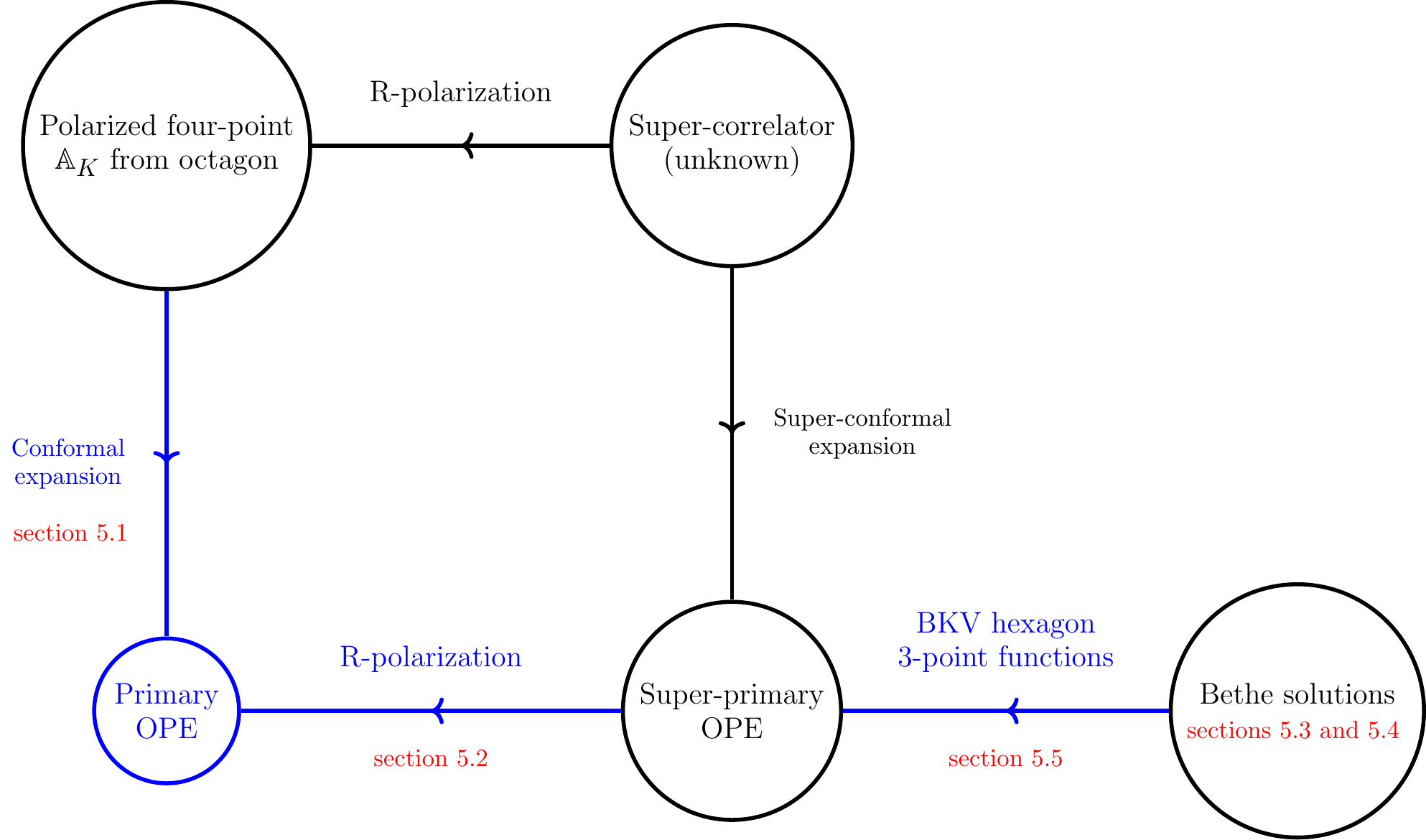}}
\caption{OPE strategy to test octagon at nine loops}
\label{fig:OPEoutline}
\end{figure}

In figure \ref{fig:OPEoutline} we present the steps we follow to perform this test, including the way we organize the remaining sections.

In section \ref{sec:New1} we follow the vertical {\color{blue}blue} arrow in figure \ref{fig:OPEoutline}. We start by performing the ordinary conformal block expansion of the nine-loop prediction \eqref{eq:OctagonNinePrediction} for the \textit{asymptotic} four-point function $\mathbb{A}_{K=22}$. In this way we obtain OPE data for leading $\tau=K$ and subleading $\tau=K+2$ twists  for the lowest spins $s=0$ and $s=2$. This gives us a list of  rational numbers that encode the mixing of OPE data of hundreds of super-conformal multiplets up to nine loops.   This list of numbers is the subject of our test and in the next sections we follow the horizontal {\color{blue}blue} arrows in figure \ref{fig:OPEoutline}  to reproduce them by other means.

In section \ref{sec:New2}, we apply the \textit{asymptotic} polarization to an Ansatz for the super-correlator composed of unknown super-OPE structure constants  and well-known super-conformal blocks. From this simple operation we learn how super-multiplets with different $\mathfrak{so}(6)$ R-charges mix in the OPE of the \textit{asymptotic} four-point function. At this stage we are only able to identity the tree level global numbers which label families of degenerate super-conformal multiplets at zero coupling $g=0$.

In section \ref{sec:New3} we identify the individual super-conformal primaries  at $g\neq 0$ by using the Bethe Ansatz which resolves the mixing problem in planar $\mathcal{N}=4$ SYM. We find the Bethe solutions which identify the {\color{red}1514} $\mathfrak{sl}(2)$ and {\color{red}2186} $\mathfrak{so}(6)$ super-conformal multiplets that mix in the OPE data of section \ref{sec:New1}. In order to find the $\mathfrak{so}(6)$ Bethe solutions, in section \ref{sec:New4}, we introduce an approach based on the simultaneous diagonalization of the higher conserved charges of the integrable spin chain. 

Finally in section \ref{sec:New5} we use the (asymptotic) BKV hexagon formula to compute the structure constants of each of the aforementioned super-conformal primaries. With this super-OPE data we reproduce the OPE-data of section \ref{sec:New1} confirming the correctness of our octagon prediction.

\subsection{The \textit{asymptotic} OPE data from hexagonalization prediction}\label{sec:New1}

Here we consider the conformal block expansion of the \textit{asymptotic} four-point function:
\beq\label{eq:asymptoticOPE}
\mathbb{A}_{K}(z,\bar{z})  \,=\, \sum_{(\Delta,s)}\,C_{12(\Delta,s)}C_{(\Delta,s)34}\;G_{\Delta,s}(z,\bar{z})
\eeq
where the sum runs over all conformal primaries, labeled by the scaling dimension and spin $(\Delta,s)$, in the OPE of $\tilde{O}_{1}$-$\tilde{O}_{2}$ and $\tilde{O}_{3}$-$\tilde{O}_{4}$. The summand contains the unknown structure constants and the well known ordinary conformal block $G_{\Delta,s}$, given in \eqref{eq:ExpandConformalBlock}.

We are interested in extracting the lowest lying OPE data given our knowledge of the correlator $\mathbb{A}_{K}$ from hexagonalization. Furthermore, since we only know the correlator at weak coupling we also need to consider the expansion of the OPE data as:
\bba\label{eq:Cgexpansion}
\Delta \,&=\, \Delta_{0} + g^{2} \, \gamma^{(1)} +\cdots \nonumber\\
C_{12(\Delta,s)}\,&=\,C^{(0)}_{12(\Delta,s)}\,+\,g^{2}\, C^{(1)}_{12(\Delta,s)}\, +\,\cdots 
\end{align}
A disadvantage of this weak coupling expansion is the huge degeneracy around the reference point $g=0$. At this point we can only organize the degenerate primary operators  into families that we name \textit{Classes} distinguished by the global tree level quantum numbers $\{\Delta_{0},s\}$. When turning on the coupling (weakly)  this degeneracy is lifted and each primary operator can be identified by its correspondent (small) anomalous dimension $\delta\Delta= \Delta-\Delta_{0}$. Yet in the OPE, as shown in more detail in appendix \ref{app:WeakOPE}, the weak coupling expansion in powers of $g^{2}$  forces the OPE data  to accumulate and appear only in sum rules $\mathsf{P}_{\{\Delta_{0},s\}}$ that mix loop corrections of  structure constants and anomalous dimensions of primaries in the same \textit{Class} $\{\Delta_{0},s\}$.  These sum rules can be neatly defined through the following generating function:
\beq\label{eq:GeneratePrimary}
\sum_{(\Delta,s)\,\in\,\textit{Class}\,\{\Delta_{0},s\}} C_{12(\Delta,s)}C_{(\Delta,s)34}\,r^{\Delta} = \,r^{\Delta_{0}}\sum_{a=0}^{\infty}\,\left(g^{2}\right)^{a} \, \sum_{b=0}^{a}\, \log(r)^{b}\, \mathsf{P}^{(a,b)}_{\{\Delta_{0},s\}}\,
\eeq
where $r$ is just a book keeping variable (or the radial coordinate of appendices \ref{app:Blocks},\ref{app:WeakOPE}) and $\textit{Class}\,\{\Delta_{0},s\}$  stands for the family of conformal primaries $(\Delta,s)$ which become degenerate at $g=0$ with tree-level charges $\{\Delta_{0},s\}$.

We use the octagon prediction \eqref{eq:OctagonNinePrediction} for the asymptotic correlator $\mathbb{A}_{K=22}$ at weak coupling and  consider the limit $z,\bar{z}\to 0$ in both sides of \eqref{eq:asymptoticOPE}  to solve for the sums rules of the lowest lying primary operators in the OPE.  We obtain all sum rules $\mathbb{P}^{(a,b)}_{\{\Delta_{0},s\}}$ with $9\geq a \geq b \geq 1$ for leading twist $\tau=22\,:\{\{\Delta_{0}=22,s=0\}\,,\,\{\Delta_{0}=24,s=2\}\}$ and subleading twist  $\tau=24\,:\{\{\Delta_{0}=24,s=0\}\,,\,\{\Delta_{0}=26,s=2\}\}$. Some instances of these sum rules are given in tables \ref{tab:predictionP90} and \ref{tab:predictionP99}.

\begin{table}[ht]
\centering
\def\arraystretch{2}
\begin{tabular}{|c|c|c|c|}
\hline
Twist-22 &   Twist-24   \\ 
\hline
\hline
$\mathsf{P}^{(9,0)}_{\{24,2\}} = -2755264512 $& $\mathsf{P}^{(9,0)}_{\{24,0\}}= -2755264512$  \\
\hline
$\mathsf{P}^{(9,0)}_{\{26,4\}} =
\text{\small $-\frac{25906812477093695541612336367}{562220051373121536}$}$&  $\mathsf{P}^{(9,0)}_{\{26,2\}}=\text{\small $-\frac{13565855042891885605834502859803364601}{100890752343750000000000000}$}$ \\
\hline
\end{tabular}
\caption{}
\label{tab:predictionP90}
\end{table}

\begin{table}[ht]
\centering
\def\arraystretch{2}
\begin{tabular}{|c|c|c|c|}
\hline
Twist-22 &   Twist-24   \\ 
\hline
\hline
$\mathsf{P}^{(9,9)}_{\{24,2\}} = \frac{4978688}{945} $& $\mathsf{P}^{(9,9)}_{\{24,0\}}= \frac{4978688}{945}$  \\
\hline
$\mathsf{P}^{(9,9)}_{\{26,4\}} =
\frac{2990537878301}{29760696}$&  $ \mathsf{P}^{(9,9)}_{\{26,2\}}=\frac{546998336876863}{1860043500}$ \\
\hline
\end{tabular}
\caption{}
\label{tab:predictionP99}
\end{table}

From the definition \eqref{eq:GeneratePrimary} it follows that only $\mathsf{P}^{(9,0)}_{\{\Delta_{0},s\}}$ contains genuine nine-loop OPE data, while the sum rules $\mathsf{P}^{(9,b)}_{\{\Delta_{0},s\}}$ with $b<9$ are composed of lower loop OPE data. In this sense we will consider that reproducing the rational numbers in table, by the methods described below, will constitute our strongest test of our hexagonalization nine-loop prediction.

\subsection{The \textit{asymptotic} OPE data from  the super-OPE data}\label{sec:New2}

In this section we identify the families of super-conformal multiplets with different $\mathfrak{so}(6)$ charges which mix under the \textit{asymptotic} $R$-charge projection and contribute to the sum rules in tables \ref{tab:predictionP90} and \ref{tab:predictionP99}. Samples of the results of this section are contained in tables \ref{tab:P90super} and \ref{tab:P99super}. 

Our analysis starts with  unpolarized external operators such that we can  organize the intermediate operators in the OPE into super-multiplets and later perform the $R$-charge projection. 

The super-OPE decomposition of the unpolarized correlator is given by:
\beq\label{eq:SuperExpansion}
\mathcal{G}_{K}(z,\bar{z},\alpha,\bar{\alpha}) = (\text{protected}) +  \sum_{(\Delta,s,m,m)}\,C^{2}_{K,K,(\Delta,s,m,n)}\, \mathcal{F}_{\Delta,s,n,m}(z,\bar{z},\alpha,\bar{\alpha})
\eeq  
where $(\text{protected})$ stands for the contributions of short and semi-short multiplets which are tree level exact. While at loop level we only need to consider the sum over long super-primaries  $(\Delta,s,n,m)$:
\beq
\Delta: \text{scaling dimension} \qquad\, s: \text{spin}\qquad \text{and}\qquad [n-m,2m,n-m]: \mathfrak{so}(6)\,\text{representation}\, 
\eeq 
The product of structure constants $C^{2}_{K,K,(\Delta,s,m,n)}$ weight the contribution of the correspondent super-conformal multiplet given by the super-conformal block:
\beq
\mathcal{F}_{\Delta,s,n,m}(z,\bar{z},\alpha,\bar{\alpha})=\frac{(z-\alpha)(z-\bar{\alpha})(\bar{z}-\alpha)(\bar{z}-\bar{\alpha})}{(z\bar{z})^{2}}\times Y_{n,m}(\alpha,\bar{\alpha})\times G_{\Delta+4,s}(z,\bar{z})
\eeq
These super-blocks are given by the simple product of the $\mathfrak{so}(6)$ $R$-charge block  $Y_{n,m}$   and the $\mathfrak{so}(4,2)$  conformal block $G_{\Delta,s}$. Explicit expressions for these functions can be found in appendix \ref{app:Blocks}.

In analogy with the ordinary OPE, at weak coupling the super-conformal expansion organizes the super-OPE data around $g=0$ accumulation points. Specifically the super-primaries data mix into sum rules $\mathcal{P}^{(a,b)}_{\{\Delta_{0},s,n,m\}}$ which can be read off from the generating function:
\beq\label{eq:GenerateSuper}
\, \sum_{(\Delta,s,m,n)\,\in\, \textit{Super-Class}\,\{\Delta_{0},s,n,m\}}   \, C^{2}_{KK(\Delta,s,m,n)} \,r^{\Delta}  \, = \,r^{\Delta_{0}}\sum_{a=0}^{\infty}\,\left(g^{2}\right)^{a} \, \sum_{b=0}^{a}\, \log(r)^{b}\, \mathcal{P}^{(a,b)}_{\{\Delta_{0},s,m,n\}}\,
\eeq
where $\textit{Super-Class}$ $\{\Delta_{0},s,n,m\}$  stands for the family of super-conformal primaries $(\Delta,s,n,m)$ which become degenerate with tree-level charges $\{\Delta_{0},s,n,m\}$ when turning off the coupling.

In order to find the dependence of the \textit{asymptotic} four-point function on the super-OPE data, we  simply perform the $R$-charge projection on the super-conformal blocks. For this we rewrite the $\mathfrak{so}(6)$ blocks in terms of cross ratios  $\sigma$, $\tau$ and perform the \textit{asymptotic} projection as:
\beq\label{eq:AprojectionOPE}
\eqref{eq:SuperExpansion}\big{|}_{\text{coefficient of}\;\sigma^{-\frac{K}{2}},\tau\to 1}\quad \Longrightarrow \quad \mathbb{A}_{K}(z,\bar{z})   = \sum_{(\Delta,s)}\,C_{12(\Delta,s)}C_{(\Delta,s)34}\;G_{\Delta,s}(z,\bar{z})
\eeq
On the right hand side we have further re-expanded the polarized super-blocks in terms of ordinary conformal blocks. This allows us to read off the \textit{asymptotic} primary OPE data in terms of the super-primary OPE data.  More specifically we find the primary sum rules $\mathsf{P}^{(a,b)}_{\{\Delta_{0},s\}}$ as linear combinations of super-sum rules $\mathcal{P}^{(a,b)}_{\{\Delta_{0},s,n,m\}}$, with the coefficients weighting this latter coming from the polarized super-block. In tables \ref{tab:P90super} and \ref{tab:P99super} we show some instances of these linear combinations relevant for our nine-loop test with $K=22$. In table \ref{tab:SuperClassesK} we show the \textit{Super-Classes} $\{\Delta_{0},s,n,m\}$ that mix into \textit{Classes} $\{\Delta_{0},s\}$ for leading and subleading twists for generic $K\geq 4$.

\begin{table}[ht]
\centering
\def\arraystretch{2}
\begin{tabular}{|c|c|c|c|}
\hline
$\tau=\Delta_{0}-s$ &$\text{\normalsize $\textit{Class}$}\atop {\,\atop \text{\normalsize $\{\Delta_{0},\,s\}\atop$}}$&$\text{\normalsize $\textit{Super-Class}$}\atop {\,\atop \text{\normalsize $\{\Delta_{0},\,s,\,n,\,m\}\atop$}}$ \\  \hline
\hline
\multirow{2}{*}{$K$} & $\{K\,,0\}$  &  {\color{blue}$\{K\,,0\,,\frac{K}{2}-1\,,\frac{K}{2}-1\}$} \\  \cline{2-3}
  & $\{K+2\,,2\}$ & $\{K+2\,,2\,,\frac{K}{2}-1\,,\frac{K}{2}-1\}$   \\ \hline 
\multirow{8}{*}{$K+2$} &  & $\{K-2\,,0\,,\frac{K}{2}-2\,,\frac{K}{2}-2\}$ \\ 
  & $\{K+2\,,0\}$ &  {\color{blue}$\{K\,,0\,,\frac{K}{2}-1\,,\frac{K}{2}-1\}$ } \\ 
  & & {\color{blue}$\{K+2\,,0\,,\frac{K}{2}\,,\frac{K}{2}\}$}  \\ \cline{2-3}
  & &  $\{K,2\,,\frac{K}{2}-2\,,\frac{K}{2}-2\}$\\ 
  & &  $\{K+2\,,0\,,\frac{K}{2}-1\,,\frac{K}{2}-1\}$\\
  & $\{K+4\,,2\}$ &  $\{K+2\,,0\,,\frac{K}{2}\,,\frac{K}{2}-2\}$\\ 
  & &  {\color{blue}$\{K+2\,,0\,,\frac{K}{2}\,,\frac{K}{2}\}$ }\\
  & &  $\{K+2\,,2\,,\frac{K}{2}\,,\frac{K}{2}\}$\\
  \hline
\end{tabular}
\caption{Twist, primaries and super-primaries. The super-multiplets mixed in the ordinary conformal expansion of the asymptotic four-point function for leading and sub-leading twists. We highlight in {\color{blue}blue} the super-primary classes which contribute to more than one primary class. }
\label{tab:SuperClassesK}
\end{table}

\begin{table}[ht]
\centering
\def\arraystretch{2}
\begin{tabular}{|c|c|c|c|}
\hline
Twist-22 &   Twist-24   \\ 
\hline
\hline
$\mathsf{P}^{(9,0)}_{\{24,2\}} = \mathcal{P}^{(9,0)}_{\{22,0,10,10\}} $& $\mathsf{P}^{(9,0)}_{\{24,0\}}=\mathcal{P}^{(9,0)}_{\{20,0,9,9\}} -\mathcal{P}^{(9,0)}_{\{22,0,10,10\}}+\mathcal{P}^{(9,0)}_{\{24,0,11,11\}}$  \\
\hline
$\mathsf{P}^{(9,0)}_{\{26,4\}} =\mathcal{P}^{(9,0)}_{\{24,2,10,10\}}$&  $\mathsf{P}^{(9,0)}_{\{26,2\}}= \left(\text{see appendix }\ref{app:LargeNineOPE}\right)$ \\
\hline
\end{tabular}
\caption{}
\label{tab:P90super}
\end{table}

\begin{table}[ht]
\centering
\def\arraystretch{2.3}
\begin{tabular}{|c|c|c|c|}
\hline
Twist-22 &   Twist-24   \\ 
\hline
\hline
$\mathsf{P}^{(9,9)}_{\{24,2\}} = \mathcal{P}^{(9,9)}_{\{22,0,10,10\}} $& $\mathsf{P}^{(9,9)}_{\{24,0\}}=\mathcal{P}^{(9,9)}_{\{20,0,9,9\}} -\mathcal{P}^{(9,9)}_{\{22,0,10,10\}}+\mathcal{P}^{(9,9)}_{\{24,0,11,11\}} $  \\
\hline
$\mathsf{P}^{(9,9)}_{\{26,4\}} =\mathcal{P}^{(9,9)}_{\{24,2,10,10\}}$&  $\text{\normalsize $\mathsf{P}^{(9,9)}_{26,2}= \mathcal{P}^{(9,9)}_{\{22,2,9,9\}} -\mathcal{P}^{(9,9)}_{\{24,0,10,10\}}+\mathcal{P}^{(9,9)}_{\{24,0,11,9\}}$}\atop \text{\normalsize $\qquad\qquad - \text{\small $\frac{526199}{350175}$}\,\mathcal{P}^{(9,9)}_{\{24,0,11,11\}} +\mathcal{P}^{(9,9)}_{\{26,2,11,11\}} $}$ \\
\hline
\end{tabular}
\caption{}
\label{tab:P99super}
\end{table}

In order to perform our nine-loop test we still need to compute the super-sum rules  before we match the results of tables \ref{tab:P90super}, \ref{tab:P99super} with tables \ref{tab:predictionP90}, \ref{tab:predictionP99}. For this we follow the steps described in the following sections. 

\subsection{Resolving mixing with Bethe Ansatz: identifying super-multiplets}\label{sec:New3}

In order to find the sum rule $\mathcal{P}^{(a,b)}_{\{\Delta_{0},s,m,n\}}$ we first need to identify the super-primaries in the \textit{Super-Class} $\{\Delta_{0},s,m,n\}$. Fortunately, in planar $\mathcal{N}=4$ SYM,  integrability  resolves this mixing of single-traces\footnote{For the twists $K$ and $K+2$  we only need to consider single-traces. At weak coupling the mixing with double-traces starts at twist $2K$.} through a Bethe Ansatz with underlying global symmetry $\mathfrak{psu}(2,2|4)$. Besides since we are dealing with long operators it suffices to consider only the asymptotic Bethe Ansatz: the Beisert-Staudacher(BS) equations. Then the mixing is resolved by identifying each super-primary or super-multiplet with  a solution of the BS equations\footnote{Taking into account that we should only consider solutions with finite and pairwise distinct roots.}.

In general a solution of BS equations consists of seven sets of Bethe roots, where each set is associated to one of the seven nodes of the $\mathfrak{psu}(2,2|4)$ Dynkin diagram. The number of roots for each node is determined by the tree level global numbers of the super-primary. 

\begin{figure}[ht]
\centering
\resizebox{1.8\totalheight}{!}{\includestandalone[width=1\textwidth]{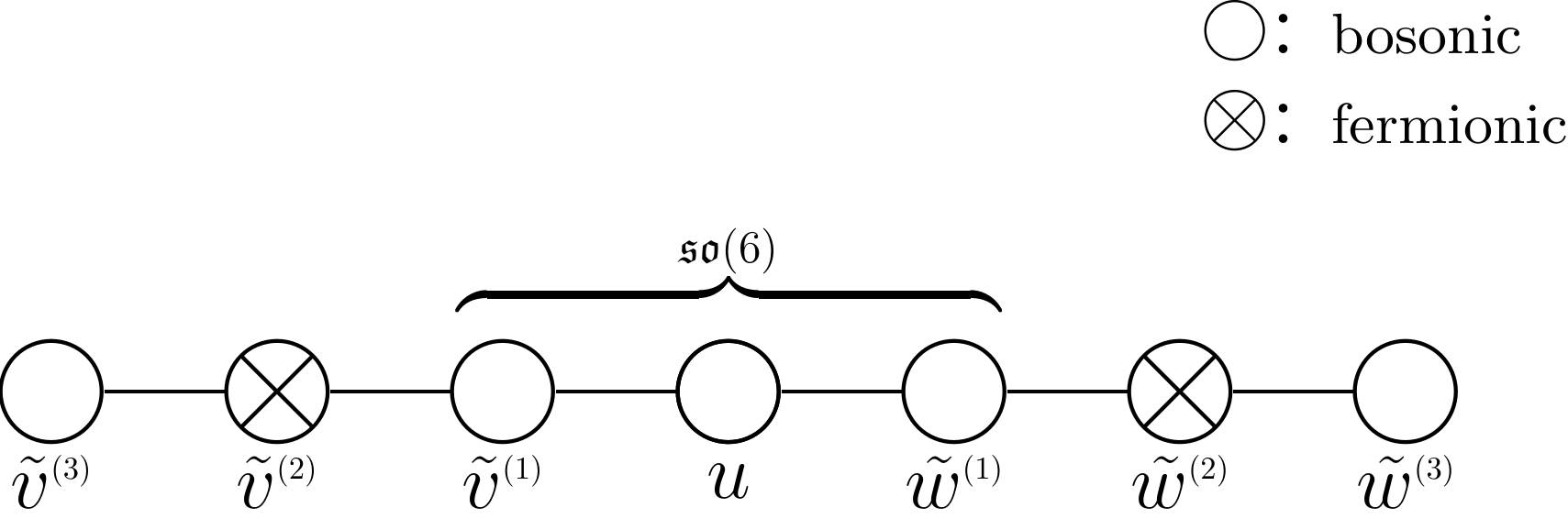}}
\caption{Beauty grading: the momentum carrying roots $u_{j}$ are associated to the $\mathfrak{su}(2)$ middle node. These roots determine the anomalous dimension.  The auxiliary roots $v^{(a)}_{j}$ and $w^{(a)}_{j}$  realized a copy of $\mathfrak{su}(2|2)$ on each wing and complete the full  $\mathfrak{psu}(2,2|4)$. The three middle bosonic nodes conform a $\mathfrak{so}(6)$ Dynkin diagram.}
\label{fig:Beauty}
\end{figure}

\begin{figure}[ht]
\centering
\resizebox{2.7\totalheight}{!}{\includestandalone[width=1\textwidth]{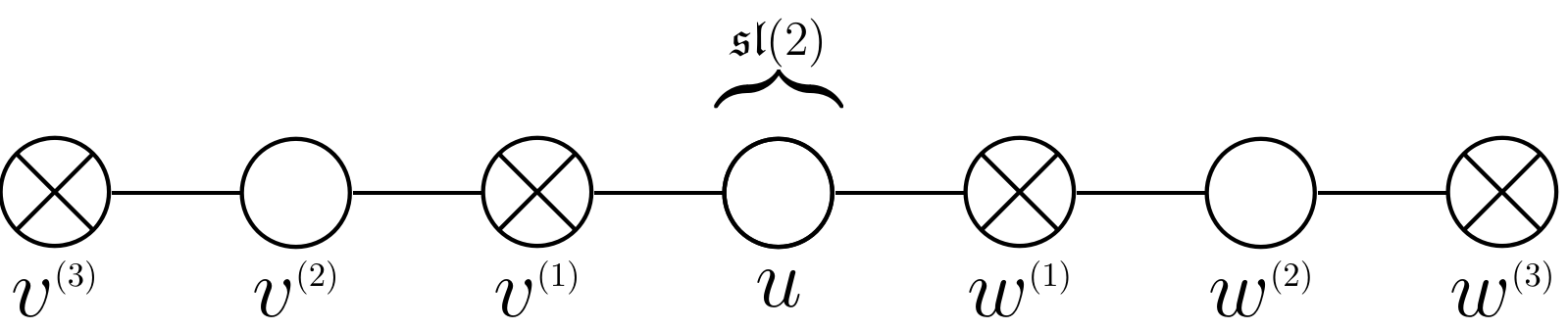}}
\caption{$\mathfrak{sl}(2)$ grading, the middle node realizes the non-compact $\mathfrak{sl}(2)$ sector, the momentum carrying roots $u_{j}$ are exactly the same as in the Beauty grading. The wing roots $v^{(a)}_{j}$ and $w^{(a)}_{j}$ are in general different in number and value with respect to the Beauty wings $\tilde{v}^{a}_{j}$ and $\tilde{w}^{(a)}_{j}$. The roots $v^{(1)}_{j}$ and $w^{(1)}_{j}$ are now fermionic and together the middle node realize the non-compact $\mathfrak{su}(1,1|2)$ sector.}
\label{fig:sl2grading}
\end{figure}

As is typical in super-groups, a $\mathfrak{psu}(2,2|4)$ super-multiplet admits various choices of superprimary or highest weight depending on the grading (Dynkin diagram) we use, see  \cite{SuperDynkin} for a review. In the super-OPE expansion \eqref{eq:SuperExpansion} we are using the so called Beauty grading, see figure \ref{fig:Beauty}. Alternatively we also have the $\mathfrak{sl}(2)$ grading, see figure \ref{fig:sl2grading}, whose super-primary has shifted global charges respect to the Beauty's. The number of Bethe roots changes accordingly with the choice of grading. Nevertheless these different descriptions of the same multiplet are related through the so called $Q$-$Q$ relations, see \cite{QQspectrumSolver}. These allow us to find the Bethe roots of any grading once we have found them in a particular one. 

For the purpose of solving the Bethe equations it is typically more convenient to first work on the  grading with the minimum number of Bethe roots. In table \ref{tab:NumberRoots} we list the number of Bethe roots in two different gradings for the \textit{Super-Classes} $\{\Delta_{0},s,m,n\}$ of table \ref{tab:SuperClassesK} with $K=22$.
\begin{table}[ht]
\centering
\def\arraystretch{1.25}
\resizebox{2.5\totalheight}{!}{
\begin{tabular}{|c|c|c|}
\hline
$\{\Delta_{0},s,n,m\}$& Beauty grading & $\mathfrak{sl}(2)$ grading\\ 
\hline
\hline
$\{20,0,9,9\}$&$L=20$\;,\;$\{0,0,1,2,1,0,0\}$\;,\;$\text{Tr}\left(Z^{19}\bar{Z}\right)$&{\color{blue}$L=20$\;,\;$\{0,0,0,2,0,0,0\}$}\;,\;$\text{Tr}\left(\mathcal{D}^{2}Z^{20}\right)$  \\ 
\hline
$\{22,0,10,10\}$&$L=22$\;,\;$\{0,0,1,2,1,0,0\}$\;,\;$\text{Tr}\left(Z^{21}\bar{Z}\right)$&{\color{blue}$L=22$\;,\;$\{0,0,0,2,0,0,0\}$}\;,\;$\text{Tr}\left(\mathcal{D}^{2}Z^{22}\right) $ \\
\hline
$\{24,0,11,11\}$&$L=24$\;,\;$\{0,0,1,2,1,0,0\}$\;,\;$\text{Tr}\left(Z^{23}\bar{Z}\right)$&{\color{blue}$L=24$\;,\;$\{0,0,0,2,0,0,0\}$}\;,\;$\text{Tr}\left(\mathcal{D}^{2}Z^{24}\right)$ \\
\hline
$\{22,2,9,9\}$&$L=20$\;,\;$\{0,1,2,4,2,1,0\}$\;,\;$\text{Tr}\left(\mathcal{D}^{2}Z^{19}\bar{Z}\right)$&{\color{blue}$L=20$\;,\;$\{0,0,0,4,0,0,0\}$}\;,\;$\text{Tr}\left(\mathcal{D}^{4}Z^{20}\right)$ \\
\hline
$\{24,2,10,10\}$&$L=22$\;,\;$\{0,1,2,4,2,1,0\}$\;,\;$\text{Tr}\left(\mathcal{D}^{2}Z^{21}\bar{Z}\right)$&{\color{blue}$L=22$\;,\;$\{0,0,0,4,0,0,0\}$}\;,\;$\text{Tr}\left(\mathcal{D}^{4}Z^{22}\right)$ \\ 
\hline
$\{26,2,11,11\}$&$L=24$\;,\;$\{0,1,2,4,2,1,0\}$\;,\;$\text{Tr}\left(\mathcal{D}^{2}Z^{23}\bar{Z}\right)$&{\color{blue}$L=24$\;,\;$\{0,0,0,4,0,0,0\}$}\;,\;$\text{Tr}\left(\mathcal{D}^{4}Z^{24}\right)$ \\
\hline
$\{24,0,11,9\}$&{\color{blue}$L=24$\;,\;$\{0,0,0,4,0,0,0\}$}\;,\;$\text{Tr}\left(Z^{21}X^{2}\bar{Z}\right)$&$L=24$\;,\;$\{0,0,2,4,2,0,0\}$\;,\;$\text{Tr}\left(\mathcal{D}^{2}Z^{22}X^{2}\right)$ \\
\hline
$\{24,0,10,10\}$&{\color{blue}$L=24$\;,\;$\{0,0,2,4,2,0,0\}$}\;,\;$\text{Tr}\left(Z^{22}\bar{Z}^{2}\right)$&$L=24$\;,\;$\{0,1,2,4,2,1,0\}$\;,\;$\text{Tr}\left(\mathcal{D}^{2}Z^{23}\bar{Z}\right)$ \\
\hline 
\end{tabular}}
\caption{The 8 \textit{Super-Classes} labelled by $\{\Delta_{0},s,n,m\}$ and their spin chain description. We specify the length $L$ of the spin chain , the seven occupation number of the nodes in the Dynkin diagram and the field content in terms of scalars and light-cone derivatives $\mathcal{D}$, in both beauty and $\mathfrak{sl}(2)$ gradings. The \textit{Super-Classes} are ordered according to the difficulty to find their correspondent Bethe solutions. We highlight in {\color{blue}blue} the descriptions with the minimun number of roots.}\label{tab:NumberRoots}
\end{table}

We recognize the first six \textit{Super-Classes}  in table \ref{tab:NumberRoots} admit a description as part of the $\mathfrak{sl}(2)$ sector. This means for all these super-multiplets  there exists a choice of super-primary containing only light-cone derivatives $\mathcal{D}$ and scalars $Z$: $\text{Tr}(\mathcal{D}^{s}Z^{L})$. For these states the number of Bethe roots is $s$, the same as the number of derivatives, and have empty wings in the $\mathfrak{sl}(2)$ grading. These Bethe roots can be found by solving the correspondent $BS$ equations of this sector:  
\bba
\text{Finite coupling\qquad\qquad\qquad\qquad} &\qquad \text{\qquad\quad  Leading order, }g=0\nonumber\\ 
\left(\frac{x^{+}_{j}}{x^{-}_{j}}\right)^{L} = \prod_{k=1\atop k\neq j}^{s} \frac{x^{-}_{j}- x^{+}_{k}}{x^{+}_{j}- x^{-}_{k}}\,\frac{1-\frac{1}{x^{+}_{k}x^{-}_{j}}}{1-\frac{1}{x^{-}_{k}x^{+}_{j}}}\,\sigma^{2}(u_{k},u_{j})\quad &\overset{g\to 0}{\longrightarrow}\quad \left(\frac{u_{j}+\frac{i}{2}}{u_{j}-\frac{i}{2}}\right)^{L} = \prod_{k=1\atop k\neq j}^{s} \frac{u_{j}-u_{k}-i}{u_{j}-u_{k}+i}
\end{align}
Here we are using the short hand notation for the Zhukovsky variable $x^{\pm}_{j}\equiv x(u_{j}\pm\frac{i}{2})$ (see \eqref{eq:Zhukovsky}) and  $\sigma(u_{k},u_{j})$ represents the dressing phase.

We are interested in solving these BS equations at weak coupling up to nine loops. For this we assume 
a loop expansion of the Bethe roots $u_{j} = u^{(0)}_{j} + g^{2} u^{(1)}_{j} +\cdots$, and 
 start by solving the leading order equations at $g=0$ for $u^{(0)}_{j}$.  At this order the BS equations  become polynomial and easier to deal with. 
 
 The leading order $\mathfrak{sl}(2)$ equations are particularly simple as they only admit solutions with real Bethe roots. Taking advantage of this feature we follow a numerical algorithm described in appendix \ref{app:BetheSolutions} which allows to find all $\mathfrak{sl}(2)$ Bethe solutions that  identify all multiplets in the first six classes of table \ref{tab:NumberRoots}. Afterwards finding the loop corrections of Bethe roots  is straightforward. The corrections $u^{(a)}_{j}$ follow from a linearized system of equations that takes as input the lower order corrections $u^{(b)}_{j}$, $0\leq b<a$. In this way we can systematically get the roots up to arbitrarily high loop order starting with the knowledge of the leading order $u^{(0)}_{j}$.

The last two \textit{Super-Classes} in table \ref{tab:NumberRoots} admit super-primaries in the $\mathfrak{so}(6)$ sector when described by the Beauty grading. This sector is spanned by operators whose elementary fields include the three complex scalars $Z,X,Y$, but it is only closed at one loop order. This means that at higher loop orders the  operators, eigenstates of the dilatation generator, can include admixtures with other elementary fields (such as covariant derivatives $\mathcal{D}\bar{\mathcal{D}}$). Nevertheless the identification of the multiplets by solving the leading $\mathfrak{so}(6)$ Bethe equations remains valid and we can use the leading order Bethe roots as a seed to find the roots at any desired loop order without caring about the specific form of the operators.

A real disadvantage of the $\mathfrak{so}(6)$ BS  equations, is that the corresponding leading order  equations ($g=0$) now admit complex solutions. This is in fact the generic case for the full sector $\mathfrak{psu}(2,2|4)$. While the real solutions can be found using the same method as for the $\mathfrak{sl}(2)$ sector, finding all complex solutions is much more challenging. Some of these complex solutions can be singular and need a regularization. There is also the issue that some solutions can be spurious and can not be identified with super-multiplets of $\mathcal{N}=4$ SYM.

 A reliable way of finding all physical complex solutions is through  the recent approach of \cite{QQspectrumSolver}, which makes use of the Q-system of $\mathfrak{psu}(2,2|4)$ exploiting consistency in all gradings at once. This method was very useful for the one-loop test performed in \cite{AsymptoticPaper} where only the lowest OPE spectrum was needed. However its efficiency drops when the length of the operators and number of roots are relatively long\footnote{It is possible to push to longer lengths if we changed $\mathsf{Solve}$ by $\mathsf{Nsolve}$ in the Mathematica notebook attached to \cite{QQspectrumSolver}.}, in particular the lengths $L$ of the operators in table \ref{tab:NumberRoots} turned out to be too long.
 
  So we needed to resort to a different approach described in section \ref{sec:New4}. We found that  a direct diagonalization of the one-loop mixing matrix was still feasible for our values of $L$. Combining this with the leading order BS equations we were able to find all correspondent $\mathfrak{so}(6)$ solutions. Then finding corrections up to nine loops followed straightforwardly as described before.  

In summary, using the methods described in appendix \ref{app:BetheSolutions} and section \ref{sec:New4} we were able to identify all multiplets in the \textit{Super-Classes} of table \ref{tab:NumberRoots}, with a total of {\color{red}1514} $\mathfrak{sl}(2)$ and {\color{red}2186} $\mathfrak{so}(6)$ Bethe solutions. The roots of these solutions constitute the input to find the OPE data of the super-primaries as we describe in the following section.

\subsection{$\mathfrak{so}(6)$ Bethe roots from diagonalization of higher conserved charges}\label{sec:New4}
In this section we address how to find the $\mathfrak{so}(6)$ Bethe roots relevant for our OPE test. This section is self-contained and may be skipped by the reader uninterested in these technical steps.

We consider the  Bethe equations of a closed and cyclic $\mathfrak{so}(6)$ spin chain of length $L$ with Dynkin diagram and Bethe roots:
\begin{figure}[ht]
\centering 
\resizebox{1\totalheight}{!}{\includestandalone[width=1\textwidth]{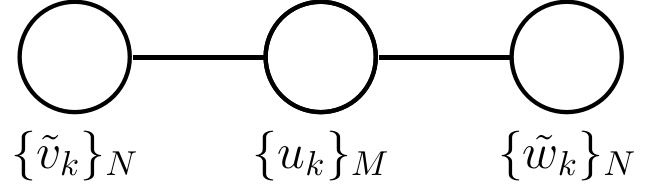}}
\caption{$\mathfrak{so}(6)$ Dynkin diagram.} 
\end{figure}
We consider the same number of roots on each wings in order to describe spaces with $\mathfrak{so}(6)$ weights $[M-2N,L-2M-2N,M-2N]$. The correspondent Bethe equations and zero-momentum condition are: 
\bba\label{eq:so6equations}
1&=\prod_{m\neq k }^{M} \frac{\tilde{v}_{k}-u_{m}-\frac{i}{2}}{\tilde{v}_{k}-u_{m}+\frac{i}{2}}\,\prod_{n\neq k}^{M} \frac{\tilde{v}_{k}-\tilde{v}_{n}+i}{\tilde{v}_{k}-\tilde{v}_{n}-i}\nonumber\\
\left(\frac{u_{k}+\frac{i}{2}}{u_{k}-\frac{i}{2}}\right)^{L}&=\prod_{m\neq k }^{M} \frac{u_{k}-u_{m}+i}{u_{k}-u_{m}-i}\,\prod_{n=1}^{N}\frac{u_{k}-\tilde{v}_{n}-\frac{i}{2}}{u_{k}-\tilde{v}_{n}+\frac{i}{2}}\,\prod_{j=1}^{N}\frac{u_{k}-\tilde{w}_{n}-\frac{i}{2}}{u_{k}-\tilde{w}_{n}+\frac{i}{2}}\nonumber\\
1&=\prod_{m\neq k }^{M} \frac{\tilde{w}_{k}-u_{m}-\frac{i}{2}}{\tilde{w}_{k}-u_{m}+\frac{i}{2}}\,\prod_{n\neq k}^{M} \frac{\tilde{w}_{k}-\tilde{w}_{n}+i}{\tilde{w}_{k}-\tilde{w}_{n}-i}\nonumber\\
1&= \prod_{m=1}^{M}\frac{u_{m}+\frac{i}{2}}{u_{m}-\frac{i}{2}}
\end{align}
As done in appendix \eqref{app:BetheSolutions} for $\mathfrak{sl}(2)$,
these $\mathfrak{so}(6)$ equations can be put in logarithmic form and then by specifying the branch mode numbers we can easily solve them numerically. In this way we can find all real solutions. However these equations also admit complex solutions and they are the hardest to find. 

The complete set of solutions could be in principle found using the $Q$-solver of \cite{QQspectrumSolver} but the lengths of our spin chains $L=24$ and the number of roots (8) turned out to be too high for this method. Fortunately we found out that for these lengths a direct diagonalization of the $\mathfrak{so}(6)$ Hamiltonian was still feasible. In what follows we describe the strategy we followed to obtain the Bethe roots from the direct diagonalization. 

The integrability of this spin chain is guaranteed by the existence of a family of local conserved charges in convolution, of which the Hamiltonian is the simplest element. Each state on the spin chain can be identified by their eingevalues $\{\lambda\}$ under these local charges. These  constitute the integrability counterpart of the Dynkin weights for the global $\mathfrak{so}(6)$ symmetry. Tipically to find this set of eigenvalues we use the knowledge of the Bethe roots obtained by solving Bethe equations:
\beq
\text{$\text{momentum carrying} \atop \text{\normalsize Bethe roots}$}:\{u\} \Longrightarrow \text{Eigenvalues}: \{\lambda\}
\eeq
In this occassion we plan to go in the oposite direction:
\beq
\{\lambda\} \Longrightarrow \{u\}
\eeq
We will first find the set of eigenvalues by a direct diagonalization of the spin chain Hamiltonian and other higher local conserved charges. In order to determine the $M$ momentum carrying Bethe roots of a space of highest weights, it will be sufficient to diagonalize the first $M$ local conserved charges in this space. 

An advantage of performing the direct diagonalization is that we only obtain the physical states corresponding to physical solutions of the Bethe equations. So we do not need to deal with any of the (singular) spurious solutions.

Here we outline the steps that we follow for this diagonalization:

\begin{enumerate}
\item \textbf{Cyclicity and highest weight space:} First we construct the coordinate basis of the space corresponding to a $\mathfrak{so}(6)$ spin chain  of length $L$ and Dynkin weights $[M-N,L-2M,M-N]$. For this we need to find all sets of $L$ scalars whose $\mathfrak{so}(6)$  charges add up to the total weights of the space. When $N$ is even one such set is:
\beq
|Z^{L-2M}(ZX)^{M-N}\left(Z\bar{Z}\right)^{\frac{N}{2}}\rangle
\eeq
Then we impose two restrictions over this basis: cyclicity and the highest weight conditions. These implement the zero-momentum and super-primary conditions of the single-trace operators in the gauge theory. 
 
In order to impose the zero-momentum condition we refine the coordinate basis by repackaging all sets into cyclic states: 
\beq
\text{Tr}(Z^{L-2M}(ZX)^{M-N}\left(Z\bar{Z}\right)^{\frac{N}{2}}) \equiv |Z^{L-2M}(ZX)^{M-N}\left(Z\bar{Z}\right)^{\frac{N}{2}}\rangle + \text{cyclic permutations}
\eeq
This reduces the number of states in the original coordinate basis by a factor of $L$ (approximately).
Then we wish to focus only on highest weight states as these are the ones described by finite Bethe roots. For this we find the linear combinations of cyclic states which are annhilated by the three $\mathfrak{so}(6)$ raising generators. 

The dimension of the new subspace is much smaller and determines the number of finite physical solutions of Bethe equations \eqref{eq:so6equations}\footnote{Sometimes we need to exclude Bethe solutions with roots at zero which are descendants of the Bethe solutions with zeros removed. See \cite{psu112Zwiebel} for instances in the $\mathfrak{psu}(1,1|2)$ sector.}.
This reduction is important as the dimension of this   space determines the sizes of the matrices we need to diagonalize. Therefore it makes this procedure more feasible.

\item \textbf{Diagonalization of local charges} In this cyclic and highest weight space we diagonalize the local conserved charges: $\mathcal{H}_{1},\,\mathcal{H}_{2}\,,\cdots \mathcal{H}_{M}$ simultaneously. Where $M$ is the number of momentum carrying Bethe roots and $\mathcal{H}_{k}$  is the $k$-th conserved charge in the family of integrable Hamiltonians in convolution.  Two of these are distinguished: $\mathcal{H}_{1}$ is the shift-operator ($e^{i\,P}$) and $\mathcal{H}_{2}$ is the nearest- neighbour Hamiltonian of the spin chain. This diagonalization identifies the Bethe states with vectors of eigenvalues: $\left(\lambda_{1},\lambda_{2}\,\cdots \lambda_{M}\right)$, where $\lambda_{1}=1$ always and $\lambda_{2}$ is the one loop anomalous dimension in the gauge theory language.

The local charges can be constructed using the transfer matrix formalism. From this we obtain the charges as operators on the spin chain. By acting upon  our coordinate basis we find the correspondent matrix representation in the cyclic highest weight space.

We start with $\mathcal{H}_{2}$, which due its nearest-neighbour interaction, has the most sparse of the matrices to diagonalize (numerically). This is the hardest step and depending on the size of the matrix it could take hours of computer time or be unfeasible. As a result of this diagonalization the states organize into subspaces that share the same eigenvalue $\lambda_{2}$. Some of these sub-spaces are one-dimensional and many others have some small degeneracy. Afterwards we act over these subspaces with the higher charges, consecutively until reaching $\mathcal{H}_{M}$. At each step we lift the degeneracy a bit more, finally obtaining the vectors of eigenvalues $(\lambda_{1},\lambda_{2},\cdots,\lambda_{M})$ which label sub-spaces that are either one-dimensional or have a two-fold degeneracy. This latter degeneracy is not lifted by diagonalizing $\mathcal{H}_{M+1}$ and as we show later it is explained by an action on the nested wing roots which leaves invariant the eigenvalues.   

\item \textbf{Middle node Bethe roots:} The momentum carrying Bethe roots can be found using the relations:
\beq\label{eq:EigenU}
\prod_{k=1}^{M}\frac{u_{k}+\frac{i}{2}}{u_{k}-\frac{i}{2}}=\lambda_{1} =1 \qquad \text{and} \qquad \sum_{k=1}^{M} \left(\frac{1}{\left(u+\frac{i}{2}\right)^{k-1}}-\frac{1}{\left(u-\frac{i}{2}\right)^{k-1}}\right) = \lambda_{k\geq 2}
\eeq
With the knowledge of the eigenvalues these equations are easy to solve for the Bethe roots (definetely easier than Bethe equations).  For each vector of eigenvalues $(\lambda_{1}\cdots \lambda_{M})$ we find a single solution of Bethe roots\footnote{Having more eigenvalues $(\lambda_{M+1}\cdots )$ give us more relations to find the correspondent roots but we found $M$ of them were sufficient to identify a single solution.} $\{u_{1},\cdots ,u_{M}\}$.

\item \textbf{Wing roots from Bethe equations:}
In order to find the nested Bethe roots on the wings we could try to find the eigenvalues of the nested local charges. Then relations analog to \eqref{eq:EigenU} will give us access to the wings $\{v\}_{N}$ and $\{w\}_{N}$. However for our case of interest we are dealing with $N=2$, so we found it was easier to solve directly the nested Bethe equation, first row in \eqref{eq:so6equations}, by inputting the middle node roots $\{u\}_{M}$. The solutions we found can be classified into two groups:
\begin{enumerate}
\item The wings are identical $\{v\}_{N} = \{w\}_{N}$, so there is only one Bethe solution $\{\{w\}_{N},\{u\}_{M},\{w\}_{N}\}$ for the vector of eigenvalues $(\lambda_{1}\cdots \lambda_{M})$.
\item The wings are different $\{v\}_{N} \neq \{w\}_{N}$ and this leads to two different solutions related by swapping the wings:  $\{\{w\}_{N},\{u\}_{M},\{v\}_{N}\}$ and $\{\{v\}_{N},\{u\}_{M},\{w\}_{N}\}$ for the same vector $(\lambda_{1}\cdots \lambda_{M})$. This explains the two-fold residual degeneracy found after diagonalization. 
\end{enumerate}
For our applications we are only interested in the  group of solutions with identical wings. Since according to the integrability selection rule found in \cite{AsymptoticPaper}, Bethe solutions with non-identical wings give a vanishing structure constant $C^{\circ\circ\bullet}=0$. 
\end{enumerate}
\subsubsection{An example: length $L=24$ and number of roots $\{2,4,2\}$}
Following this diagonalization recipe we found the (physical) bethe solutions for spaces with various lengths $L=6,8,\cdots, 24$ and $\mathfrak{so}(6)$  weights determined by $M=4$ momentum carrying roots  and $N=2$ auxiliary roots on each wing.  We report the number of solutions for each case in table \ref{tab:so6numbersolutions}.
\begin{table}[ht]
\begin{tabular}{|c|c|c|c|c|c|c|c|c|c}
\hline
L & 6 & 8 & 10 & 12 & 14 & 16 &  24  \\ 
\hline
\hline
\{\text{sym}\,,\text{no-sym}\} & \{8,1\} & \{24,5\}  & \{52,14\}  & \{100,28\}  & \{166,53\}  & \{260,87\}  &  \{966,371\} \\
\hline
\end{tabular}
\caption{Number of solutions of the $\mathfrak{so}(6)$ spin chain  of length  $L$ and roots $\{2,4,2\}$. We classify them according to \{symmetric wings, non-symmetric wings\}.}
\label{tab:so6numbersolutions}
\end{table}

For the lowest lengths in  table \ref{tab:so6numbersolutions} we compared our results against the Q-solver of \cite{QQspectrumSolver}, finding perfect agreement.

 Furthermore, we were able to reach up to $L=24$ and find all $\mathfrak{so}(6)$ Bethe solutions that enter in our OPE consistency check (see last two rows in table \ref{tab:NumberRoots}). We used $\mathsf{Mathematica}$ to perform the diagonalization numerically with about 150 decimal digits of precision. In this way we could find all Bethe solutions from the numerical eigenvalues. These included almost singular solutions whose deviation from exact strings would only come at order $10^{-45}$, see table \ref{tab:NearStrings}. Although this deviation seems negligible, taking it into account is very important!\footnote{In particular some factors in the BKV formula in \eqref{eq:BKV} diverge for singular Bethe solutions.}
\begin{table}[ht]
\centering
\def\arraystretch{1.2}
\resizebox{3.8\totalheight}{!}{\begin{tabular}{|c|c|c|c|}
\hline
$u_{1}^{(0)}$ & $u_{2}^{(0)}$ & $u_{3}^{(0)}$ & $u_{4}^{(0)}$\\
\hline
\hline
{\color{blue} $-3.775756759$} & {\color{blue}$-0.07391374823$} & {\color{blue}$0.0151079602-5\times10^{-10}\,i$} & {\color{blue}$0.0151079602+5\times10^{-10}\,i$ } \\
 \hline
 $-1.829135907$ & $-0.1511209049$ & $0.0266840879-5\times10^{-10}\,i$ & $0.0266840879+5\times10^{-10}\,i$ \\
 \hline
 $-1.162800880$ & $-0.2355067318$ & $0.0341089969-5\times10^{-10}\,i$ & $0.0341089969+5\times10^{-10}\,i$ \\
 \hline
 $-0.8184936558$ & $-0.3323810079$ & $0.0383217165-5\times10^{-10}\,i$ & $0.0383217165+5\times10^{-10}\,i$ \\
 \hline
 -0.6026135491 & -0.4499760493 & $0.0402087653-5\times10^{-10}\,i$ & $0.0402087653+5\times10^{-10}\,i$ \\
 \hline
\end{tabular}}
\caption{Momentum carrying roots at leading order in $g^{2}$: $\{u_{1}^{(0)},u_{2}^{(0)},u_{3}^{(0)},u_{4}^{(0)}\}$ corresponding to almost singular solutions for $L=24$ and $\mathfrak{so}(6)$ root numbers $\{2,4,2\}$. We only present them with 10 digits of precision. To see their deviation from the singular position $\frac{i}{2}$ we need to go up to $10^{-35}$ and in particular $10^{-45}$ for the solution highlighted in {\color{blue}blue}.}
\label{tab:NearStrings}
\end{table}

\begin{figure}[ht]
\centering
\includegraphics[scale=.5]{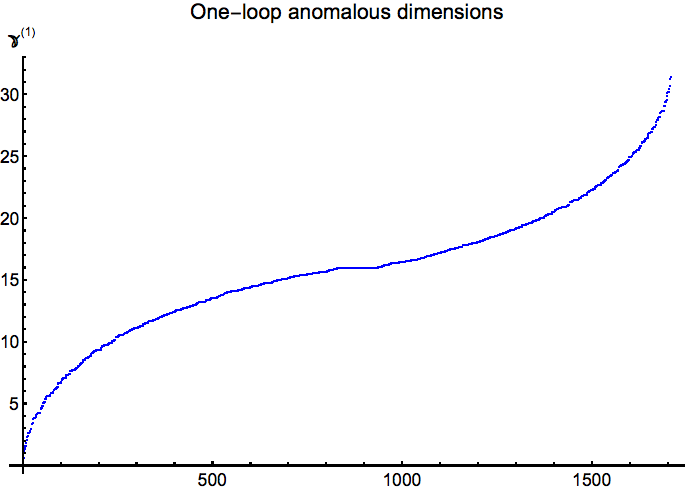}
\caption{A $\mathsf{ListPlot}$ of the anomalous dimensions $\gamma^{(1)}$ of the 1708 $\mathfrak{so}(6)$ super-primaries in the \textit{Super-Class} $\{\Delta_{0}=24,s=0,n=10,m=10\}$. The flat region in the middle corresponds to 90 degenerate super-primaries with anomalous dimension $\gamma^{(1)}=16$. This degeneracy is (partially) lifted when considering the higher-conserved charges.}
\end{figure}

\subsection{Reconstructing the \textit{asymptotic} OPE data with BKV hexagon formula}\label{sec:New5}

Having identified the super-primaries of table \ref{tab:NumberRoots} with Bethe solutions we can now reconstruct the leading terms in the OPE series of the \textit{asymptotic} four-point function. To do so we need to compute the data of each super-primary: anomalous dimensions and structure constants. Both these quantities follow from the knowledge of Bethe roots, the former is computed with formula \eqref{eq:AnomalousDimension} and the latter with the BKV hexagon formula \eqref{eq:BKV}. 

Furthermore there is a simplification, as we argued in  \cite{AsymptoticPaper},  the \textit{asymptotic} four-point function can be reconstructed ignoring wrapping corrections to the OPE data of each (super)primary. This again thanks to the limit $K\gg 1$ and  our choice of polarizations. These impose the bound on the twist $\tau\geq K$ of the exchanged operator $\mathcal{O}_{\Delta,s}$ and force the the three bridges in  $\langle \tilde{O}_{1}\tilde{O}_{2}\mathcal{O}_{\Delta,s}\rangle$ and $\langle \mathcal{O}_{\Delta,s}\tilde{O}_{3}\tilde{O}_{4}\rangle$ to be large($\geq K/2$).
Under these conditions wrapping corrections to the correspondent anomalous dimensions and structure constants kick in not earlier than $(g^{2})^{\frac{K}{2}-1}$. Therefore in this nine-loop exercise where we set $K=22$ it is safe to completely ignore wrapping corrections.

\subsubsection{Anomalous dimensions and BKV formula}
\begin{figure}[ht]
\centering
\resizebox{3.8\totalheight}{!}{\includestandalone[width=1\textwidth]{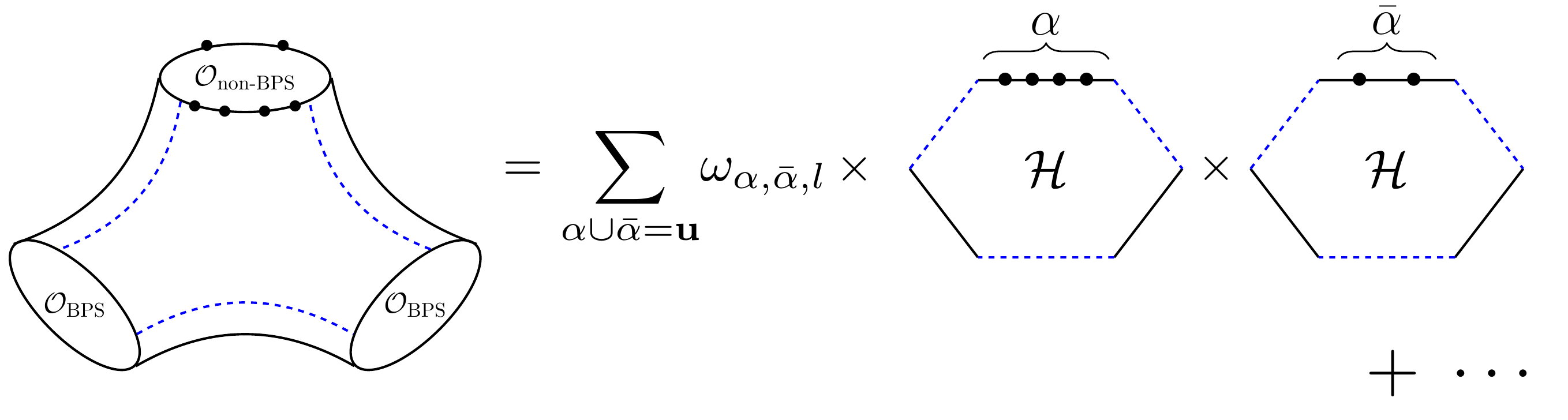}}
\caption{The asymptotic part of the BKV formula for structure constants of two protected and one non-protected operators, consists on a sum over bi-partitions  of the set of Bethe roots describing the non-BPS operator: $\alpha\cup\bar{\alpha}=\mathbf{u}$.  The summand is given by a product of two hexagon form factors, each evaluated in one of the bi-partitions: $\mathcal{H}_{\alpha}$ and $\mathcal{H}_{\bar{\alpha}}$, and a weight factor depending on the partitions and the length $l$ of the bridge connecting the non-protected operator and one of protected ones. }
\label{fig:BKV}
\end{figure}

 For a single trace super-primary with middle node Bethe roots $\mathbf{u}\equiv\{u_{1},u_{2},\cdots\}$ the asymptotic part of the anomalous dimension is given by:
\beq\label{eq:AnomalousDimension}
\Delta(\mathbf{u})-\Delta^{(0)} \,=\, \gamma(\mathbf{u}) = \sum_{u_{j}\in\mathbf{u}}\left(\frac{1}{x^{[+]}(u_{j})} -\frac{1}{x^{[-]}(u_{j})} \right) 
\eeq
In order to compute the structure constants we only consider the asymptotic part of the  BKV formula depicted in figure \ref{fig:BKV}.
The squared of the normalized structure constant for any non-BPS operator was given in \cite{AsymptoticPaper} as:
\beq\label{eq:BKV}
C^{2}_{KK(\Delta,s,n,m)}(\mathbf{u};\mathbf{v})=\frac{\langle{\bf v} |{\bf v}\rangle^2\, \prod_{u_{j}\in\mathbf{u}}\mu(u_{j})}{\langle{\bf u}|{\bf u} \rangle\prod_{i<j}S(u_i,u_j)}\times\mathcal{A}^2(\mathbf{u}\,;\mathbf{v})\,.
\eeq
where now $\mathbf{v}$ represents the three sets of Bethe roots on the left(or right) wing of the Dynkin diagram. 

For explicit expressions of the components in \eqref{eq:BKV} we refer to \cite{AsymptoticPaper} (where we used the $\mathfrak{sl}(2)$ grading). These include: the normalization given by the squared of the Gaudin norm $\langle \mathbf{u}|\mathbf{u} \rangle$ and the abelian Beisert $S$-matrix phase. The physical measure $\mu$ and the sum over partitions $\mathcal{A}$, depicted in figure \ref{fig:BKV}, which contains the hexagon form factors and depends on both types of roots $\mathbf{u}$ and $\mathbf{v}$. Also the wing-Gaudin norm  $\langle \mathbf{v}|\mathbf{v} \rangle$ which is set to one in the absence of wing roots and to zero  when the wings left and right are not identical. This latter fact refines the selection rule of the global symmetry, with an integrability-based selection rule: only wing-symmetric Bethe solutions appear in the OPE of two identical protected operators. 

\subsubsection{Summing over Bethe solutions and perfect match with OPE of octagon prediction}

Finally using the formulae \eqref{eq:AnomalousDimension} and \eqref{eq:BKV} and the Bethe solutions we constructed the super-sum rules $\mathcal{P}^{(9,a)}_{\{\Delta_{0},s,m,n\}}$. We report them in table \ref{tab:Final9sums} including the number of super multiplets we identified on each \textit{Super-Class} $\{\Delta_{0},s,m,n\}$. 
\begin{table}[ht]
\centering
\def\arraystretch{1.2}
\resizebox{2.5\totalheight}{!}{\begin{tabular}{|c|c|c|c|}
\hline
$\{\Delta_{0},s,n,m\}$& \Large $ \#\; \text{ of Bethe solutions} \atop  \text{or super-multiplets} $ & $\mathcal{P}^{(9,9)}$ & $\mathcal{P}^{(9,0)}$ \\ 
\hline \hline
$\{20,0,9,9\}$& $10$ & $\frac{4978688}{945}$ & $-2755264512$ \\
\hline
$\{22,0,10,10\}$& $11$ &  $\frac{4978688}{945}$ & $-2755264512$ \\
\hline
$\{24,0,11,11\}$ & $12$ &   $\frac{4978688}{945}$ & $-2755264512$ \\
\hline
$\{22,2,9,9\}$& $369$ & $\frac{632354992917077}{7440174000}$  & $-\frac{3880272703256846017844314261170552101}{100890752343750000000000000}$\\
\hline
$\{24,2,10,10\}$& $486$ & $\frac{2990537878301}{29760696}$ & $-\frac{25906812477093695541612336367}{562220051373121536}$\\
\hline
$\{26,2,11,11\}$& $626$ & $\frac{1012348238189803}{8630601840}$ & $-\frac{24206493839806735477486708172518503089}{445073004000599941716074496}$\\
\hline
$\{24,0,11,9\}$&  $478$ & $\frac{1344109856}{19845}$  & $-41148586528$  \\
\hline
$\{24,0,10,10\}$& $966 + 2\times {\color{red}371}$& $\frac{181961001311}{1086750}$ & $-\frac{1561516641881791110594032737679}{17968750000000000000}$ \\
\hline 
\end{tabular}}
\caption{The number of Bethe solutions or number of long super-multiplets on each of the 8 super-classes involved in our consistency check. The highlighted {\color{red} 371} is number of $\mathfrak{so}(6)$ Bethe solutions with non-identical wings, which have vanishing structure constants according to equation \eqref{eq:BKV}.}
\label{tab:Final9sums}
\end{table}

With this super-data we were able to reconstruct the primary sum rules $\mathsf{P}^{(9,0)}_{\{\Delta_{0},s\}}$ in tables \ref{tab:P90super} and  \ref{tab:P99super}, finding a perfect match with the hexagonalization predictions in tables \ref{tab:predictionP90} and \ref{tab:predictionP99} (see also  appendix \ref{app:LargeNineOPE}). We have also  reconstructed all sum rules $\mathsf{P}^{(a,b)}_{\{\Delta_{0},s\}}$  with $9\geq a \geq b \geq 1$ , finding perfect agreement with the hexagonalization predictions.

Some final (technical) comments regarding this nine-loop test:
\begin{itemize}
\item  We needed to work with a high  numerical precision for the Bethe roots. This allowed us to obtain the sum rules with many decimal digits and be able to recognize the (complicated) rational numbers they represent.

In particular the existence of the nearly-exact strings of table \ref{tab:NearStrings} demands us to work with very high numerical precision when finding the loop corrections of the Bethe roots:
\beq
u^{(0)}_{j}\,\to\,u^{(1)}_{j}\,\to\,\cdots\,\to \, u^{(9)}_{j}
\eeq
The reason being that when we go from $u^{(0)}_{j}$ to $u^{(9)}_{j}$ the numerical precision severily drops and it is important that the decreased precision of $u^{(9)}_{j}$ is higher than $10^{-45}$ so it can still recognize the deviation from $\frac{i}{2}$ of the nearly-singular positions of $u^{(0)}_{j}$. This is important to not lose this information when using the BKV formula where combinations of the form $u^{(0)}_{j}\times u^{(9)}_{j}$ can appear.

\item When constructing the super sum rules $\mathcal{P}^{(9,a)}_{\{24,0,10,10\}}$ we excluded the $371$ $\mathfrak{so}(6)$ Bethe solutions with non-identical wings. This can be considered as a test of the vanishing of those structure constants according to the integrability selection rule found in \cite{AsymptoticPaper}.  
\item Regarding the octagon, in detailed, this perfect matching at nine loops constitutes a successful test of the three-particle mirror integral $\mathcal{I}_{n=3,l=0}$ in \eqref{eq:n3l0integral} and with that also a test of the prescription of \cite{HexagonalizationI} for the $\mathcal{Z}$-markers on the mirror basis, which is manifest in \eqref{eq:Waverage} and the character \eqref{eq:Xn}.
\end{itemize}

\section{Conclusion and future directions}\label{sec:Conclusions}
In this paper we have made used of hexagonalization in the asymptotic regime $(K\gg 1)$ at weak coupling $(g^{2}\to 0)$ to compute four-point functions of one-half BPS single-trace operators in planar $\mathcal{N}=4$ SYM. Accompanied by specific choices of polarizations we have shown the efficiency of this method to provide explicit results, in principle up to arbitrary loop order. In particular the correlators we computed: \textit{simplest} $\mathbb{S}$ and \textit{asymptotic} $\mathbb{A}$, were shown to be  given by sums of squared octagons.  These octagon form factors were obtained from the gluing of two hexagons and at weak coupling explicitly evaluated in terms of Ladder integrals giving expressions of uniform and maximal transcendentality at every loop order. 

Having these explicit results the next step is to study their different kinematical limits. We have done so and will be reporting our findings in our upcoming papers \cite{FrankShort} and \cite{TillPedroFrankVascoWorkInProgress}.  In the present draft we have restrained to present only the (Euclidean) OPE of the \textit{asymptotic} correlator in the $1$-$2$ channel and we have used it to perform a test of our results. This channel is dominated by single traces and its OPE series can be reconstructed by identifying the exchanged operators with solutions of Beisert-Staudacher equations and using the  BKV hexagon formula to compute the correspondent structure constants. This reconstruction is very challenging due to mixing and the difficulty to solve BS equations, however we have been able to reconstruct the first few terms of the OPE series up to nine-loops and successfully matched them with our octagon prediction. We regard this as a strong check of the correctness of our use of hexagonalization.  Furthermore,
this test has also served to do further checks, specifically in the $\mathfrak{so}(6)$ sector, of the integrability selection rule for three point functions found in \cite{AsymptoticPaper}. 

\begin{figure}[ht]
\centering
\resizebox{5.8\totalheight}{!}{\includestandalone[width=.8\textwidth]{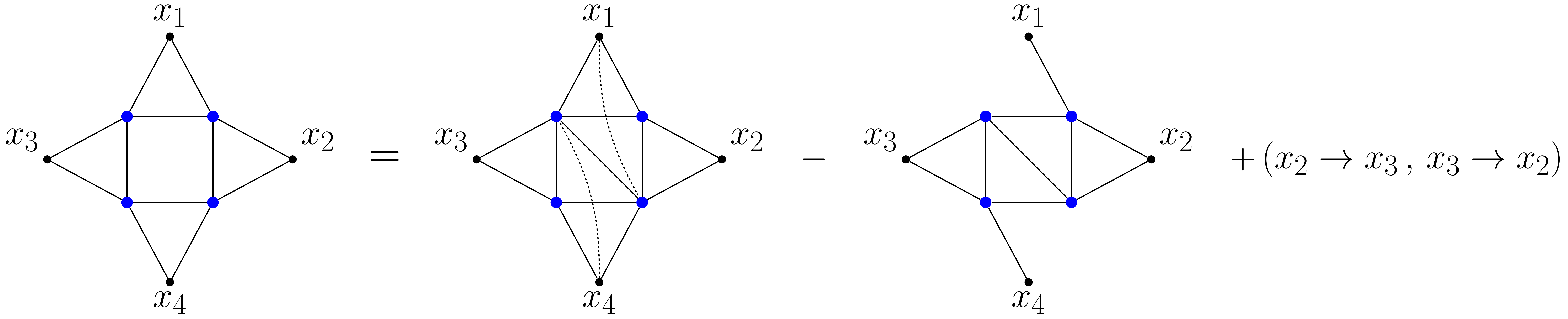}}
\caption{A new magic identity for four-loop conformal integrals. Solid and dashed lines represent scalar propagators and the inverse respectively.}
\label{fig:NewMagic}
\end{figure}

 Another source of confidence for the correcteness of our computation is the comparisson with the explicit three-loop results of \cite{AllThreeLoop} in terms of Ladder integrals. At this order we found a perfect match for both \textit{simplest} and \textit{asymptotic} correlators.  At four loops we compared with the results of \cite{AllFiveLoop} but only in the OPE limit, finding good agreement as well. By doing this latter comparisson we were able to recognize a new identiy between conformal integrals\footnote{Again we verified it using the OPE expansion provided in the ancillery of \cite{AllFiveLoop}. A proof of this identity is in progress by Vasco Gon\c{c}alves .}, see figure \ref{fig:NewMagic}. Finally, at five loops, our results were already used as input to fix some coefficients in the Ansatz of \cite{AllFiveLoop}.  

As a future direction it would be interesting to identify the basis of conformal integrals that appear in our polarized correlators at higher loop orders. This could allow us to  find explicit expresions for unknown conformal integrals in terms of sums of products of Ladders.  An instance of such identification was given in \cite{GluingLadders}, where it was conjectured that a family of fishnet  diagrams  evaluate to determinants of matrices whose entries are given by Ladders. These fishnets also appear in the Feynman diagrammatic expansion of our correlators and we can recognize their determinant expressions are contained in the leading loop order $(g^{2})^{n(n+l)}$ coefficient of the mirror integrals $\mathcal{I}_{n,l}$, for instance:
\beq
\mathcal{I}_{n=3,l=0}\,\big{|}_{\left(g^{2}\right)^{9}} \,=\, \text{Det}\begin{pmatrix}
F_{1}\,& \frac{1}{6}\,F_{2}\,& \frac{1}{240}\,F_{3}\\
2\,F_{2}\,& F_{3}\,& \frac{1}{20}\,F_{4}\\
12\,F_{3}\,& 12\,F_{4}\,& F_{5}\\
\end{pmatrix} \,\propto\,\raisebox{-3557787sp}{\resizebox{.4\totalheight}{!}{\includestandalone[width=.8\textwidth]{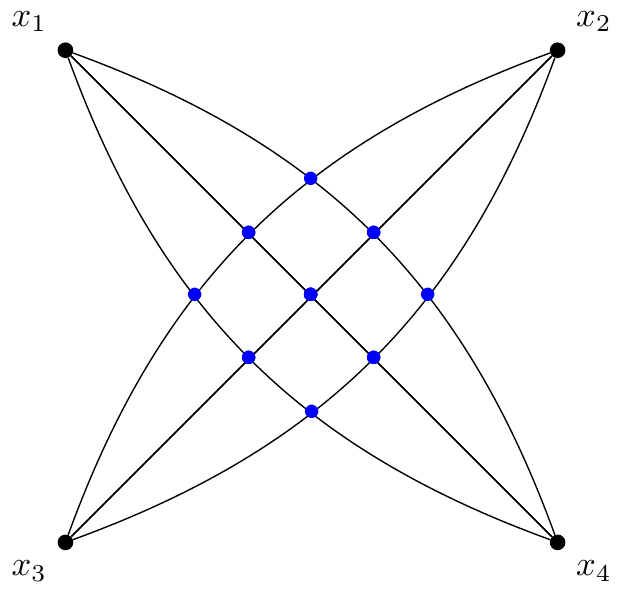}} } 
\eeq

It would be also important to attempt a resummation of the series of Ladders in our mirror integrals following  \cite{ResumLadders,ResumPentaLadders}, as this could give us access to the finite coupling octagon(s).  For this it would be helpful to find the closed form of the coefficients in formula \eqref{eq:SumLadders}.

\begin{figure}[ht]
\centering
 \resizebox{1\totalheight}{!}{\includestandalone[width=.8\textwidth]{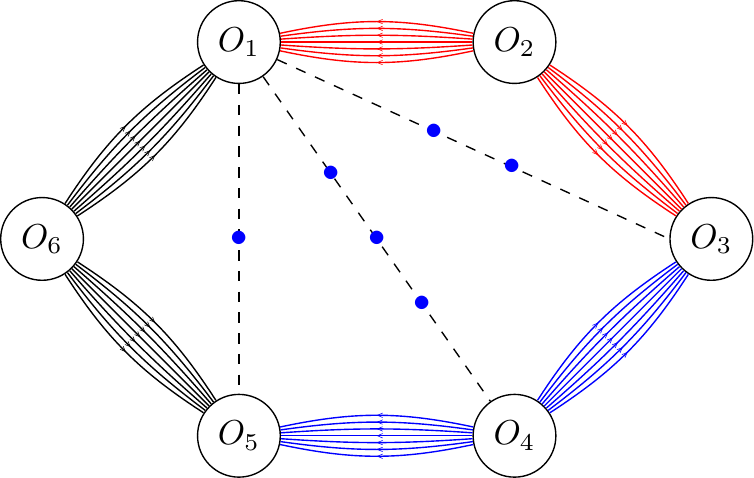}}
\caption{The \textit{simplest} six-point function. We make full use of the three complex scalars to adjust the external polarizations. For large charges this correlator is given by the product of two dodecagons. The dodecagon is obtained by gluing four hexagons  including complicated multi-particle strings on zero bridges.}
 \label{fig:SixSimplest}
\end{figure}

Another direction would be to extend the strategies presented here to compute polarized higher-point functions. The situation for these correlators is not as simple since the number of bridges is bigger and our choices of external R-charge polarizations are limited to the three complex scalars in $\mathcal{N}=4$ SYM. Hence it is harder to achieve optimal simplifications such as factorization of the correlators into octagons. The best one can do is to obtain a factorization into higher polygons as shown in figure \ref{fig:SixSimplest} for a polarized six-point function.

It is as well interesting to consider non-planar corrections to our polarised four-point functions following \cite{NonPlanar0,NonPlanar}\footnote{See also \cite{NonPlanarEden} for non-protected two-point functions at non-planar level}. We are addressing this problem in our upcoming paper \cite{TillPedroFrankToAppear}.

\begin{figure}[ht]
\centering
 \resizebox{1.6\totalheight}{!}{\includestandalone[width=.8\textwidth]{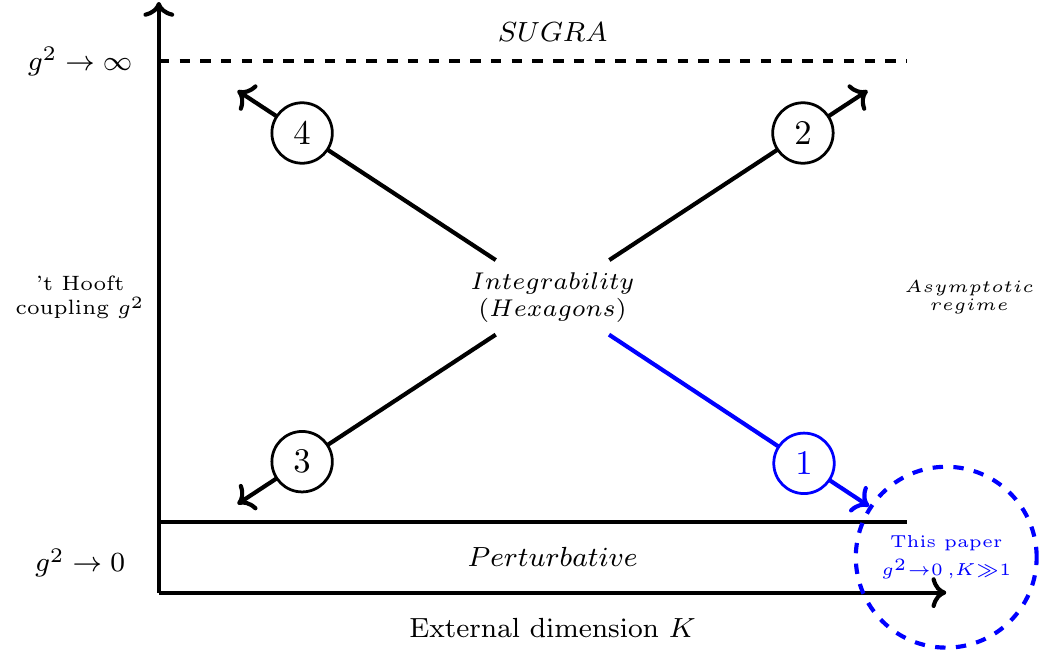}}
\caption{}
 \label{fig:Regimes}
\end{figure}

We also look forward to exploit the non-perturbative nature of  hexagonalization to study other regimes of the coupling $g^{2}$ and the external dimension $K$, see figure  \ref{fig:Regimes}.

In figure \ref{fig:Regimes} we enumerate the corners according to the degree of difficulty of accesability for hexagonalization:

\newcommand*\circled[1]{\tikz[baseline=(char.base)]{%
            \node[shape=circle,draw,inner sep=2pt] (char) {#1};}}      
\begin{enumerate}[label=\protect\circled{\arabic*}]
\item This is studied in this paper and constitutes the most accesible corner of figure \ref{fig:Regimes} for hexagonalization. One could still try to extend this work for generic polarizations for which other mirror contribuitions are needed. 
\item This regime corresponds to $g^{2}\to \infty$ and $K\to \infty$. There are two different subregimes depending on whether it is the `t Hooft coupling the largest parameter or if it is the external scaling dimension. We will be addressing this regime in a future publication \cite{TillPedroFrankVascoWorkInProgress}.
\item This is the perturbative regime for finite small $K$ and was addressed in the original hexagonalization paper at one loop. Although the weak coupling limit keeps the number of contributions finite at a given loop order, dealing with some of these contributions is technically hard. As they come with complicated matrix parts. These complications already start at two loops.
\item In this regime, $g^{2}\to \infty$ for arbitrary $K$,  is known to be described by supergravity in the bulk dual and the four-point correlator is known in Mellin space in terms of simple rational functions and in position space in terms of $D$-functions. Yet from the point of view of hexagonalization seems to be the hardest regime to access as there is no truncation in the number of mirror states. It would be interesting to reproduce these known results and  study how the decoupling of non-protected single traces becomes apparent in hexagonalization.
\end{enumerate}

Finally, we would like to highlight the method described in section \ref{sec:New4} to find all numerical solutions of $\mathfrak{so}(6)$ Bethe equations.  This is based on direct diagonalization of the local higher-conserved charges of the spin chain and although is conceptually standard in integrability, we have not found its practical use in the literature\footnote{Similar works perform a diagonalization of the transfer matrix $T$ for numerical values of the spectral parameter and then use the Baxter $T$-$Q$ equations to find the Bethe roots. One of these methods is described in detailed in section 2.2 of \cite{XXZnepomechie} where is refered as  ``McCoy's method" since its original use in \cite{McCoyPotts}. We learned about this method after completing our computation.} in the way presented here. An efficient implementation of this method for other sectors such as $\mathfrak{psu}(1,1|2)$ or the full $\mathfrak{psu}(2,2|4)$ sector could serve to extend the database of $\mathcal{N}=4$ Bethe solutions provided in \cite{QQspectrumSolver}.

\section*{Acknowledgements}

I am grateful to Shota Komatsu and Pedro Vieira for invaluable discussions about hexagonalization and OPE that inspired this project. I thank Till Bargheer and  Vasco Gon\c{c}alves  for discussions and collaboration in related projects. I also thank Thiago Fleury for discussions and comments on the draft.

\appendix

\section{More details on the octagon}\label{app:MoreOctagon}

In order to decode the sum in \ref{eq:OctagonSumPsi}, in this appendix, we  introduce the  mirror states $\psi$ and the hexagon form factors $\langle \mathcal{H}|\psi\rangle$ more explicitly.

\subsection{The mirror states $\psi$}\label{app:mirrobasis}
The spectrum on a mirror edge is given by the multi-particle states:
\beq\label{eq:multi-particle}
|\psi\rangle \equiv |u_{1} \cdots u_{n}\rangle^{(A_{1}\dot{A}_{1})\cdots (A_{n}\,\dot{A}_{n})} 
\eeq
where each particle has a charged $(A\,\dot{A})$ under the residual symmetry of the BMN vacuum $\mathfrak{su}(2|2)_{L}\otimes \mathfrak{su}(2|2)_{R}$. The admissible representations of this group are label by the bound state number $a$. The elementary particle $a=1$ lies in the bi-fundamental representation of  $\mathfrak{su}(2|2)_{L}\otimes \mathfrak{su}(2|2)_{R}$, in this case $A$ and $\dot{A}$ can take four different flavors: two bosonic and two fermionic, making a total of 16 possible states. While the general $a$ bound state particle lies in the $a$-th anti-symmetric representation of each $\mathfrak{su}(2|2)$ copy and can be obtained by fusion of $a$ elementary particles.

The rapidity $u$ is a continuous variable running from $-\infty$ to $\infty$.  In terms of this variable the measure $\mu_{a}$, energy $E_{a}$ and momentum $p_{a}$ of the $a$-th bound state particle can be written as:
\bba\label{eq:weakMeasure}
\mu_{a}(u) &= \frac{a}{g^{2}}\;\frac{\left(x^{[-a]}\,x^{[a]}\right)^{2}}{(1-(x^{[-a]})^{2})\,\left(1-x^{[-a]}\,x^{[a]}\right)^{2}\,(1-\left(x^{[a]}\right)^{2})} \,,
\nonumber\\
e^{-E_{a}(u)} &= \frac{1}{x^{[-a]}\,x^{[a]}} 
\qquad\text{and}\qquad 
p_{a}(u) =  u - g\,\left(\frac{1}{x^{[-a]}}+\frac{1}{x^{[a]}}\right) 
\end{align}
where $x$ is the Zhukovsky given in terms of the rapidity and the coupling as:
\beq
x^{[\pm a]}(u)= x(u\pm\frac{i\,a}{2})\,,\qquad x(u) = \frac{u+\sqrt{u^{2}-4g^{2}}}{2g}\,,\qquad \frac{u}{g} = x(u)+\frac{1}{x(u)}
\eeq
For our asymptotic multi-particle states \eqref{eq:multi-particle} the total energy and momentum are computed by simply adding up the contributions of the composites, while the total measure is given by the simple product:  
\beq
E_{\psi} = \sum_{j=1}^{n} E_{a_{j}}(u_{j})\,,\qquad p_{\psi} = \sum_{j=1}^{n} p_{a_{j}}(u_{j}) \quad \text{and} \quad \mu_{\psi} =  \prod_{j=1}^{n}\,\mu_{a_{j}}(u_{j})
\eeq
This description of the mirror states allows us to make the sum  in \eqref{eq:OctagonSumPsi} more precise:
\beq
\sum_{\psi}\,\to \, \sum_{n=0}^{\infty}\,\frac{1}{n!}\sum_{a_{1}=1}\cdots \sum_{a_{n}=1}^{\infty}\,\int du_{1} \,\cdots \,\int du_{n} \sum_{\text{internal flavour indexes}\atop\text{for each $a_{k}$ bound state}}
\eeq
This sum includes: a sum over the number of particles $\sum_{n=0}^{\infty}$, where $n=0$ corresponds to the mirror BMN vacuum and the factorial comes from Bose statistics. A sum over the bound state number $\sum_{a_{i}=1}^{\infty}$ and an integral $\int du_{i}$ for each particle. Finally a sum over all possible flavors ($A\dot{A}$) within each bound state representation.
\subsection{The hexagon form factor $\langle \mathcal{H}|\psi\rangle$}\label{app:hexagon}
The hexagon form factor for a  multi-particle state is given by:
\beq\label{eq:multiH}
\langle \mathcal{H} |u_{1}\,u_{2}\cdots u_{n}\rangle^{\left(A_{1}\dot{A}_{1}\right)\,\cdots\, \left(A_{n}\dot{A}_{n}\right)} =  \prod_{i<j}h_{a_{i}a_{j}}(u,v) \, \times\,\mathcal{H}_{mat}^{\left(A_{1}\dot{A}_{1}\right)\,\cdots\, \left(A_{n}\dot{A}_{n}\right)}(u_{1},\cdots, u_{n})
\eeq
where each $A_{j}$ and $\dot{A}_{j}$ group a $a_{j}$ number of flavor indexes corresponding to a bound-state in a $a_{j}$-anti-symmetric representation of $\mathfrak{psu}(2|2)_{L}$ and $\mathfrak{psu}(2|2)_{R}$ respectively.

The first factor in \eqref{eq:multiH} is given by a product of abelian two-particle hexagon form factors. For the elementary particles ($a=1$), the two-particle form factor is:
\beq
h(u,v) = \frac{x^{[-]}(u)-x^{[-]}(v)}{x^{[-]}(u)-x^{[+]}(v)}\frac{1-\frac{1}{x^{[-]}(u)x^{[+]}(v)}}{1-\frac{1}{x^{[+]}(u)x^{[+]}(v)}}\,\frac{1}{\sigma(u,v)}
\eeq
where $\sigma$ is the root-square of the BDS dressing phase. 

By using fusion relations we can build the two-particle form factor for any bound state $h_{ab}(u,v)$, starting with $a$ and $b$ fundamental particles respectively. In our computation we do not need the explicit form of this function,  but just the simpler symmetric product:
\bba\label{eq:symProd}
P_{ab}(u,v) &= h_{ab}(u,v) h_{ba}(v,u)\nonumber\\
&= \mathcal{K}^{++}_{ab}(u,v) \,  \mathcal{K}^{+-}_{ab}(u,v) \,  \mathcal{K}^{-+}_{ab}(u,v) \,  \mathcal{K}^{--}_{ab}(u,v)
\end{align}
with the $\mathcal{K}$ function given in terms of Zchukovsky variables as:
\beq\label{eq:Kinteraction}
\mathcal{K}^{\pm \pm}_{ab}(u,v) = \frac{ x^{[\pm a]}(u) - x^{[\pm b]}(v) }{1 - x^{[\pm a]}(u)\,x^{[\pm b]}(v)}
\eeq
Fortunately in this symmetric product the dressing phase, with complicated analytic properties, drops out thanks to the relation $\sigma(u,v)\sigma(v,u)=1$.

\begin{figure}[ht]  
\centering       
	 \resizebox{2.5\totalheight}{!}{\includestandalone[width=.8\textwidth]{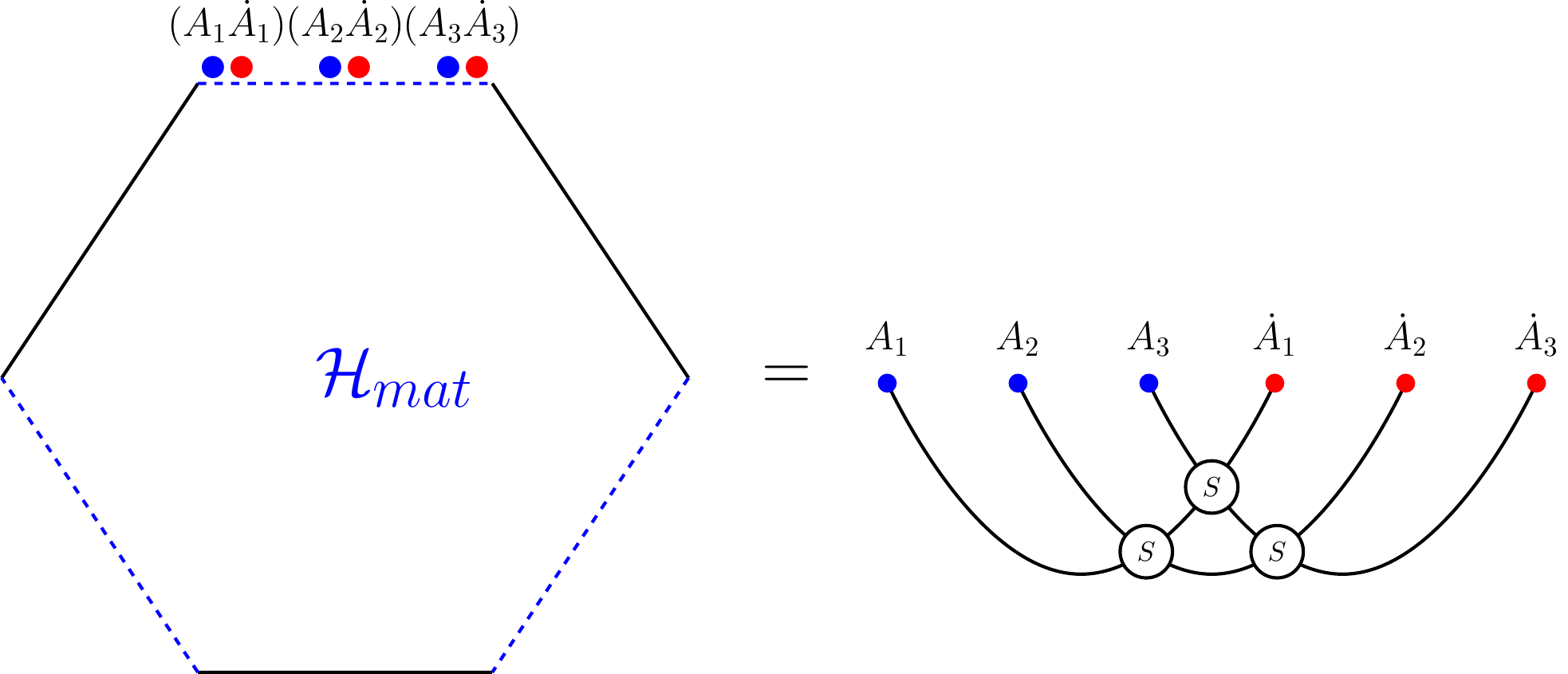}}
\caption{The matrix part of the multi-particle hexagon form factor}
\label{fig:MatrixHex}
\end{figure}

 The matrix part  $\mathcal{H}_{mat}$ in \eqref{eq:multiH}  takes the incoming indexes in the $\mathfrak{psu}(2|2)_{L}$ and $\mathfrak{psu}(2|2)_{R}$ representation and combines them into a tensor invariant under a diagonal symmetry group  $\mathfrak{psu}(2|2)_{D}$  (this is the residual symmetry preserved by the hexagon). To construct this tensor we use an arrangement of Beisert's $\mathfrak{su}(2|2)$ S-matrices realizing a scattering process with the $\mathfrak{psu}(2|2)_{L}$ particles as incoming and the $\mathfrak{psu}(2|2)_{R}$ as outgoing states. For a three-particle state this can be graphically depicted in figure \ref{fig:MatrixHex}.

The S-matrix for the fundamental representations can be found in \cite{SmatrixBeisert}, while the case of general mirror bound states has been worked out in \cite{HexagonalizationII} (see also \cite{BoundS}). In this paper we do not need the explicit form of the S-matrix but we just have to take into account the unitarity of the S-matrix $S_{ab}(u,v) S_{ba}(v,u)=1$ which is graphically depicted as:
\beq
\raisebox{-3577787sp}{\resizebox{.7\totalheight}{!}{\includestandalone[width=.8\textwidth]{UnitarityS}}}
\eeq
\subsection{Summing over flavor indexes}

We reorganize the summand by distinguishing between the elements that depend on the internal flavor and the ones which do not.  We repackage this latter group into the weight $W^{(\text{flavour})}$ and rewrite the octagon as:
\bba\label{eq:OctagonSum3}
\mathbb{O}_{l}&=\sum_{n=0}^{\infty}\frac{1}{n!}\,\sum_{a_{1}=1}^{\infty}\cdots \sum_{a_{n}=1}^{\infty}\,\int du_{1} \,\cdots \,\int du_{n}\,\prod_{j=1}^{n}\mu_{a_{j}}(u_{j})\,e^{-E_{a_{j}}(u_{j})\,l}\,e^{i\,p_{a_{j}}(u_{j})\,\log z\bar{z}}\, \prod_{i<j} \, P_{a_{i} a_{j}}(u_{i},u_{j})
 \nonumber\\
&\qquad\qquad\qquad\qquad\qquad\qquad\qquad\qquad\qquad \times W^{(\text{flavour})}_{a_{1}\cdots a_{n}}(u_{1},\cdots\,u_{n})
\end{align}
The symmetric product $P_{ab}(u,v)$ in \eqref{eq:symProd}, appears from the product of the abelian part of hexagon form factor of $\psi$ and  its conjugate $\bar{\psi}$. The flavor weight $W^{(\text{flavour})}$ encodes the sum over internal flavor states for the subspace of $n$ particles in representations $\{a_{1}\cdots a_{n}\}$ and rapidities $\{u_{1},\cdots,u_{n}\}$ respectively, we label this subspace as $V_{a_{1}\cdots a_{n}}$.  The correspondent summand  contains the matrix part of the hexagon form factors and the chemical potentials that depend on the angles $\theta,\phi,\varphi$:
\beq
W^{(\text{flavour})}_{a_{1}\cdots a_{n}}(u_{1},\cdots , u_{n}) \,=\, \sum_{\psi \in V_{a_{1}\cdots a_{n}} } \langle  \mathcal{H}_{mat}|\psi\rangle\,e^{i\,L_{\psi}\,\phi}\,e^{i \,R_{\psi}\, \theta}\,e^{i \,J_{\psi}\, \varphi}\, \langle \psi |\mathcal{H}_{mat}\rangle 
\eeq
In order to perform this sum we need to use the explicit $\mathfrak{su}(2|2)_{L}\otimes \mathfrak{su}(2|2)_{R}$ basis of the bound state representations.  The cases $n=1$ and $n=2$ where worked out in great detail in \cite{HexagonalizationI} and \cite{NonPlanar}\footnote{see appendix D.2 therein.} respectively.  In these references the choice of basis includes a prescription on how to include $\mathcal{Z}$ markers. We are instructed to consider two ways of dressing the mirror basis with $\mathcal{Z}$-markers (two subspaces: $V_{a}= V_{a}^{+}\oplus V_{a}^{-}$), perform the sum for each and then average over the two choices. Then the weight factor is composed as:
\beq\label{eq:Waverage}
W^{(\text{flavor})}_{a_{1}\cdots a_{n}} \,=\frac{ W^{(\text{flavor})+}_{a_{1}\cdots a_{n}}\,+\, W^{(\text{flavor})-}_{a_{1}\cdots a_{n}}}{2}
\eeq
Both $W^{(\text{flavour})\pm}_{a_{1}\cdots a_{n}}$ can be computed as shown in figure \ref{fig:3characters} for the case $n=3$.

Thanks to the unitary of the $\mathfrak{su}(2|2)$ S-matrix the flavour weight factorises into one-particle contributions which were computed in \cite{HexagonalizationI} and shown to be given by a $\mathfrak{su}(2|2)$ character:
\beq\label{eq:finalWflavour}
W^{(\text{flavour})\pm}_{a_{1}\cdots a_{n}} = \prod_{j=1}^{n}\left(\cos\phi-\cosh(\varphi\pm\theta)\right)\frac{\sin a_{j}\phi}{\sin\phi}
\eeq
notice the dependence on the rapidities totally drops.

Finally using \eqref{eq:OctagonSum3}, \eqref{eq:Waverage} and \eqref{eq:finalWflavour} we arrive to \eqref{eq:FinalOctagon} in the main text.

\section{An efficient way to evaluate the mirror integrals}\label{app:EfficientResidue}

In this appendix we show how to evaluate the integrals $\mathcal{I}_{n,l}$, see \eqref{eq:MirrorI}, at weak coupling. For this it is convenient to introduce the basis of integrals:
\beq\label{eq:Ibasis}
I_{m,n} = \sum_{a=1}^{\infty}\,{\color{blue}\frac{\sin(a\phi)}{z-\bar{z}}\;\,a}\;\,\int_{-\infty}^{\infty}\,du\,\frac{{\color{blue}(z\bar{z})^{-i\,u}}}{(u-\frac{i}{2}\,a)^{m}\,(u+\frac{i}{2}\,a)^{n}}
\eeq
For any non-negative $m$,$n$ these integrals can be simply evaluated by taking the residue at the pole  $u=\frac{i}{2}\,a$. Then performing the sum over the bound state number $a$ we obtain the closed form expresion:
\beq\label{eq:polIbasis}
I_{m,n} = (i)^{m+n}\,\sum_{k=0}^{m-1}\,(-1)^{n+k}\;\frac{\binom{m+n-k-2}{n-1}}{k!}\times \log(z\bar{z})^{k}\times \frac{\text{Li}_{m+n-k-2}(z)-\text{Li}_{m+n-k-2}(\bar{z})}{z-\bar{z}}
\eeq
We want to show that at any loop order the mirror integrals $\mathcal{I}_{n,l}$ can be expressed as a sum of products of the more elementary integrals in \eqref{eq:polIbasis}. To make this evident we explain the steps we follow to perform the loop expansion. We start by repeating some of the formulae of the main text to make this section self-contained.

The $n$-particle  mirror integral $\mathcal{I}_{n,l}$ is given by:
\bba\label{eq:MirrorI}
\mathcal{I}_{n,l}(z,\bar{z}) &= \frac{1}{n!}\sum_{a_{1}=1}^{\infty}\cdots \sum_{a_{n}=1}^{\infty}\,\int du_{1}\cdots \int du_{n}\, \prod_{j=1}^{n}\bar{\mu}_{a_{j}}(u_{j},l,z,\bar{z}) \times \prod_{j<k}^{n}\,P_{a_{j}a_{k}}(u_{j},u_{k})
\end{align}
the effective measure packages the chemical potentials potentials:
\beq\label{eq:EffectiveU}
\bar{\mu}_{a}(u,l,z,\bar{z}) = \,{\color{blue}\frac{\sin(a\phi)}{z-\bar{z}} }\times\,\mu_{a}(u) \times e^{-E_{a}(u)\,l}\times (z\bar{z})^{-i\,p_{a}(u)}
\eeq
and these are explicitly given in terms of the Zhukovsly variable:
\beq
x^{[\pm a]}(u)= x(u\pm\frac{i\,a}{2})\,,\qquad x(u) = \frac{u+\sqrt{u^{2}-4g^{2}}}{2g}\,,\qquad \frac{u}{g} = x(u)+\frac{1}{x(u)}
\eeq
as:
\bba
\mu_{a}(u) &= \frac{{\color{blue}a}}{g^{2}}\;\frac{\left(x^{[-a]}\,x^{[a]}\right)^{2}}{(1-(x^{[-a]})^{2})\,\left(1-x^{[-a]}\,x^{[a]}\right)^{2}\,(1-\left(x^{[a]}\right)^{2})} \label{eq:chemical1} \\
e^{-E_{a}(u)\,l} &= \left(\frac{1}{x^{[-a]}\,x^{[a]}}\right)^{l} \label{eq:chemical2} \\
(z\bar{z})^{-i\,p_{a}(u)} &= {\color{blue} (z\bar{z})^{-i\,u}}\,(z\bar{z})^{i\, g\,\left(\frac{1}{x^{[-a]}}+\frac{1}{x^{[a]}}\right)} \label{eq:chemical3}
\end{align}
and the two-particle symmetric hexagon form factor is:

\bba\label{eq:Pab}
P_{ab}(u,v)= \, \mathcal{K}^{++}_{ab}(u,v)\mathcal{K}^{+-}_{ab}(u,v) \mathcal{K}^{-+}_{ab}(u,v) \,  \mathcal{K}^{--}_{ab}(u,v) 
\end{align}
with:
\beq
\mathcal{K}^{\pm \pm}_{ab}(u,v) = \frac{ x^{[\pm a]}(u) - x^{[\pm b]}(v) }{1 - x^{[\pm a]}(u)\,x^{[\pm b]}(v)}
\eeq

To perform the weak coupling expansion of the integrand we need the expansion of the Zhuckovsky variable:
\beq\label{eq:AppXpansion}
x^{[\pm a]} = \frac{u\pm\frac{i}{2}\,a}{g} - \frac{g}{u\pm\frac{i}{2}\,a} - \frac{g^{3}}{(u\pm\frac{i}{2}\,a)^{3}} - \frac{2\,g^{5}}{(u\pm\frac{i}{2}\,a)^{5}} - \frac{5 \, g^{7}}{(u\pm\frac{i}{2}\,a)^{7}}\, + \mathcal{O}(g)^{9}
\eeq
This exhibits poles whose degree increases with each loop order. Likewise the integrand inherits these poles for each of the variables of integration. In particular we do not obtain extra poles coupling two rapidities. Differences of rapidities comming from \eqref{eq:Pab} only appear on the numerator so they can be easily expanded out to. 

In order to make more explicit the pole structure of the integrand we propose the following change of variables:
\beq\label{eq:repMathA}
\left(u-\frac{i}{2}\,a \right) \,\to\, \frac{1}{\mathcal{A}_{-}} \qquad  \text{and} \qquad \left(u+\frac{i}{2}\,a\right) \,\to \, \frac{1}{\mathcal{A}_{+}}
\eeq 
Similarly we use other letters for other pairs of rapidity-bound state number, for instance: $\mathcal{B}$ for $(v,b)$, $\mathcal{C}$ for $(w,c)$, etc. 

Under this new notation the expansion in \eqref{eq:Xpansion} looks like: 
\beq\label{eq:AXpansion}
x^{[\pm a]}(u) = \frac{1}{g\,\mathcal{A}_{\pm}} \, - \, g\,\mathcal{A}_{\pm} \, - \, g^{3}\,\mathcal{A}_{\pm}^{3}\, - \, 2\,g^{5}\,\mathcal{A}_{\pm}^{5} \, - \, 5\,g^{7}\,\mathcal{A}_{\pm}^{7} \, + \, \mathcal{O}(g)^{9}
\eeq
Plugging in this latter expansion for each rapidity in the components \eqref{eq:chemical1},\eqref{eq:chemical2},\eqref{eq:chemical3} and \eqref{eq:Pab} we find the mirror integrand takes the schematic form:
\beq\label{eq:IntegrandMath}
\text{\color{blue} stripped integrand} \,=\, \sum_{m=0}^{\infty}\, \left(g^{2}\right)^{m} \, \sum_{k=0}^{m} \log(z\bar{z})^{k} \,\times \mathsf{Polynomial}(\mathcal{A},\mathcal{B},\mathcal{C}\,\cdots)
\eeq
where \text{\color{blue} stripped integrand} is the integrand after we have stripped out the {\color{blue}blue factors} in \eqref{eq:EffectiveU},\eqref{eq:chemical1},\eqref{eq:chemical2}  for each rapidity. The expansion on $\log(z\bar{z})$ comes from the loop expansion of \eqref{eq:chemical3}. The function $\mathsf{Polynomial}$ is a polynomial  on the variables  \eqref{eq:repMathA}. 
Schematically for the $n=3$ integrand it has the form:
\beq\label{eq:Polynomial}
\mathsf{Polynomial}(\mathcal{A},\mathcal{B},\mathcal{C}) = \text{coef}\,\times \mathcal{A}_{-}^{m_{1}}\,\mathcal{A}_{+}^{n_{1}}\,\mathcal{B}_{-}^{m_{2}}\,\mathcal{B}_{+}^{n_{2}}\,\mathcal{C}_{-}^{m_{3}}\,\mathcal{C}_{+}^{n_{3}} + \cdots
\eeq
where the dots represent analog terms with different coefficients and exponents ($m_{k},n_{k}$).

Now to go from the integrand \eqref{eq:IntegrandMath} to the integral we just need to perform a replacement. Whenever we see a couple $\mathcal{A}_{-1}\,\mathcal{A}_{1}$ in \eqref{eq:Polynomial} we replace it by the basis in \eqref{eq:polIbasis} as:
\bba\label{eq:ReplaceI}
\text{\color{blue}stripped integrand}\quad &\overset{{\color{white}\text{restore blue factors},\atop \text{perform sums and integrals}}}{\Longrightarrow} \qquad \mathcal{I}_{n,l}\nonumber\\
\mathcal{A}_{-}^{m_{1}}\,\mathcal{A}_{+}^{n_{1}}\,\mathcal{B}_{-}^{m_{2}}\,\mathcal{B}_{+}^{n_{2}}\,\mathcal{C}_{-}^{m_{3}}\,\mathcal{C}_{+}^{n_{3}} \quad &\overset{\text{restore {\color{blue}blue factors}},\atop \text{perform sums and integrals}}{\Longrightarrow}\quad I_{m_{1},n_{1}}\,I_{m_{2},n_{2}}\,I_{m_{3},n_{3}}
\end{align}

\begin{itemize}
\item \textbf{Maximal trascendentality:} using the explicit form of the basis \eqref{eq:polIbasis}  and collecting in $\log$s we can make manifest the uniform and maximal transcendentality of the mirror integrals:
\beq\label{eq:structureI}
\mathcal{I}_{n,l} = \sum_{J=n(n+l)}^{\infty}\, \left(g^{2}\right)^{J} \, \sum_{k=0}^{J} \log(z\bar{z})^{k} \,\times \sum_{j_{1}+\cdots +j_{n}=2J-k} \,\text{coe}_{j_{1}\cdots j_{n}}\; \text{pl}_{j_{1}}\text{pl}_{j_{2}}\cdots \text{pl}_{j_{n}} 
\eeq
where all indexes $j$ are positive and $\text{pl}_{j}\equiv\frac{\text{Li}_{j}(z)-\text{Li}_{j}(\bar{z})}{z-\bar{z}}$. Some of the coefficients  $\text{coe}_{j_{1},\cdots,j_{n}}$ could be zero.
\item \textbf{Single-valuedness:} The mirror integrals are single-value when $z$ and $\bar{z}$ are complex conjugates (euclidean regime). 

Considering the integrand is invariant under the exchange $a \to -a$ we can argue that whenever we have the  term with $\mathcal{A}_{-}^{m_{1}}\,\mathcal{A}_{+}^{n_{1}}$ we should also have a term with $\mathcal{A}_{-}^{m_{1}}\,\mathcal{A}_{+}^{n_{1}}$ and the same coefficients. Therefore there is a refinement of \eqref{eq:ReplaceI} which can allow us to express $\mathcal{I}_{n,l}$ into sums of products of the basis $I_{m,n}+I_{n,m}$. One can verified that this symmetric basis is single valued.

Taking into account this property we present our explicit results for some mirror integrals in terms of Ladder integrals. These belong to the family of single-valued polylogarithms. 
\item \textbf{Mathematica:} The loop expansion sketch in this appendix can be easily implemented in $\mathsf{Mathematica}$. The key steps are the replacements \eqref{eq:AXpansion} and \eqref{eq:ReplaceI}. In our implementation of this algorithm we could obtain 17 loops in few minutes. 
\end{itemize}

\section{Mirror integrals up to nine loops}\label{app:NineLoopIntegrals}
In this appendix we present some of the mirror integrals in a loop expansion. 

The series expansion of the one-particle for any bridge parameter is given in terms of ladder integrals as:
\beq
\mathcal{I}_{n=1\,,\,l}(z,\bar{z}) = \sum_{k=l+1}^{\infty} \left(g^{2}\right)^{k}\,(-1)^{k-l-1}\,\binom{2k-2}{k-l-1}\, F_{k}(z,\bar{z})
\eeq
For higher number of particles we were unable to find a closed form for the coefficients of the ladders. Here we only provide the expansion of the mirror integrals entering the nine loop \textit{asymptotic} prediction in \eqref{eq:OctagonNinePrediction2}.

For $n=2$ these include $l=0,1,2$ (we leave $z,\bar{z}$ dependence implicit):
\bba
\mathcal{I}_{n=2\,,\,l=0}\,&=\,g^{8}\left(F_{1}F_{3}-\frac{1}{3}
F_{2}^{2}\right)+g^{10}\left(-6\,F_{1}F_{4}+F_{2}F_{3}\right) + g^{12}\left(28\,F_{1}F_{5}+\frac{4}{5}\,F_{2}F_{4}-\frac{9}{5}\,F_{3}^{2}\right)\nonumber\\
&\qquad\;+ g^{14}\left(-120\,F_{1}F_{6}-16\,F_{2}F_{5}+\frac{36}{5}\,F_{3}F_{4}\right)\nonumber\\
&\qquad\;\;+g^{16}\left(495\,F_{1}F_{7}+\frac{690}{7}\,F_{2}F_{6}-\frac{9}{7}\,F_{3}F_{5}-\frac{486}{35}\,F_{4}^{2}\right)\nonumber\\
&\qquad\quad + g^{18}\,\left(-2002 F_{1}F_{8}-\frac{979}{2}\,F_{2}F_{7}-\frac{1203}{14}\,F_{3}F_{6}+\frac{456}{7}\,F_{4}F_{5}\right)+\mathcal{O}(g^{20})
\end{align}

\bba
\mathcal{I}_{n=2\,,\,l=1}\,&=\,g^{12}\left(F_{2}F_{4}-\frac{1}{2}\,F_{3}^{2}\right) + g^{14}\left(-8F_{2}F_{5}+\frac{12}{5}\,F_{3}F_{4}\right)\nonumber\\
&\qquad+g^{16}\left(45\, F_{2}F_{6}+2\,F_{3}F_{5}-\frac{33}{5}F_{4}^{2}\right)\nonumber\\
&\qquad\;+g^{18}\left(-220\,F_{2}F_{7}-\frac{380}{7}\,F_{3}F_{6}+\frac{240}{7}\,F_{4}F_{5}\right)+\mathcal{O}(g^{20})
\end{align}

\beq
\mathcal{I}_{n=2,l=2} \,=\, g^{16}\left(F_{3}F_{5}-\frac{3}{5}\,F_{4}^{2}\right)\,+\,g^{18}\left(-10\,F_{3}F_{6}+4\,F_{4}F_{5}\right)\,+\,\mathcal{O}(g^{20})\qquad\qquad\qquad\qquad\qquad\qquad\qquad\qquad\quad
\eeq
Finally the only $n=3$ integral contributing at nine loops has $l=0$:
\beq\label{eq:n3l0integral}
\mathcal{I}_{n=3,l=0} \,=\ g^{18}\left(F_{1}F_{3}F_{5}-\frac{3}{5}\,F_{1}F_{4}^{2}-\frac{1}{3}\,F_{2}^{2}F_{5}+\frac{1}{5}\,F_{2}F_{3}F_{4}-\frac{1}{20}\,F_{3}^{3}\right)+\mathcal{O}(g^{20})
\eeq
\section{Cross ratios}\label{app:CrossRatios}
\subsubsection*{Space-time cross ratios}
We use the cross ratios:
\beq
u = \frac{x_{12}^{2}x_{34}^{2}}{x^{2}_{13} x^{2}_{24}}\,\qquad\text{and}\qquad v = \frac{x_{14}^{2}x_{23}^{2}}{x^{2}_{13} x^{2}_{24}}\
\eeq
as well as the light-cone cross ratios:
\beq
z\,\bar{z} =u \qquad \text{and} \qquad (1-z)(1-\bar{z})=v
\eeq
or in radial coordinates:
\beq
 r^{2} = u \qquad \text{and}\qquad e^{i\phi}= \sqrt{\frac{z}{\bar{z}}}
\eeq
\subsubsection*{$R$-charge cross ratios}
Likewise for the $R$-space we use the cross ratios:
\beq
\sigma =\frac{(y_{1}.y_{2})(y_{3}.y_{4})}{(y_{1}.y_{3})(y_{2}.y_{4})} \qquad \text{and} \qquad \tau= \frac{(y_{1}.y_{4})(y_{2}.y_{3})}{(y_{1}.y_{3})(y_{2}.y_{4})}
\eeq
or
\beq
\alpha\bar{\alpha} = \sigma  \qquad \text{and} \qquad (1-\alpha)(1-\bar{\alpha}) =\tau
\eeq
and radial coordinates:
\beq\label{eq:Rradial}
\rho^{2} = \sigma  \qquad\text{and} \qquad  e^{i\theta} =\sqrt{\frac{\alpha}{\bar{\alpha}}} 
\eeq

\section{The super-block of a long multiplet}\label{app:Blocks}
The long multiplets appearing on the OPE decomposition of identical $\frac{1}{2}$-BPS operators $\mathcal{O}_{[0,p,0]}$ have superprimaries with scaling dimension $\Delta$, spin $s$ and $R$-symmetry representations:
\beq
\mathcal{O}_{\Delta,s,[n-m,2m,n-m]}\qquad \text{with} \qquad 0\leq m \leq n \leq p-2
\eeq
The super-conformal block for these long representations is given by:
\beq\label{eq:superblock}
 \mathcal{F}_{\Delta,s,n,m}(z,\bar{z},\alpha,\bar{\alpha}) = \frac{(z-\alpha)(z-\bar{\alpha})(\bar{z}-\alpha)(\bar{z}-\bar{\alpha})}{(z\bar{z})^{2}}\,\,G_{\Delta+4,s}(z,\bar{z})  \times Y_{n,m}(\alpha,\bar{\alpha})
\eeq
where $G_{\Delta,s}$ is the $\mathfrak{so}(4,2)$ scalar conformal block in \eqref{eq:ExpandConformalBlock} and   the $\mathfrak{so}(6)$ harmonic function $Y_{n,m}$ is given in \eqref{eq:ExpandRBlock}.

Similarly we can define $\mathcal{F}^{(d)}$ by replacing $G$ by the derivative $G^{(d)}$ (see equation \eqref{eq:Gderivative}):
\beq\label{eq:Fderivative}
 \mathcal{F}^{(d)}_{\Delta,s,n,m}(z,\bar{z},\alpha,\bar{\alpha}) = \frac{(z-\alpha)(z-\bar{\alpha})(\bar{z}-\alpha)(\bar{z}-\bar{\alpha})}{(z\bar{z})^{2}}\times \, G^{(d)}_{\Delta+4,s}\times Y_{n,m}  
\eeq
These derivatives are useful to express the perturbative OPE expansion. When performing this expansion we are interested in the small cross-ratio series expansion of the conformal block. In what follows we provide such expansions in radial coordinates for both the $\mathfrak{so}(4,2)$ conformal block and the $\mathfrak{so}(6)$ R-symmetry block.

\subsection{$\mathfrak{so}(4,2)$ conformal blocks}
A 4D scalar conformal block for a exchanged primary of dimension $\Delta$ and spin $s$ admits the radial expansion
\beq\label{eq:ExpandConformalBlock}
 G_{\Delta,s}(r,\phi) = \,r^{\Delta}\,\sum_{i=0}^{\infty}\,\sum_{j=0}^{i}\,\mathsf{A}^{(i,j)}_{\Delta,s}\, r^{i}\,\frac{\sin((s+i-2j+1)\,\phi)}{\sin\,\phi}  
\eeq
where each power of the radial cross ratio $r$ corresponds to the scaling dimension of a descendant, see \cite{RadialCoordinates}. 

The coefficients $\mathsf{A}$ weighting each descendant contribution in \eqref{eq:ExpandConformalBlock}  depend on the primary charges $(\Delta,s)$ and the level $(i,j)$ (number of  boxes $\square\equiv \mathcal{D}\bar{\mathcal{D}}$ and derivatives $\mathcal{D}$). An explicit expression for these coefficients is given in terms of Pochammers as:
\beq\label{eq:Acoe}
\mathsf{A}^{(i,j)}_{\Delta , s} \,=\, \frac{1}{j!(i-j)!}\, \frac{\left(\left(\frac{\Delta -s-2}{2}\right)_{j}\right)^{2}}{\left(\Delta-s-2\right)_{j}} \,\frac{\left(\left(\frac{\Delta + s}{2}\right)_{i-j}\right)^{2}}{\left(\Delta+s\right)_{i-j}} 
\eeq
For the perturbative OPE expansion we define the following derivatives of the conformal blocks:
\beq\label{eq:Gderivative}
 G^{(d)}_{\Delta_{0},s}(r,\phi) =\,r^{\Delta_{0}} \,\sum_{i=0}^{\infty}\,\sum_{j=0}^{i}\,\mathsf{A}^{(d)\,(i,j)}_{\Delta_{0},s}\, r^{i}\,\frac{\sin((s+i-2j+1)\,\phi)}{\sin\,\phi}  
\eeq
In this expansion the coefficients are given by derivatives  of \eqref{eq:Acoe} with respect to the scaling dimension:
\beq
\mathsf{A}^{(n)\,(i,j)}_{\Delta,s} = \frac{\partial^{n}}{\partial\Delta^{n}} \, \mathsf{A}^{(i,j)}_{\Delta,s}
\eeq

\subsection{$\mathfrak{so}(6)$ R-symmetry blocks}
In radial coordinates \eqref{eq:Rradial} the $R$-symmetry blocks are given by the finite series:
\beq\label{eq:ExpandRBlock}
 Y_{n,m}(\rho,\theta) = \frac{m!^{2} \, (n+1)!^{2}}{2m! \,(2n+2)!}\,\sum_{j=0}^{n+1}\sum_{k=0}^{m} \, \mathsf{B}^{(j,k)}_{n,m}\,\rho^{-j-k-3}\,\frac{\sin\left((j-k)\theta\right)}{\sin\theta} 
\eeq
This block corresponds to a $\mathfrak{so}(6)$ multiplet with highest weight $[n-m,2m,n-m]$, with $n\geq m$. The coefficients $\mathsf{B}$ are: 
\beq
\mathsf{B}^{(j,k)}_{n,m} = \mathsf{M}_{n+1,j}\,\mathsf{M}_{m,k} \qquad \text{with}\quad \mathsf{M}_{m,k} = (-1)^{m-k} \, \frac{(m+k)!}{(k!)^{2}\,(m-k)!}
\eeq
We find this expansion convenient when extracting OPE data from the supercorrelator. However when performing a $R$-charge projection on these blocks it is better to use the cross ratios $(\alpha,\bar{\alpha})$ or $(\sigma,\tau)$.

\section{Perturbative OPE expansion}\label{app:WeakOPE}
The confomal block expansion of a four-point function depends on the OPE data: scaling dimension and structure constants of the intermediate operators. When working in a perturbative regime these dynamical data admits an expansion in the coupling $g^{2}$ as:
\bba\label{eq:weakOPEdata}
\Delta &= \Delta_{0} + g^{2}\, \gamma^{(1)} + \, g^{4} \, \gamma^{(2)} + \cdots  \nonumber\\
C_{pp\Delta} &= C^{(0)} + g^{2} \, C^{(1)} + g^{4}\,C^{(2)} + \cdots 
\end{align} 
Similarly the reduced four-point function admits the coupling expansion:
\bba
\langle pppp \rangle \,&=\, \langle pppp\rangle^{(0)} + g^{2} \, \langle pppp\rangle^{(1)}  + g^{4} \,\langle pppp \rangle^{(2)} \,+\, \cdots  \nonumber\\
& =   \sum_{a=0}^{\infty}\,(g^{2})^{a} \, \langle pppp \rangle^{(a)} 
\end{align}
We want to match this latter expansion against the comformal block decomposition:
\beq\label{eq:blockDecomposition}
\langle pppp \rangle = (\text{protected})\,+\,\sum_{\{\Delta,l,n,m\}}\, C^{2}_{\Delta,l,n,m}\,\mathcal{F}_{\Delta,l,n,m}
\eeq
where protected stands for short-multiplets which do not acquire anomalous dimensions. The sum is perform over all long multiplets with charges  $\{\Delta,l,n,m\}$.

 Since the scaling dimensions appear in the exponents of the radial cross ratio, in the limit $g^{2}\to 0$, the correlator develops $\log(r)$ terms:
\beq\label{eq:rlog}
r^{\Delta} = r^{\Delta_{0}}+g^{2}\,\gamma^{(1)}\,\log(r)\,r^{\Delta_{0}}+g^{4}\,\left(\gamma^{(2)}\,\log(r)+\frac{1}{2}\left(\gamma^{(1)}\right)^{2}\,\log(r)^{2}\,\right)r^{\Delta_{0}} + O(g)^{6}
\eeq
By plugging the expansions \eqref{eq:weakOPEdata} into \eqref{eq:blockDecomposition} we  can find the loop corrections $\langle pppp \rangle^{(a)}$ in terms of the OPE data and the derivatives of the conformal blocks \eqref{eq:Fderivative}.
The low loop corrections of the four-point function, organized according to their $\log(r)$-singularities, are given by:
\bba\label{eq:firstloops}
\langle pppp \rangle^{(0)} &= \,\left(\text{protected multiplets}\right) +
\sum_{\{\Delta\}}\left(C^{(0)}\right)^{2} \,\mathcal{F}^{(0)} \nonumber\\
\langle pppp\rangle^{(1)} &=\,\sum_{\{\Delta\}}\left( 2\, C^{(0)}\,C^{(1)}\mathcal{F}^{(0)} + \gamma^{(1)}\,\left(C^{(0)}\right)^{2}\,\mathcal{F}^{(1)}\right)\, + \left(\,\gamma^{(1)}\,\left(C^{(0)}\right)^{2}\,\mathcal{F}^{(1)}\right)\,{\color{blue}\log(r)} \nonumber\\
\langle pppp\rangle^{(2)} &=\, \sum_{\{\Delta\}} \text{\smaller $\, \left(\left(C^{(1)}\right)^{2}+ 2\,C^{(0)}\,C^{(2)}\right)\,\mathcal{F}^{(0)} + \left(2\,C^{(0)}\,C^{(1)}\,\gamma^{(1)} +\left(C^{(0)}\right)^{2}\,\gamma^{(2)}\right)\,\mathcal{F}^{(1)} \, + \, \frac{1}{2}\left(C^{(0)}\,\gamma^{(1)}\right)^{2}\,\mathcal{F}^{(2)} $}\nonumber\\
&\qquad  + \left(\left(2\,	C^{(0)}\,C^{(1)}\,\gamma^{(1)}\,+\,\left(C^{(0)}\right)^{2}\,\gamma^{(2)}\right)\,\mathcal{F}^{(0)}\,+\, \left(C^{(0)}\gamma^{(1)}\right)^{2}\,\mathcal{F}^{(1)}\right)\,{\color{blue}\log(r)} \nonumber\\
&\qquad\; + \left(\frac{1}{2}\,\left(C^{(0)}\,\gamma^{(1)}\right)^{2}\,\mathcal{F}^{(0)}\right)\,{\color{blue}\log(r)^{2}}  
\end{align}
The sum on $\{\Delta\}$( short for $\{\Delta,s,m,n\}$) runs over all long multiplets. The loop corrections $\gamma^{(a)}$ and $C^{(a)}$ should have labels $(\Delta)$. While the blocks $\mathcal{F}^{(a)}$ depend only on the tree level ($g=0$) charges $(\Delta_{0})$, fact that follows from \eqref{eq:rlog}.

We can then reorganize the sum over multiplets as follows:
\beq
\sum_{\{\Delta\}} \to  \sum_{\{\Delta_{0}\}}\,\sum_{(\Delta)\in\textit{Super-Class}\,{\{\Delta_{0}\}}}
\eeq
where a \textit{Class} or \textit{Super-Class} denotes a family of long multiplets that become degenerate with charges $\{\Delta_{0}\}$ when turning off the coupling $g=0$. On the right hand side, the first sum runs over all admissible tree level charges $\{\Delta_{0}\}$ in the OPE. While the innermost sum runs over all multiplets $(\Delta)$ within a \textit{Class} labeled by $\{\Delta_{0}\}$. In the OPE expansion we can absorb this second sum within the definition of the sum rules:
\bba
\mathcal{P}^{(0,0)}_{\{\Delta_{0}\}} &= \sum_{(\Delta)\in \textit{Super-Class}\,\{\Delta_{0}\}} \, \left(C^{(0)}\right)^{2} \nonumber\\
\mathcal{P}^{(1,0)}_{\{\Delta_{0}\}} &= \sum_{(\Delta)} \,2\, C^{(0)}\,C^{(1)} \,,\qquad \mathcal{P}^{(1,1)}_{\{\Delta_{0}\}} = \sum_{(\Delta)} \,\gamma^{(1)}\,\left(C^{(0)}\right)^{2}\nonumber\\
\mathcal{P}^{(2,0)}_{\{\Delta_{0}\}} &= \sum_{(\Delta)} \left(C^{(1)}\right)^{2}+ 2\,C^{(0)}\,C^{(2)}\,,\qquad \mathcal{P}^{(2,1)}_{\{\Delta_{0}\}} = \sum_{(\Delta)}2\,C^{(0)}\,C^{(1)}\,\gamma^{(1)} +\left(C^{(0)}\right)^{2}\,\gamma^{(2)}\,,\nonumber\\
 \mathcal{P}^{(2,2)}_{\{\Delta_{o}\}} &= \sum_{(\Delta)} \frac{1}{2}\left(C^{(0)}\,\gamma^{(1)}\right)^{2}
\end{align}
Using these sum rules we can rewrite \eqref{eq:firstloops} as:
\bba\label{eq:newfirstloops}
\langle pppp \rangle^{(0)} &= \,\left(\text{protected multiplets}\right) +
\sum_{\{\Delta_{o}\}} \,\mathcal{P}^{(0,0)} \,\mathcal{F}^{(0)} \nonumber\\
\langle pppp\rangle^{(1)} &=\,\sum_{\{\Delta_{o}\}}\left( \mathcal{P}^{(1,0)}\mathcal{F}^{(0)} + \mathcal{P}^{(1,1)}\,\mathcal{F}^{(1)}\right)\, + \left(\,\mathcal{P}^{(1,1)}\,\mathcal{F}^{(1)}\right)\,{\color{blue}\log(r)} \nonumber\\
\langle pppp\rangle^{(2)} &=\, \sum_{\{\Delta_{o}\}}\,\left( \mathcal{P}^{(2,0)}\,\mathcal{F}^{(0)} + \mathcal{P}^{(2,1)}\,\mathcal{F}^{(1)} \, + \, \mathcal{P}^{(2,2)}\,\mathcal{F}^{(2)} \right)\nonumber\\
&\qquad\quad+ \left(\mathcal{P}^{(2,1)}\,\mathcal{F}^{(0)}\,+\, 2\,\mathcal{P}^{(2,2)}\,\mathcal{F}^{(1)}\right)\,{\color{blue}\log(r)} \nonumber\\
&\qquad\quad+  \left(\mathcal{P}^{(2,2)}\,\mathcal{F}^{(0)}\right)\,{\color{blue}\log(r)^{2}}
\end{align}
The general definition of the sum rules in terms of the OPE data can be read off from the following generating function (originally defined in equation (21) of \cite{NonCompactTailoring} ):
\beq
\, \sum_{(\Delta)\,\in\, \textit{Super-Class}\,\{\Delta_{0}\}}   \, C^{2}_{pp\Delta} \,r^{\Delta}  \, = \,r^{\Delta_{0}}\sum_{a=0}^{\infty}\,\left(g^{2}\right)^{a} \, \sum_{b=0}^{a}\, \log(r)^{b}\, \mathcal{P}^{(a,b)}_{\{\Delta_{0}\}}\,
\eeq
with the OPE data $C$ and $\Delta$ expanded as in \eqref{eq:weakOPEdata}.

With these definitions we can finally express the loop corrections $\langle pppp \rangle^{(a)}$ in terms of the OPE data and organized according to the $\log$ exponents as:
\beq
\langle pppp \rangle^{(a)}  = \sum_{\{\Delta_{0}\}}\sum_{b=0}^{a} \, \log(r)^{b}\,\sum_{k=b}^{a} \binom{k}{b}\,\mathcal{P}^{(a,k)}_{\{\Delta_{0}\}} \, \mathcal{F}^{(a-k)}_{\{\Delta_{0}\}}
\eeq

\section{Prediction nine loop complete ope data of check}\label{app:LargeNineOPE}
In this appendix we provide the primary and super-primary sum rules that could not fit in tables \ref{tab:P90super} and \ref{tab:Final9sums}.

The missing primary sum rule in table \ref{tab:P90super}  is given in terms of the following super-primary sum rules:
\bba\label{eq:RelationHardSums9}
\mathsf{P}^{(9,0)}_{\{26,2\}}&= \mathcal{P}^{(9,0)}_{\{22,2,9,9\}}+\mathcal{P}^{(9,0)}_{\{24,0,10,10\}}-\mathcal{P}^{(9,0)}_{\{24,0,11,9\}}-\frac{526199}{350175}\,\mathcal{P}^{(9,0)}_{\{24,0,11,11\}}+\mathcal{P}^{(9,0)}_{\{26,2,11,11\}}\nonumber\\
&\qquad-\frac{27}{2102500}\,\mathcal{P}^{(9,1)}_{\{24,0,11,11\}}+\frac{2191}{1524312500}\,\mathcal{P}^{(9,2)}_{\{24,0,11,11\}} -\frac{59373}{276281640625}\,\mathcal{P}^{(9,3)}_{\{24,0,11,11\}}\nonumber\\
&\qquad+\frac{8059143}{200304189453125}\, \mathcal{P}^{(9,4)}_{\{24,0,11,11\}}-\frac{263012022}{29044107470703125}\,\mathcal{P}^{(9,5)}_{\{24,0,11,11\}}\nonumber\\
&\qquad+\frac{50158623078}{21056977916259765625}\,\mathcal{P}^{9,6}_{\{24,0,11,11\}}-\frac{10951243453584}{15266308989288330078125}\,\mathcal{P}^{(9,7)}_{\{24,0,11,11\}}\nonumber\\
&\qquad+\frac{2694497074981488}{11068074017234039306640625}\,\mathcal{P}^{(9,8)}_{\{24,0,11,11\}}\nonumber\\
&\qquad\;\;-\frac{737870670163918368}{8024353662494678497314453125}\,\mathcal{P}^{(9,9)}_{\{24,0,11,11\}}
\end{align}
These super sum rules on the right hand side can be computed by identifying the super-multiplets (Bethe solutions) and using the BKV formula. These are our results:
\bba\label{eq:HardSuperSum9}
\mathcal{P}^{(9,0)}_{\{22,2,9,9\}}&= -\frac{3880272703256846017844314261170552101}{100890752343750000000000000}\,, \quad &\mathcal{P}^{(9,1)}_{\{24,0,11,11\}}&= 4175618304\,,\nonumber\\
\mathcal{P}^{(9,0)}_{\{24,0,10,10\}}&= -\frac{1561516641881791110594032737679}{17968750000000000000}\,,\quad &\mathcal{P}^{(9,2)}_{\{24,0,11,11\}}&= -3028140544\,,\nonumber\\
\mathcal{P}^{(9,0)}_{\{24,0,11,9\}}&= -41148586528\,,\quad &\mathcal{P}^{(9,3)}_{\{24,0,11,11\}}&= 1382002336\,,\nonumber\\
\mathcal{P}^{(9,0)}_{\{24,0,11,11\}}&= -2755264512\,,\quad &\mathcal{P}^{(9,4)}_{\{24,0,11,11\}}&= -438371648\,,\nonumber\\
\mathcal{P}^{(9,0)}_{\{26,2,11,11\}}&= -\frac{24206493839806735477486708172518503089}{445073004000599941716074496}\,,\quad &\mathcal{P}^{(9,5)}_{\{24,0,11,11\}}&= \frac{1506986624}{15}\,,\nonumber\\
\, &\,\quad &\mathcal{P}^{(9,6)}_{\{24,0,11,11\}} &= -\frac{250252288}{15}\,,\nonumber\\
\, &\,\quad &\mathcal{P}^{(9,7)}_{\{24,0,11,11\}} &= \frac{29139968}{15}\,,\nonumber\\
\, &\,\quad &\mathcal{P}^{(9,8)}_{\{24,0,11,11\}} &= -\frac{45540352}{315}\,,\nonumber\\
\, &\,\quad &\mathcal{P}^{(9,9)}_{\{24,0,11,11\}} &= \frac{4978688}{945}\,
\end{align}
Plugging these sum rules \eqref{eq:HardSuperSum9} into  \eqref{eq:RelationHardSums9} we obtain the primary sum:
\beq
\mathsf{P}^{(9,0)}_{\{26,2\}}= -\frac{13565855042891885605834502859803364601}{100890752343750000000000000}
\eeq
which perfectly matches the sum rule listed in table \ref{tab:predictionP90} obtained from the nine-loop hexagonalization predicition of the \textit{asymptotic} four point function.
\section{Solving leading order $\mathfrak{sl}(2)$ Bethe equations}\label{app:BetheSolutions}
This sector is described by gauge invariant operators with field content $\text{Tr}\left(\mathcal{D}^{s}Z^{L}\right)$. In the spin chain description we have $s$ magnons moving in a cyclic and periodic chain made out of $L$ scalars. The correspondent leading order ($g=0$) Bethe equations  are:
\beq\label{eq:sl2leading}
\left(\frac{u_{m}+\frac{i}{2}}{u_{m}-\frac{i}{2}}\right)^{L}=\prod_{n\neq m }^{s} \frac{u_{m}-u_{n}-i}{u_{m}-u_{n}+i}
\eeq
accompanied with the zero-momentum condition imposed by cyclicity of the trace:
\beq\label{eq:sl2zeromomentum}
\prod_{m=1}^{s} \frac{u_{m}+\frac{i}{2}}{u_{m}-\frac{i}{2}} =1
\eeq
An important simplification in this $\mathfrak{sl}(2)$ sector is that their Bethe equations only admit real Bethe roots.  This is advantegeous and allows us to efficiently solve them, at least numerically, by first transforming the equations \eqref{eq:sl2leading} to a manifestly real form. This can be achieved by taking their log-form and using a trigonometic identity that leads to:
\beq\label{eq:sl2log}
2\,L\,\arctan\left(2\,u_{m}\right) +\sum_{n=1}^{s}\,2\,\arctan\left(u_{m}-u_{n}\right) = 2\,\pi\,Q_{m}
\eeq
The integer or semi-integer $Q_{m}$ is a mode number that specifies the branch of the correspondent Bethe root $u_{m}$. All mode numbers should be taken distinct to guaranteed all Bethe roots are different. Besides, considering the limits of the $\arctan$ we can determine  the bounds of the mode numbers. For even $s$ these are:
\beq\label{eq:Qbound}
| Q_{m} | \leq \frac{L+s-3}{2}
\eeq
While the zero-momentum condition \eqref{eq:sl2zeromomentum} translates into the following condition for the mode numbers:
\beq\label{eq:Qsum}
\sum_{m=1}^{s} Q_{m} =0
\eeq
every set of integers or semi-integers $\{Q_{1},\cdots,Q_{s}\}$ satisfying \eqref{eq:Qbound} and \eqref{eq:Qsum}, specifies a solution of Bethe the $\mathfrak{sl}(2)$ equations. In order to find the correspondent Bethe roots  we can plug in this set of mode numbers on the right-hand side of \eqref{eq:sl2log} and solve. This is better accomplished by a numerical search of the roots which can be effeciently performed using $\mathsf{FindRoot}$ in $\mathsf{Mathematica}$. 

An explicit realization of the procedure described in this section can be found in a $\mathsf{Mathematica}$ notebook available in \cite{NorditaPedroLectures}. Using this algorithm it takes few minutes to find all $\mathfrak{sl}(2)$ Bethe solutions in table \ref{tab:NumberRoots}.

\end{document}